\documentclass[
aps,
prx,
superscriptaddress,
amsmath,amssymb,
floatfix,
]{revtex4-2}

\usepackage{graphicx}
\usepackage{dcolumn}
\usepackage{bm}

\usepackage{xcolor}
\definecolor{linkblue}{RGB}{0, 70, 140}
\definecolor{nicepink}{RGB}{200, 70, 120}
\definecolor{brightpink}{RGB}{220, 50, 120}

\usepackage[colorlinks=true,
            linkcolor=blue,
            citecolor=red,
            urlcolor=blue]{hyperref}

\begin{document}


\title{Collisionless Phase Mixing Mimics Diffusive Transport in Radiation Belt Observations}

\author{Adnane Osmane}\email{adnane.osmane@helsinki.fi}
\affiliation{Department of Physics, University of Helsinki, Helsinki, Finland}
\author{Xin An}
\affiliation{Department of Earth, Planetary, and Space Sciences, University of California, Los Angeles, CA, USA}
\author{Anton Artemyev}
\affiliation{Department of Earth, Planetary, and Space Sciences, University of California, Los Angeles, CA, USA}
\affiliation{Department of Physics, University of Texas, Arlington, TX, USA}
\author{Oliver Allanson}
\affiliation{Space Environment and Radio Engineering, Electronic, Electrical and Systems Engineering, School of Engineering, University of Birmingham, Birmingham, UK}
 \affiliation{Faculty of Environment, Science and Economy, University of Exeter, Exeter, UK}
\author{Jay Albert}
\affiliation{Air Force Research Laboratory Space Vehicles Directorate, Kirtland, NM, USA}
\author{Miroslav Hanzelka}
\affiliation{Institute of Atmospheric Physics of the Czech Academy of Sciences, Prague, Czech Republic}

\date{\today}

\begin{abstract}
Since the dawn of the space age, observations of energetic particles in planetary radiation belts have been interpreted within a diffusive transport framework, even though the dominant processes that populate and deplete these belts—such as injections and moon-driven absorption—produce particle distributions that are highly structured and spatially localized. This exposes a fundamental question: how can coherent phase-space structures evolving under collisionless dynamics give rise to observational signatures that appear consistent with diffusion-based transport? Here we show that diffusion-like behaviour inferred from radiation-belt observations can arise solely from an observational phase-mixing effect, independent of stochastic wave–particle transport. As orbiting spacecraft sweep across neighbouring drift shells while trapped particles undergo electromagnetic drifts, measurements inevitably sample regions with slightly different drift frequencies. This converts spatially localized drift-phase structures into rapidly decorrelating temporal signals, making them observationally indistinguishable from those produced by stochastic wave–particle processes. We derive the associated correlation function analytically and show that the effective lifetime of these structures is only a few drift periods. Consequently, even highly localised injections rapidly lose coherence, preventing spacecraft from resolving fine-scale structure in the distribution function. These results show that collisionless dynamics can produce observational signatures that mimic diffusive transport on timescales shorter than those expected from radial transport, limiting the inference of particle acceleration processes and leading to biased transport rates and inaccurate long-term flux predictions. This calls for a reassessment of diffusion-based interpretations inferred from sparse in-situ measurements of radiation belts at Earth, across the solar system, and in the recently discovered radiation belts of ultra-cool brown dwarfs. 
\end{abstract}

\maketitle


\section{Introduction}\label{sec1}
Radiation belts are among the most efficient natural particle accelerators in the Universe and are a ubiquitous feature of magnetized planetary systems. First identified around Earth during the early space age \citep{VanAllen}, they extend over more than ten planetary radii and accelerate electrons to energies of about 15 MeV at Earth and several hundred MeV at Jupiter. Subsequent spacecraft observations established that radiation belts are common throughout the Solar System, including at the giant planets \citep{Bolton_Jupiter, Kollmann_Saturn, Allen_Uranus}. Even comparatively small magnetized bodies can sustain trapped energetic particle populations: Ganymede, for example, hosts a compact radiation belt generated by its intrinsic magnetic field \citep{Clark_Ganymede, Kollman_Ganymede}. More recently, ultra-cool, rapidly rotating brown dwarfs with kilo-gauss magnetic fields have revealed even more extreme radiation-belt environments \citep{Kao23_BD, Climent_BD}.

Across this wide range of systems, radiation belts provide natural laboratories for studying the acceleration, transport, and loss of energetic particles in strongly magnetized, weakly collisional plasmas. Yet despite decades of observations and modelling, a fundamental question remains unresolved: how to reconcile the strongly structured, often highly localised dynamics that populate these systems with the predominantly diffusive behaviour inferred from spacecraft measurements. Understanding this connection is central to the interpretation of radiation-belt observations and, more broadly, to the physics of transport in magnetised plasmas. This apparent tension is rooted in the historical development of radiation-belt transport models.

At the time of their discovery, two central difficulties immediately emerged. Radial transport was understood to arise from violations of the third adiabatic invariant (the magnetic flux) \cite{Kellogg59, Davis_Chang_1962, northrop1963adiabatic, Birmingham67, Cary09}, yet the fluctuating electric and magnetic fields responsible for these violations were essentially unconstrained, as only ground-based magnetometer measurements were available. At the same time, computational resources were insufficient to resolve particle drift motion and radial transport over the long timescales relevant to radiation-belt dynamics. Explaining the presence and persistence of energetic particles therefore required a transport framework that depended only weakly on the detailed structure of the electromagnetic fluctuations while remaining tractable for numerical implementation.

In response to these challenges, early theoretical work by Parker \cite{Parker60} and F\"althammar \cite{Falthammar65} led independently to a Fokker–Planck description in which the complexity of the fluctuating fields is reduced to a single radial diffusion coefficient. In this framework, inward diffusion, combined with conservation of the first adiabatic invariant—the magnetic moment—naturally produces betatron acceleration and the relativistic energies observed in radiation belts. Radial diffusion has since remained the primary framework for describing radial transport in global radiation-belt models and in case studies of geomagnetic events \citep{Brizard_Chan, Shprits_2006, Shprits_2009, Ozeke12, Thorne13, Reeves13, Weichao_2013, Baker14, Ozeke14, Jaynes15, Mann_2016, Cunningham16, Ali15, Ali_radial_diffusion, Jaynes18, Olifer19, Weichao_2019, Lejosne20, Staples_2022, George_losses_2022, Lejosne22, George_Vlasiator, Olifer24, Sarris2024, Bentley_2024, Ozeke25}.

However, recent observations show that some of the most striking radiation-belt dynamics occur under conditions that depart from the assumptions underlying radial diffusion \cite{osmane2023radial}. High-resolution measurements at Earth, Jupiter, and Saturn reveal abrupt injections, sharp spatial gradients, and structured loss regions, indicating that particle distributions can remain strongly structured in magnetic local time and radial distance \cite{Thomsen1977, vanAllen1980Mimas, 1993AdSpR..13j.221S, Roussos07, Turner17, Olifer18, Kim2021_injection, Kim2023_injection, Senlin_injection, Kamaletdinov25b}. Such features arise from processes that are inherently localized, including bursty injections and moon-magnetosphere interactions. These processes operate in different combinations across Earth’s radiation belts and the gas giants, producing spatially inhomogeneous sources and losses that evolve under largely collisionless dynamics \cite{Wentworth_lifetimes_trapped_Coulomb, Walt_Nuclear_Explosion_Coulomb, Abel_Thorne_1998, Selesnick_2012}. Yet despite the coexistence of intermittent injections and coherent, spatially structured loss processes, spacecraft observations often exhibit signatures that are commonly interpreted in terms of radial diffusion.

This apparent tension reflects a more fundamental limitation in the interpretation of spacecraft measurements. Because observations are obtained along trajectories that sweep across neighbouring drift shells, the measured signal does not directly correspond to the temporal evolution of a fixed particle population. Instead, it represents a convolution of spatial structure and sampling geometry. As a result, it is not \textit{a priori} clear to what extent temporal variability in spacecraft data reflects intrinsic dynamical processes, such as diffusion, or arises from the way spatially structured populations are sampled.

In this work, we show that even in the absence of diffusion or dissipation, the combined effects of differential drift and spacecraft motion give rise to an intrinsic observational phase-mixing process. Because particles on adjacent drift shells rotate with slightly different angular frequencies, spatially localised structures progressively lose phase coherence as they evolve. When sampled by an orbiting spacecraft, successive measurements correspond to regions characterised by different drift phases rather than a single coherent structure, so that the resulting temporal signal undergoes rapid decorrelation even though the underlying particle dynamics remain strictly ballistic and conserve phase-space density, as required by Liouville’s theorem. We demonstrate that, under such conditions, the effective observational lifetime of spatially localised particle populations is limited to only a few drift periods, making fine-scale spatial and temporal structure in the distribution function difficult to resolve in spacecraft data.

This mechanism has important implications for the interpretation of radiation-belt observations. Ballistic (i.e., adiabatic, non-diffusive) phase mixing, namely the progressive dephasing of particles drifting at different azimuthal frequencies, can produce temporal signatures that closely resemble those expected from radial diffusion, implying that diffusion-like behaviour inferred from spacecraft measurements may not uniquely diagnose stochastic transport driven by electromagnetic fluctuations. Instead, part of the apparent diffusive evolution of radiation-belt populations may arise from observational filtering imposed by spacecraft trajectories. These results place fundamental limits on the recoverability of spatially localised dynamics in radiation belts and have broader implications for interpreting particle transport in planetary and substellar magnetospheres.

To quantify this phase-mixing mechanism, the remainder of this paper is organised as follows. In Section~\ref{sec2} we introduce the theoretical framework used to quantify the ballistic evolution of drift-phase structures. Without loss of generality, we derive analytical solutions for a particle population that is spatially localised in magnetic local time and drift shell, representative of impulsive injection events and other drift-phase structures. The same formalism can also be applied to drift echoes, which correspond to linear solutions of the radial transport equation. Observations indicate that injection morphologies span a range of azimuthal extents, from relatively localised structures to enhancements occupying several hours of magnetic local time (e.g., \citet{Kavanagh07, Gabrielse19}). In the present work, we consider an initially MLT-localised population as a representative example of a drift-phase structure whose subsequent observational evolution can be treated analytically. 

Using these solutions, we first compute the signal measured by a spacecraft sweeping across the structure with velocity $V_s$, and subsequently derive the corresponding two-time correlation function analytically. The autocorrelation function compares measurements obtained at different spatial locations in the radiation belts. From an observational perspective, it can therefore be interpreted as quantifying the degree to which two points measurements sampling neighbouring drift shells remain coherent in time. This defines an observational coherence timescale associated with the progressive loss of recognisable structure due to drift-phase shearing and spacecraft sweeping, rather than physical dissipation (i.e., irreversible entropy-increasing processes such as collisional or diffusive transport). In Section~\ref{sec3}, we use the correlation function to extract effective lifetimes for both the drift-averaged component and the drift-structured spatial and temporal fluctuations of the distribution function. Section~\ref{sec4} then examines the limitations of the model and the implications of our results for the interpretation and modelling of radiation belts at Earth, at the giant planets, and in ultra-cool brown dwarfs. Finally, Section~\ref{Conclusion} summarises the main results, discusses their implications for the interpretation of spacecraft observations, and outlines directions for future work.

\section{Methodology}\label{sec2}
In this section we solve the bounce-averaged drift-kinetic equation for particles trapped in a static, spatially inhomogeneous magnetic field with dipolar topology. Because the field is time independent, all three adiabatic invariants are conserved. Our objective is to determine the ballistic evolution of the bounce-averaged distribution function for an injection localised in magnetic local time (MLT) and drift shell. We emphasise that the assumption of MLT localisation is adopted for analytical convenience and should not be interpreted as implying that all injections are strongly localised. Observations indicate that injection regions may range from highly localised structures to broad, azimuthally expanding populations occupying several hours of MLT \cite{Kavanagh07, Gabrielse19}. In the present formulation this variability is captured by the concentration parameter $\kappa$ defined hereafter, allowing the description of drift-phase structures ranging from MLT-homogeneous populations ($\kappa=0$) to strongly localised enhancements ($\kappa\gg 1$).

We then model how such an MLT-localised structure is sampled by a spacecraft traversing the radiation belts. As the spacecraft moves radially, it intersects neighbouring drift shells associated with slightly different azimuthal drift frequencies. The resulting time series therefore reflects both the intrinsic evolution of the injection and the sampling of adjacent drift shells along the trajectory.

Measurements are treated as instantaneous point samples. Under typical radiation-belt measurement cadences, the radial displacement during the accumulation time is small compared with both the radial width of the injection and the characteristic $L$-scale over which the drift frequency $\langle\dot\varphi\rangle_b$ varies. The loss of observable coherence is therefore attributed primarily to drift-phase shear combined with sampling geometry, rather than to instrumental averaging.

Combining the ballistic solution with this sampling model allows us to determine the effective observational lifetime of MLT-localised injections. The derivation of the bounce-averaged drift-kinetic equation, magnetic field model, guiding-centre dynamics, and bounce-averaging procedure is provided in Appendix~\ref{_Appendix_A}. While many of the quantities used here appear in the literature \cite{schulz1974, Roederer1970, Lyons, osmane2023radial}, we present a consolidated derivation to make the treatment self-contained and to explicitly delineate the limits of validity of the model.

\subsection{Bounce-averaged drift kinetic equation}
We consider the bounce-averaged distribution function $
F=F(\varphi, L, \alpha_{\rm eq}, t)$,  which depends on the azimuthal angle \(\varphi\), the drift shell \(L\), the equatorial pitch angle \(\alpha_{\rm eq}\), and time \(t\). After averaging over the fast bounce motion, its evolution is governed exactly by the bounce-averaged drift-kinetic equation
\begin{equation}
\label{_EQ:Final_Bounce_averaged_Drift_Kinetic_}
\frac{\partial F}{\partial t}
+
\left\langle
\frac{v_\varphi(\lambda)}{r(\lambda)\cos\lambda}
\right\rangle_b
\frac{\partial F}{\partial \varphi}
=0,
\end{equation}
where \(\langle\cdots\rangle_b\) denotes the average along the bounce orbit. The coefficient multiplying \(\partial F/\partial\varphi\) is the bounce-averaged azimuthal drift frequency,
\begin{equation}
\langle \dot{\varphi}\rangle_b
=
\left\langle
\frac{v_\varphi(\lambda)}{r(\lambda)\cos\lambda}
\right\rangle_b.
\end{equation}

Here \(\lambda\) is the magnetic latitude, \(\lambda_m\) is the mirror latitude, $v_\varphi(\lambda)$ is the azimuthal drift speed, and \(s\) denotes the arc length measured along the magnetic field line. For a dipolar magnetic field the geometry is described by
\[
r(\lambda)=L R_E\cos^2\lambda,
\qquad
\frac{ds}{d\lambda}=L R_E\cos\lambda\,D(\lambda),
\qquad
D(\lambda)=\sqrt{1+3\sin^2\lambda},
\]
with \(R_E\) the Earth's radius. Using these relations, the bounce-averaged drift frequency can be written explicitly as
\begin{equation}
\begin{aligned}
\label{_Eq:Final_bounce_averaged_drift}
\langle \dot{\varphi}\rangle_b
&=
\frac{4}{\tau_b}\int_0^{\lambda_m}
\frac{v_\varphi(\lambda)}{r(\lambda)\cos\lambda}\,
\frac{1}{|v_\parallel(\lambda)|}\,
\frac{ds}{d\lambda}\,d\lambda  \\
&=
\frac{4}{\tau_b}\int_0^{\lambda_m}
v_\varphi(\lambda)\,
\frac{D(\lambda)}{\cos^2\lambda}\,
\frac{1}{|v_\parallel(\lambda)|}\,d\lambda,
\end{aligned}
\end{equation}
where \(\tau_b\) is the bounce period,
\begin{equation}
\label{_EQ:Final_bounce_period}
\tau_b
=
4\int_0^{\lambda_m}
\frac{1}{|v_\parallel(\lambda)|}\,
\frac{ds}{d\lambda}\,d\lambda
=
4L R_E\int_0^{\lambda_m}
\frac{\cos\lambda\,D(\lambda)}{|v_\parallel(\lambda)|}\,d\lambda.
\end{equation}

The azimuthal drift velocity \(v_\varphi\), which contains both the grad-\(B\) and curvature drifts and enters directly into Equation~(\ref{_Eq:Final_bounce_averaged_drift}), is
\begin{equation}
\label{_EQ:Final_azimuthal_drfit}
v_\varphi(L, \lambda)
=
-\frac{3}{q_s}\left[
\frac{m_s\gamma v_\parallel^2 L^2}{B_E R_E}\,
\frac{\cos^5\lambda\left(1+\sin^2\lambda\right)}{D^4(\lambda)}
+
\frac{\mu_s}{\gamma L R_E}\,
\frac{1+\sin^2\lambda}{\cos\lambda\,D^2(\lambda)}
\right].
\end{equation}
Here \(m_s\) and \(q_s\) are respectively the mass and charge of particle species \(s\), \(\gamma\) is the relativistic Lorentz factor, and \(\mu_s\) is the first adiabatic invariant. In the relativistic formulation adopted here,
\[
\mu_s=\frac{p_\perp^2}{2m_s B},
\]
where \(p_\perp=m_s\gamma v_\perp\) is the relativistic perpendicular momentum. The magnetic-field magnitude in a dipole is
\begin{equation}
B(L,\lambda)=\frac{B_E}{L^3}
\frac{\sqrt{1+3\sin^2\lambda}}{\cos^6\lambda}.
\label{_eq:dipole_magnitude_B_L_lambda}
\end{equation}
The constant \(B_E\) denotes the Earth's magnetic field strength at the surface on the magnetic equator, corresponding to \(r=R_E\), \(\lambda=0\), and \(L=1\).

For later use, it is convenient to express the bounce-averaged drift frequency in terms of the kinetic energy \(E_k=m_s c^2(\gamma-1)\) and the equatorial pitch angle \(\alpha_{\rm eq}\). In that form, one obtains
\begin{equation}\label{eq:phidot_b_factored_}
\begin{aligned}
\langle \dot\varphi\rangle_b
&=
-\frac{3 m_s c^2 L}{q_sB_E R_E^2}\,
\frac{\frac{E_k}{m_sc^2}\left(\frac{E_k}{m_s c^2}+2\right)}{\frac{E_k}{m_sc^2}+1}\,
\mathcal{M}(\alpha_{\rm eq}).
\end{aligned}
\end{equation}
The dimensionless function \(\mathcal{M}(\alpha_{\rm eq})\) is defined by
\begin{equation}
\label{eq:M_compact_}
 \mathcal{M}(\alpha_{\rm eq})=
 \frac{
\displaystyle
\int_{0}^{\lambda_m}
\frac{D(\lambda)}{\cos^2\lambda}\;
\frac{\mathcal{V}(\lambda;S_{\rm eq})}
{\sqrt{1-S_{\rm eq}\,D(\lambda)/\cos^6\lambda}}\,d\lambda
}{
\displaystyle
\int_{0}^{\lambda_m}
\cos\lambda\,D(\lambda)\;
\frac{1}{\sqrt{1-S_{\rm eq}\,D(\lambda)/\cos^6\lambda}}\,d\lambda
},
\end{equation}
where
\[
S_{\rm eq}=\sin^2\alpha_{\rm eq}.
\]
The kernel \(\mathcal{V}\) is then given by
\begin{equation}
\mathcal{V}(\lambda;\alpha_{\rm eq})
=
\left(1+\sin^2\lambda\right)\left[
\frac{\cos^5\lambda}{D^4(\lambda)}
+\frac{S_{\rm eq}}{2}\,
\frac{D(\lambda)-2}{\cos\lambda\,D^3(\lambda)}
\right].
\label{eq:V_compact_}
\end{equation}

All of the above equations are derived in Appendix~\ref{_Appendix_A}. The appendix is included mainly for completeness and for readers wishing to verify the intermediate steps; in what follows, we use Equations~(\ref{_EQ:Final_Bounce_averaged_Drift_Kinetic_}), (\ref{_Eq:Final_bounce_averaged_drift}), (\ref{_EQ:Final_azimuthal_drfit}), (\ref{eq:phidot_b_factored_}), (\ref{eq:M_compact_}), and (\ref{eq:V_compact_}) as the starting point for constructing the radial-transport solution \footnote{While exact analytical solutions can be obtained for equatorially trapped particles, approximate expressions are also available for particles confined close to the equatorial plane, where $v \simeq v_\perp$, $\lambda \ll 1$ (so that $\sin\lambda \simeq \lambda$ and $\cos\lambda \simeq 1$). In the general case, however, one must first determine the mirror latitude $\lambda_m$ as a function of the equatorial pitch angle $\alpha_{\rm eq}$, and subsequently evaluate numerically the integrals appearing in Equations~(\ref{_Eq:Final_bounce_averaged_drift}) and (\ref{_EQ:Final_bounce_period}). Tabulated numerical evaluations and accurate analytical fits for the bounce and drift frequencies as functions of pitch angle in a dipole field are available in the literature (e.g., \citet{Hamlin1961}), and may be used to bypass explicit numerical integration.}.

\subsection{Solutions for spatially localised injections and drift phase structures}
We then proceed as follows to solve Equation (\ref{_EQ:Final_Bounce_averaged_Drift_Kinetic_}) with the bounce-drift frequency expressed in terms of Equation (\ref{eq:phidot_b_factored_}). The drift frequency $\langle \dot{\varphi}\rangle_b$  is independent of the phase $\varphi$, so we seek solutions of $F$ in terms of the following Fourier sum: 
\begin{equation}
\label{EQ:Fourier_Sum}
F=\sum_m F_{m}(L, t) e^{im\varphi}, 
\end{equation}
where $F_m$ is a Fourier coefficient of degree $m\in \mathbb{Z}$, with the consequence that $F_m^*=F_{-m}$, since the bounce-averaged distribution function $F$ is a real quantity. Inserting Equation (\ref{EQ:Fourier_Sum}) into Equation (\ref{_EQ:Final_Bounce_averaged_Drift_Kinetic_}), we find the following differential equation for $F_m$: 
\begin{equation}
    \frac{\partial F_m}{\partial t}=-im\langle \dot{\varphi}\rangle_b F_m \Longrightarrow F_m=F_m(t=0)e^{-im\langle \dot{\varphi}\rangle_b t}
\end{equation}
The general solution for Equation (\ref{_EQ:Final_Bounce_averaged_Drift_Kinetic_}) is therefore given by 
\begin{equation}\label{EQ:solution_1}
    F=\sum_mF_m(t=0)e^{im\varphi-im\langle \dot{\varphi}\rangle_b t}. 
\end{equation}
The next step is to complement Equation~(\ref{EQ:solution_1}) with an initial condition that is localised in both MLT and drift shell. To this end, we employ the von~Mises distribution \cite{abramowitz1968handbook} to represent an initial profile confined to a finite range of MLT angles. The von~Mises distribution may be regarded as the circular analogue of a Gaussian and has the satisfying property, for radial transport problems in particular \cite{Osmane25_Radial, Osmane2025}, that it admits a Fourier-series representation:
\begin{eqnarray}\label{Eq:vonMises}
    \mathcal{Q}(\varphi; \kappa)
    &=&
    \frac{e^{\kappa \cos\varphi}}{2\pi I_0(\kappa)}
    =
    \frac{1}{2\pi}\sum_{m}
    \psi_m(\kappa)\, e^{i m \varphi}.
\end{eqnarray}
In Equation (\ref{Eq:vonMises}), the von-Mises distribution $\mathcal{Q}$ is written in terms of the modified Bessel function of the first kind with degree zero $I_0(\kappa)$ \cite{abramowitz1968handbook} and the parameter $\kappa \in [0, \infty )$, which determines how localised in $\varphi$ (or equivalently in magnetic local time, MLT), the distribution is. The coefficient $\psi_m$ is a ratio of the modified Bessel function of degree $m$ to the modified Bessel function of degree $0$: 
\begin{equation}
\psi_m(\kappa) = \frac{I_m(\kappa)}{I_0(\kappa)}.
\end{equation}
The von Mises distribution is $2\pi$-periodic, normalised to unit area, and exhibits an exponential decay away from a single peak. Its width, expressed in radians, decreases with increasing $\kappa$. In this work, we consider values of $\kappa$ in the range $0 \leq \kappa \leq 10$, corresponding to structures localised over angular scales no smaller than approximately $20\%$ of a full drift orbit. Moreover, the modified Bessel function of the first kind satisfies the symmetry $I_{-m}(\kappa) = I_m(\kappa)$, implying $\psi_{-m} = \psi_m$. For $m=0$, $\psi_0 = 1$, while for all $m \neq 0$, $\psi_m < 1$. In the limiting case $\kappa = 0$, one has $\psi_m = 0$ for all $m \neq 0$, so that only the $m=0$ mode contributes.

The localisation in drift shell is represented here by a Gaussian profile centred at $L=b$ with characteristic width $\sigma$:
\begin{equation}\label{EQ:Gaussian_L}
    \mathcal{G}(L)
    =
    n_s \exp\left[
    -\left(\frac{L-b}{2\sigma}\right)^2
    \right].
\end{equation}
The constant $n_s$ is a normalisation factor obtained by integrating Equation~(\ref{EQ:Gaussian_L}) over the physically relevant domain in $L$, for example from the inner boundary of the radiation belts at $L\simeq 1$ \cite{Walton2024} to an outer limit $L=L_{\max}$ corresponding to drift shells intersecting the magnetopause. For the purposes of the present analysis, however, the explicit form of $n_s$ is not required, as we are concerned solely with the ballistic evolution of the distribution function and transport across drift-shells is not considered. If particle losses or the absolute particle content were to be included, the precise normalisation would become necessary. We therefore omit its explicit expression here.

\begin{figure}[ht]
\centering
\includegraphics[width=1\textwidth]{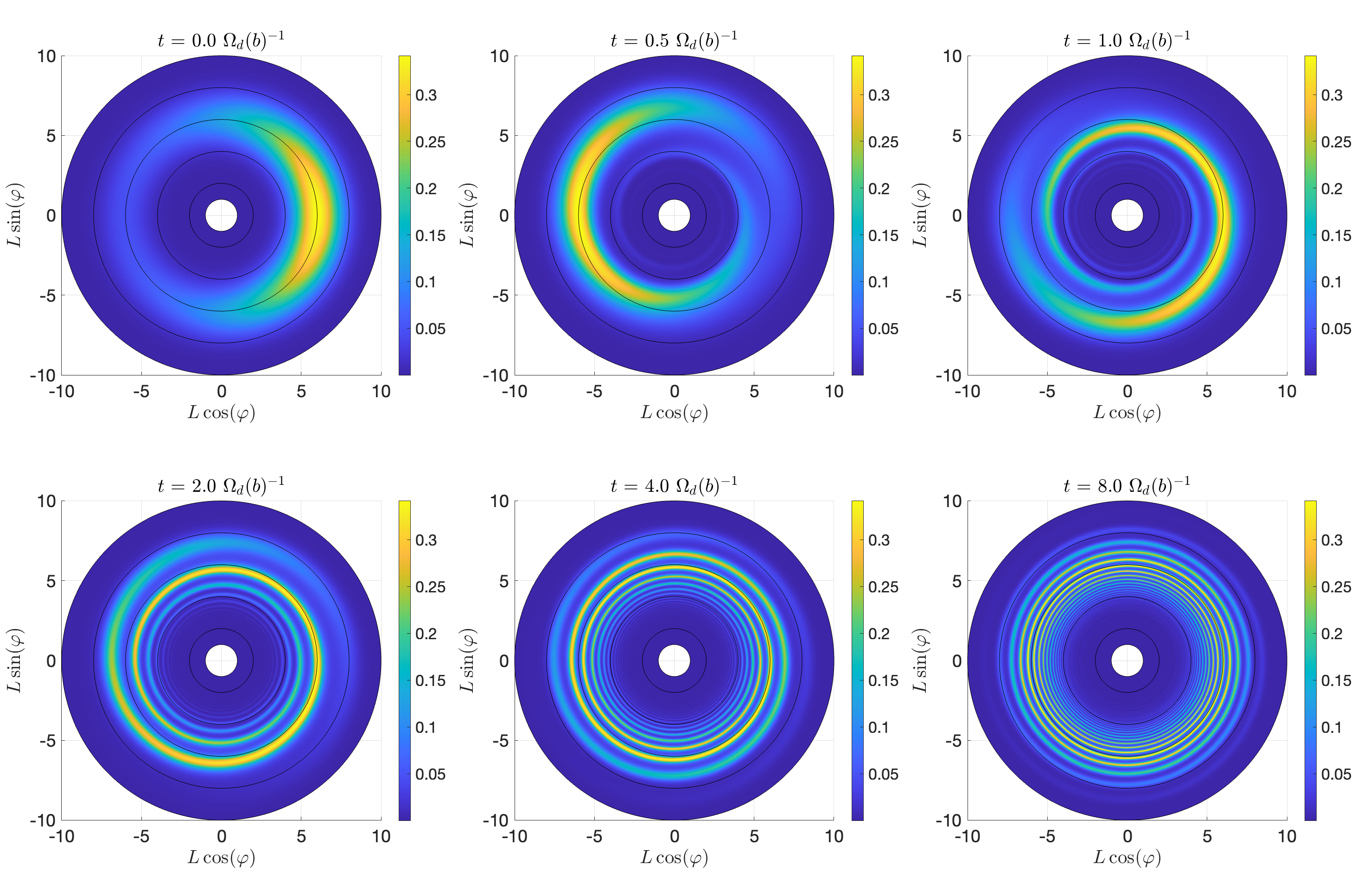}
\caption{Evolution of the ballistic solution given by Equation~(\ref{EQ:final_solution_MLT_injection}) for a drift-shell-centred injection with $b=6$ and $\sigma=1.5$, using $\kappa=1$. At $t=0$ the MLT localisation spans approximately half of the MLT range. The results are shown for equatorially trapped particles ($\alpha_{\rm eq}=90^\circ$) with an energy of $1\,\mathrm{MeV}$. The six panels illustrate the temporal evolution relative to the drift period at $L=6$. The top row (from left to right) corresponds to $t=0$, $t=\tfrac{1}{2}$ drift period, and $t=1$ drift period. The bottom row (from left to right) shows $t=2$, $t=4$, and $t=8$ drift periods. The evolution reflects the ballistic azimuthal phase translation governed by the bounce-averaged drift motion along separate drift-shells.}\label{fig1}
\end{figure}

\begin{figure}[ht]
\centering
\includegraphics[width=1\textwidth]{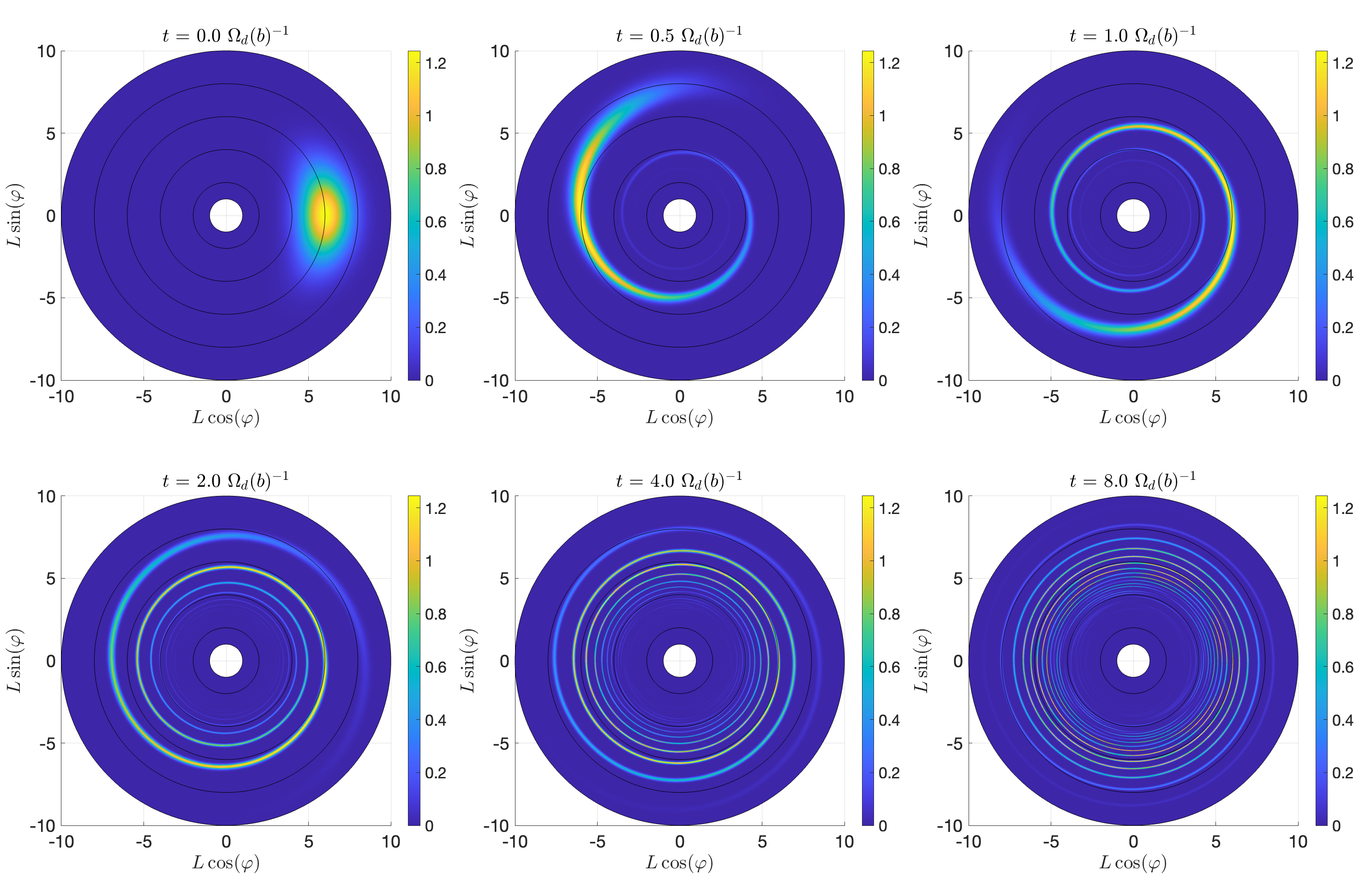}
\caption{Same as Figure (\ref{fig1}), but with $\kappa=10$, corresponding to a much narrower MLT localisation at $t=0$. All other parameters are identical ($b=6$, $\sigma=1.5$, $\alpha_{\rm eq}=90^\circ$, $\mathcal{E}=1\,\mathrm{MeV}$), and the six panels show the evolution at $t=0$, $\tfrac{1}{2}$, $1$, $2$, $4$, and $8$ drift periods at $L=6$.}\label{fig2}
\end{figure}

We therefore impose the initial condition at $t=0$,
\begin{equation}
F(0,\varphi,L)=\mathcal{Q}(\varphi)\,\mathcal{G}(L),
\end{equation}
so that the corresponding Fourier coefficients satisfy $F_m(0,L)=\frac{\mathcal{G}(L)\,\psi_m(\kappa)}{2\pi}$. The full solution describing an MLT- and drift-shell-localised injection is then given by
\begin{equation}\label{EQ:final_solution_MLT_injection}
    F(\varphi,L,t)
    =
    \frac{n_s}{2\pi}
    \exp\left[
    -\left(\frac{L-b}{2\sigma}\right)^2
    \right]
    \sum_{m}
    \psi_m(\kappa)\,
    e^{i m \varphi - i m \langle \dot{\varphi}\rangle_b t} \ .
\end{equation}
The real part of Equation~(\ref{EQ:final_solution_MLT_injection}) describes the ballistic evolution of an injection for particles with arbitrary equatorial pitch angles $\alpha_{\rm eq}$ and energy $\mathcal{E}$. Representative examples of such solutions are shown in Figures~(\ref{fig1}) and ~(\ref{fig2}) to illustrate the progressive shearing of the initial injection under purely ballistic evolution. Although the azimuthal profile on each individual drift shell undergoes a rigid phase translation, the $L$-dependence of the bounce-averaged drift frequency $\langle \dot{\varphi} \rangle_b$ causes neighbouring drift shells to rotate at slightly different rates. This differential rotation leads to a systematic azimuthal shearing of the initially localised structure. 

As time increases, the injection is progressively stretched into increasingly narrow azimuthal stripes. These filaments become thinner and more tightly wound with each successive drift period, reflecting the cumulative phase separation between adjacent drift shells. The effect is more pronounced for the more strongly localised case ($\kappa=10$), where the initially sharper MLT profile generates finer-scale structure as the shearing proceeds. 

\subsection{Mapping phase-space structures to satellite trajectories}
In this section we quantify what a satellite measures as it sweeps across different drift shells. The spacecraft sampling can be modelled as a radial sweep anywhere in the equatorial plane, and we therefore neglect the dependence of the sampling on magnetic latitude $\lambda$. This simplification applies only to the measurement geometry and not to the particle dynamics themselves. The solutions derived above remain valid for arbitrary equatorial pitch angle $\alpha_{\rm eq}$, since the effects of mirror motion enter through the bounce-averaged quantities, such as $\langle \dot{\varphi}\rangle_b$ and $\tau_b$. In this sense, the full pitch-angle dependence of the distribution is retained, even though latitudinal sampling effects are omitted. The more general problem, in which a spacecraft simultaneously samples variations in both drift shell and magnetic latitude, is deferred to future work. Neglecting the latitudinal excursion is nevertheless appropriate for comparison with experiments such as those onboard the Van Allen Probes \cite{Mauk2013}, which remain within approximately $\pm10$ degrees of the magnetic equator for most of their orbits.

In what follows we compute two observational diagnostics derived from the spacecraft time series. The first is the measured signal itself,
\begin{equation}\label{eq:St_def}
S(t)=F\!\big(\varphi_s(t),L_s(t),t\big),
\end{equation}
which quantifies the instantaneous amplitude of the injection sampled along the spacecraft trajectory in the equatorial plane at the location $(\varphi_s(t),L_s(t))$.

The second diagnostic is the finite-window autocorrelation of the measured time series \cite{BendatPiersol2010},
\[
R(\tau;T)=\frac{1}{T-\tau}\int_{0}^{T-\tau} S(t)\,S^\ast(t+\tau)\,dt,
\]
together with its normalised form $C(\tau;T)=R(\tau;T)/R(0;T)$. This quantity measures the degree to which the signal at time $t$ remains statistically correlated with the signal at a later time $t+\tau$, within an observation window of duration $T$. 

If the spacecraft were stationary, so that $(\varphi_s,L_s)$ remained fixed, the correlation function would correspond to a single-point, two-time correlation of the distribution function at a given location in phase space. In that case the decay of $C(\tau;T)$ would directly reflect the intrinsic temporal evolution of the drift-phase structure at that location.

In contrast, a real spacecraft moves through the radiation belts and therefore samples different drift shells as time progresses. In the present sweeping model the lag $\tau$ corresponds not only to a temporal separation but also to a radial displacement $\Delta L = V_s \tau$, since the spacecraft crosses neighbouring drift shells while particles simultaneously undergo azimuthal drift. Consequently, the autocorrelation effectively compares measurements obtained at different spatial locations in phase space. From an observational perspective, the correlation function may therefore be interpreted as quantifying the degree to which probes sampling neighbouring drift shells, such as for the Van Allen Probes \cite{Mauk2013} or the Arase mission \cite{Miyoshi2018}, can still observe the same drift-phase structure after a time delay $\tau$.

The window duration $T$ denotes the total time interval over which the correlation is evaluated and therefore determines how much of the injected structure is included in the analysis. In practice, $T$ may be given a direct observational interpretation. If the lag $\tau$ represents the time separation between two spacecraft sampling the radiation belts on similar orbits, then the window duration can be naturally associated with a characteristic orbital timescale. For instance, choosing $T$ comparable to one orbital period corresponds to evaluating the correlation over a full pass through the radiation belts, during which the spacecraft repeatedly samples neighbouring drift shells as it sweeps radially. 

The associated normalised autocorrelation $C(\tau;T)$ therefore quantifies the persistence of phase coherence of the injection as sampled along the spacecraft trajectory, and defines an observational coherence time that reflects the loss of recognisable structure due to drift-phase shear and spacecraft sweeping, rather than physical dissipation. The associated decorrelation time, defined for example by the lag at which $|C(\tau;T)|$ falls below a prescribed threshold (e.g.\ $1/e$), characterises the timescale over which the injection remains recognisable in the measured time series.

\subsection{Signal measured by an orbiting spacecraft}\label{sec:sweeping_highm}

We model the spacecraft measurement as an instantaneous sample of the bounce-averaged distribution along its trajectory in $(L,\varphi)$. In this subsection we neglect the explicit azimuthal motion of the spacecraft and treat its phase as approximately constant over the interval of interest, $\varphi_s(t)\simeq \varphi_0$. This approximation is justified because the phase decorrelation we describe is governed by the much faster magnetic drift of trapped particles who orbit the Earth on timescales of minutes to a few hours, while the azimuthal motion of the spacecraft is comparatively slow and does not contribute significantly to the evolution of the measured phase. Instead, phase mixing is shown below to arise from the radial motion of the spacecraft across neighbouring drift shells, which sample particles with slightly different drift frequencies. The spacecraft is therefore assumed to sweep radially across drift shells at constant speed $V_s$, so that
\begin{equation}\label{eq:Ls_linear}
L_s(t)=L_0+V_s t .
\end{equation}
where $V_s$ has units of inverse time, i.e., $1/s$. The measured time series is then $S(t)=F\!\big(\varphi_0,L_s(t),t\big)$.
Using the ballistic solution for an MLT- and drift-shell-localised injection,
\begin{equation}\label{eq:F_solution_recall}
F(\varphi,L,t)
=
\frac{n_s}{2\pi}
\exp\!\left[
-\left(\frac{L-b}{2\sigma}\right)^2
\right]
\sum_{m}
\psi_m(\kappa)\,
\exp\!\left[
i m \varphi - i m \langle \dot{\varphi}\rangle_b\, t
\right],
\end{equation}
together with the linear-in-$L$ expression for the bounce-averaged drift frequency,
\begin{equation}\label{eq:phidot_linear}
\langle \dot\varphi\rangle_b(L)
=
-A(\alpha_{\rm eq},E_k)\,L,
\qquad
A(\alpha_{\rm eq},E_k)=
\frac{3 E_k}{q_sB_E R_E^2}\,
\frac{\left(\frac{E_k}{m_s c^2}+2\right)}{\left(\frac{E_k}{m_sc^2}+1\right)}\mathcal{M}(\alpha_{\rm eq}),
\end{equation}
we obtain an explicit expression for the spacecraft time series:
\begin{equation}\label{eq:St_full}
S(t)=
\frac{n_s}{2\pi}
\exp\!\left[
-\left(\frac{L_0+V_s t-b}{2\sigma}\right)^2
\right]
\sum_{m}
\psi_m(\kappa)\,
\exp\!\left[
i m \varphi_0 + i m A (L_0+V_s t)\, t
\right].
\end{equation}

Equation~(\ref{eq:St_full}) shows that, in the absence of losses, the effect of spacecraft sweeping enters in two distinct ways: (i) through the radial envelope, which modulates the amplitude as the spacecraft traverses the injection region, and (ii) through the drift phase, which becomes time dependent along the trajectory. Indeed, the phase of the $m$-th azimuthal harmonic along the trajectory is
\begin{equation}\label{eq:Phi_m_general}
\Phi_m(t)=m\Big[\varphi_0 + A L_0 t + A V_s t^2\Big],
\end{equation}
so that the instantaneous angular frequency of the $m$-th mode in the measured time series is
\begin{equation}\label{eq:omega_m_general}
\omega_m(t)= \frac{d\Phi_m}{dt}
=
m\Big(A L_0 + 2A V_s t\Big).
\end{equation}
Thus, spacecraft sweeping ($V_s\neq 0$) converts each azimuthal Fourier component into a chirped contribution whose oscillation frequency increases linearly with time, with the rate of increase proportional to $m$.

An interesting, albeit fortuitous, configuration arises when the spacecraft encounters the peak of the injected enhancement, that is, when its initial drift shell coincides with the centre of the Gaussian envelope, $L_0=b$. In this limit, Equations~(\ref{eq:St_full})--(\ref{eq:omega_m_general}) reduce to
\begin{equation}\label{eq:St_L0eqb}
S(t)=
\frac{n_s}{2\pi}
\exp\!\left[
-\left(\frac{V_s t}{2\sigma}\right)^2
\right]
\sum_{m}
\psi_m(\kappa)\,
\exp\!\left[
i m \varphi_0 + i m A (b+V_s t)\, t
\right],
\end{equation}
\begin{equation}\label{eq:Phi_m_L0eqb}
\Phi_m(t)=m\Big[\varphi_0 + A b\, t + A V_s t^2\Big],
\end{equation}
and
\begin{equation}\label{eq:omega_m_L0eqb}
\omega_m(t)=m\Big(A b + 2A V_s t\Big).
\end{equation}
In this case the amplitude modulation is purely Gaussian in time, with characteristic width $2\sigma/V_s$, reflecting the time required for the spacecraft to traverse the radial extent of the injection. Simultaneously, the drift-phase contribution remains chirped, with an instantaneous frequency that grows linearly in time at a rate proportional to both the radial sweep speed $V_s$ and the azimuthal mode number $m$.

The phase of the $m$-th azimuthal harmonic along the spacecraft trajectory is therefore
\begin{equation}
\Phi_m(t)=m\left[\varphi_0 + A L_0 t + A V_s t^2\right],
\end{equation}
so that the instantaneous angular frequency of that mode in the time series is
\begin{equation}
\omega_m(t)=\frac{d\Phi_m}{dt}
=
m\left(A L_0 + 2A V_s t\right).
\end{equation}

Thus, spacecraft sweeping ($V_s\neq 0$) converts each azimuthal Fourier component into a chirped signal whose temporal frequency increases linearly with time and is proportional to the mode number $m$. Although the underlying phase-space distribution remains perfectly coherent and without dissipation, high-$m$ modes are progressively transferred to increasingly high temporal frequencies.

For a measurement with finite cadence $\Delta t$, the Nyquist frequency is
\[
\omega_N=\frac{\pi}{\Delta t}.
\]
Modes satisfying $\omega_m(t)>\omega_N$ become unresolvable in the observed time series. Frequencies exceeding the Nyquist frequency may also fold back into the resolvable range through aliasing, but without preserving a faithful representation of the underlying structure. Consequently, high-$m$ structure may disappear from the measurement even in the absence of diffusion or spatial averaging \footnote{The measurement cadence of particle fluxes is typically very short, of the order of $\sim 10\,\mathrm{s}$ or less, though operational space weather products can be distributed at 1-5 minute resolution. The corresponding Nyquist limit is therefore unlikely to impose a practical restriction in the present context. Indeed, as shown in the following section, the dominant limitation arises not from aliasing due to chirping, but from a more severe loss of coherence associated to phase-mixing. Nevertheless, the above result implies that over sufficiently long times, even the low-$m$ modes will \textit{appear} to increase in frequency, and may eventually become unresolvable within a finite-resolution measurement.}. The apparent smoothing of localised injections therefore arises from the finite temporal bandwidth of the observation, rather than from physical dissipation.

\subsection{Analytical solution for the correlation function}\label{sec:single_sc_autocorr}
While $S(t)$ indicates when the spacecraft encounters elevated flux, the autocorrelation determines for how long the sampled structure remains statistically correlated along the trajectory. The associated decorrelation time defined as the lag at which $|C(\tau;T)|$ falls below a prescribed threshold provides a quantitative measure of the effective observational lifetime of the injection as inferred from a single-satellite time series.

We therefore consider linear motion in both coordinates,
\begin{equation}\label{eq:traj_single_linear}
L_s(t)=L_0+V_s t,
\qquad
\varphi_s(t)=\varphi_0+\omega_s t,
\end{equation}
and using Equation~(\ref{EQ:final_solution_MLT_injection}) and $\langle\dot\varphi\rangle_b(L)=-A\,L$, write the measured signal as
\begin{equation}\label{eq:St_single_explicit}
S(t)=
\frac{n_s}{2\pi}
\exp\!\left[-\left(\frac{L_0+V_s t-b}{2\sigma}\right)^2\right]
\sum_m \psi_m(\kappa)\,
\exp\!\left[i m\varphi_0 + i m \Theta(t)\right],
\end{equation}
where
\begin{equation}\label{eq:Theta_def}
\Theta(t)= \omega_s t + A(L_0+V_s t)\,t
=
(\omega_s + A L_0)t + A V_s t^2 .
\end{equation}
Substituting Equation~(\ref{eq:St_single_explicit}) gives
\begin{equation}\label{eq:R_double_sum_single_conj}
\begin{aligned}
R(\tau;T)
&=
\left(\frac{n_s}{2\pi}\right)^2
\frac{1}{T-\tau}
\sum_m\sum_{m'}
\psi_m(\kappa)\,\psi_{m'}^{\ast}(\kappa)\,
e^{i(m-m')\varphi_0}
\\
&\hspace{0.8cm}\times
\int_{0}^{T-\tau}
\mathcal{W}(t)\,\mathcal{W}(t+\tau)\,
\exp\!\Big\{i m\Theta(t)-i m'\Theta(t+\tau)\Big\}\,
dt.
\end{aligned}
\end{equation}
where we define the radial envelope along the trajectory as,
\begin{equation}\label{eq:G_single_def}
\mathcal{W}(t)=
\exp\!\left[-\left(\frac{L_0+V_s t-b}{2\sigma}\right)^2\right].
\end{equation}
A phase-independent measure, to make our estimate independent of the initial phase at which the encounter with the injected plasma took place, is obtained by averaging over $\varphi_0$,
\begin{equation}\label{eq:Rbar_def_conj}
\big{\langle}{R}(\tau;T)\big{\rangle}_{\varphi_0}= \frac{1}{2\pi}\int_{0}^{2\pi} R(\tau;T)\,d\varphi_0.
\end{equation}
Since $\frac{1}{2\pi}\int_{0}^{2\pi} e^{i(m-m')\varphi_0}\,d\varphi_0=\delta_{m,m'}$, with $\delta_{m, m'}$ the Kronecker delta,
the phase average enforces $m'=m$ and yields
\begin{equation}\label{eq:Rbar_final_conj}
\big{\langle}{R}(\tau;T)\big{\rangle}_{\varphi_0}=
\left(\frac{n_s}{2\pi}\right)^2
\sum_m |\psi_m(\kappa)|^2\,
\frac{1}{T-\tau}
\int_{0}^{T-\tau}
\mathcal{W}(t)\,\mathcal{W}(t+\tau)\,
\exp\!\Big\{i m\Delta\Theta(t,\tau)\Big\}\,
dt,
\end{equation}
where the phase difference is now
\begin{equation}\label{eq:DeltaTheta_def_conj}
\Delta\Theta(t,\tau)= \Theta(t)-\Theta(t+\tau)
=
-(\omega_s + A L_0)\tau
- A V_s\big(2t\tau+\tau^2\big).
\end{equation}
We shall use $\big{\langle}{R}(\tau;T)\big{\rangle}_{\varphi_0}$, or rather its normalised form ${C}(\tau;T)=\big{\langle}{R}(\tau;T)\big{\rangle}_{\varphi_0}/\big{\langle}{R}(0;T)\big{\rangle}_{\varphi_0}$, as a robust estimate of the coherence time of localised structure, independent of the initial MLT at which the injection is encountered.

Equation~(\ref{eq:Rbar_final_conj}) can be solved numerically in the general case. 
However, a simple analytical solution can be obtained by considering the limiting case 
$L_0=b$ and $\omega_s=0$. This corresponds to a configuration in which the spacecraft 
initially encounters the injection at its peak and the relative motion is primarily 
radial. Such a configuration provides the most favourable conditions for sampling the 
injection and therefore represents an upper bound on the observable correlation signal. Then
\begin{equation}
\mathcal{W}(t)=\exp\!\left[-\left(\frac{V_s t}{2\sigma}\right)^2\right],
\qquad
\Theta(t)=A(b+V_s t)t = A b\,t + A V_s t^2,
\end{equation}
and therefore
\begin{equation}\label{eq:DeltaTheta_L0b}
\Theta(t)-\Theta(t+\tau)
=
-A b\,\tau
-A V_s\big(2t\tau+\tau^2\big).
\end{equation}
The product of envelopes becomes
\begin{equation}\label{eq:GG_product_L0b}
\mathcal{W}(t)\mathcal{W}(t+\tau)
=
\exp\!\left[-\frac{V_s^2\tau^2}{8\sigma^2}\right]\,
\exp\!\left[-\frac{V_s^2}{2\sigma^2}\left(t+\frac{\tau}{2}\right)^2\right].
\end{equation}
Hence the phase-averaged autocorrelation takes the form
\begin{equation}\label{eq:Rbar_L0b_form}
\big{\langle}{R}(\tau;T)\big{\rangle}_{\varphi_0}=
\left(\frac{n_s}{2\pi}\right)^2
\sum_m |\psi_m(\kappa)|^2\,
e^{-i m A b\,\tau}\,
e^{-i m A V_s\tau^2}\,
\exp\!\left[-\frac{V_s^2\tau^2}{8\sigma^2}\right]\,
\frac{1}{T-\tau}\,
\mathcal{I}_m(\tau;T),
\end{equation}
where
\begin{equation}\label{eq:Im_def_L0b}
\mathcal{I}_m(\tau;T)=
\int_{0}^{T-\tau}
\exp\!\left[-a\left(t+\frac{\tau}{2}\right)^2\right]\,
\exp\!\left[-i\,\beta_m(\tau)\,t\right]\,dt,
\qquad
a= \frac{V_s^2}{2\sigma^2},
\qquad
\beta_m(\tau)= 2mA V_s\tau .
\end{equation}
This integral admits the closed form
\begin{equation}\label{eq:Im_closed_L0b}
\begin{aligned}
\mathcal{I}_m(\tau;T)
&=
\frac{\sqrt{\pi}}{2\sqrt{a}}\,
\exp\!\left[
\frac{i}{2}\beta_m(\tau)\tau
-\frac{\beta_m(\tau)^2}{4a}
\right]
\\
&\hspace{0.8cm}\times
\left[
\operatorname{erf}\!\left(\sqrt{a}\left(T-\frac{\tau}{2}\right)+\frac{i\,\beta_m(\tau)}{2\sqrt{a}}\right)
-
\operatorname{erf}\!\left(\sqrt{a}\left(\frac{\tau}{2}\right)+\frac{i\,\beta_m(\tau)}{2\sqrt{a}}\right)
\right].
\end{aligned}
\end{equation}
Since $\sqrt{a}=V_s/(\sqrt{2}\sigma)$, this may be written explicitly as
\begin{equation}\label{eq:Im_closed_explicit_L0b}
\begin{aligned}
\mathcal{I}_m(\tau;T)
&=
\frac{\sqrt{\pi}\,\sigma}{\sqrt{2}\,V_s}\,
\exp\!\left[
i\,mA V_s\tau^2
-2 m^2 A^2\sigma^2\tau^2
\right]
\\
&\hspace{0.8cm}\times
\left[
\operatorname{erf}\!\left(
\frac{V_s}{\sqrt{2}\sigma}\left(T-\frac{\tau}{2}\right)
+i\,\sqrt{2}\,mA\sigma\,\tau
\right)
-
\operatorname{erf}\!\left(
\frac{V_s\tau}{2\sqrt{2}\sigma}
+i\,\sqrt{2}\,mA\sigma\,\tau
\right)
\right].
\end{aligned}
\end{equation}
Substituting Equation~(\ref{eq:Im_closed_explicit_L0b}) into Equation~(\ref{eq:Rbar_L0b_form}) shows that the phase factor $e^{-i m A V_s\tau^2}$ cancels exactly, leaving
\begin{equation}\label{eq:Rbar_closed_L0b}
\begin{aligned}
\big{\langle}{R}(\tau;T)\big{\rangle}_{\varphi_0}=
\left(\frac{n_s}{2\pi}\right)^2
\sum_m |\psi_m(\kappa)|^2\,
e^{-i m A b\,\tau}\,
\exp\!\left[-\frac{V_s^2\tau^2}{8\sigma^2}\right]\,
\frac{1}{T-\tau}\,
\frac{\sqrt{\pi}\,\sigma}{\sqrt{2}\,V_s}\,
e^{-2 m^2 A^2\sigma^2\tau^2}
\\
\hspace{1.1cm}\times
\left[
\operatorname{erf}\!\left(
\frac{V_s}{\sqrt{2}\sigma}\left(T-\frac{\tau}{2}\right)
+i\,\sqrt{2}\,mA\sigma\,\tau
\right)
-
\operatorname{erf}\!\left(
\frac{V_s\tau}{2\sqrt{2}\sigma}
+i\,\sqrt{2}\,mA\sigma\,\tau
\right)
\right].
\end{aligned}
\end{equation}
We finally compute the normalised  autocorrelation ${C}(\tau;T)$ and find
\begin{equation}\label{eq:Cbar_closed_L0b_simpler}
\begin{aligned}
{C}(\tau;T)
&=
\exp\!\left[-\frac{V_s^2\tau^2}{8\sigma^2}\right]\,
\frac{T}{T-\tau}\,
\frac{
\displaystyle \sum_m |\psi_m(\kappa)|^2\,e^{-i m A b\,\tau}\,
e^{-2 m^2 A^2\sigma^2\tau^2}\,
\Delta_m(\tau;T)
}{
\displaystyle \operatorname{erf}\!\left(\frac{V_s T}{\sqrt{2}\sigma}\right)
\sum_m |\psi_m(\kappa)|^2
},
\end{aligned}
\end{equation}
where we have introduced the compact notation
\begin{equation}\label{eq:Delta_m_def}
\Delta_m(\tau;T)=
\operatorname{erf}\!\left(
\frac{V_s}{\sqrt{2}\sigma}\left(T-\frac{\tau}{2}\right)
+i\,\sqrt{2}\,mA\sigma\,\tau
\right)
-
\operatorname{erf}\!\left(
\frac{V_s\tau}{2\sqrt{2}\sigma}
+i\,\sqrt{2}\,mA\sigma\,\tau
\right).
\end{equation}
Equation~(\ref{eq:Cbar_closed_L0b_simpler}) shows that the decay of the normalised correlation arises from two distinct effects: (i) the reduction of radial-envelope overlap, captured by the prefactor $\exp\!\left[-V_s^2\tau^2/(8\sigma^2)\right]$, which is purely geometric but must be accounted for, and (ii) the progressive loss of phase coherence due to drift-phase mixing, encapsulated by the factor $\exp\!\left[-2m^2A^2\sigma^2\tau^2\right]$. The latter term, which we analyse more carefully in the next section, leads to a characteristic decorrelation time
\begin{equation}
\tau_c
=
\frac{1}{\sqrt{2}\,m\,\sigma\,\langle \dot\varphi\rangle_b(b)/b}
.
\end{equation}
This timescale is shorter for increasingly localised structures ($m\gg 1$) and for more energetic particles ($E_k/m_sc^2\gg 1$), implying that fine-scale MLT structure associated with energetic populations loses coherence more rapidly than large-scale components carried by sub-relativistic particles.

It is nevertheless important to treat the error-function contribution in $\Delta_m(\tau;T)$ with care, since its argument generally acquires a large imaginary part on timescales much greater than the drift period of the particles. In the present problem, however, this does not lead to any divergence: the potentially large imaginary-axis behaviour is regularised by the accompanying Gaussian factor, and the combined expression remains bounded. This motivates the asymptotic treatment developed below.

\section{Results}\label{sec3}
\subsection{Effective lifetime of localised injections}
In this section, we use Equation (\ref{eq:Cbar_closed_L0b_simpler}) to quantify the effective lifetime of spatially localised structures as perceived from a satellite sweeping across the radiation belts. Because the normalised correlation function measures the degree to which the sampled distribution retains memory of its initial configuration, its decay directly quantifies the timescale over which a localised structure remains observationally coherent, thereby providing a natural definition of its effective lifetime. The analytical form of the normalised correlation function allows us to separate two physically distinct contributions to the decay: (i) the gradual loss of radial overlap due to spacecraft motion, and (ii) the intrinsic phase mixing arising from gradients in the drift frequency across drift shells. 

We first compute the correlation function for the $m=0$ contribution, which characterises the lifetime of the drift-averaged distribution function. Since the phase-mixing decorrelation scales as $1/m$, the $m=0$ component does not decay through phase mixing and is therefore independent of the drift period. We then consider the $m\neq 0$ contributions. A strongly localised initial injection necessarily requires representation as a superposition of many Fourier modes $m$. The problem of determining how long such localised structures remain observable is therefore equivalent to establishing the timescale over which a given mode $m$ survives the combined effects of phase mixing and satellite sweeping.
\subsubsection{Correlation function for the drift-averaged component ($m= 0$)}
Extracting from Equation (\ref{eq:Cbar_closed_L0b_simpler}) the $m=0$ component in the limit where $T\longrightarrow \infty$, we find that the autocorrelation of the drift-averaged component of the distribution function is given by: 
\begin{equation}
    \begin{aligned}
{C}_{m=0}(\tau)
&=
\exp\!\left[-\frac{V_s^2\tau^2}{8\sigma^2}\right]\ \left(1-\displaystyle \operatorname{erf}\!\left(\frac{V_s \tau}{\sqrt{8}\sigma}\right)\right).
\end{aligned}
\end{equation}
Thus, the drift-averaged lifetime is entirely controlled by the satellite motion, and decorrelation sets in once $V_s\tau/(\sqrt{8}\,\sigma)\simeq 1/2$. At $\tau=0$, $C_{m=0}=1$, and in the limit $V_s\tau/(\sqrt{8}\,\sigma)\ll 1$ the correlation function decays linearly,
\begin{equation}
C_{m=0}(\tau)
\simeq
1 - \frac{1}{\sqrt{2\pi}}\,\frac{V_s}{\sigma}\,\tau,
\qquad
\text{for}
\qquad
\frac{V_s\tau}{\sqrt{8}\,\sigma} \ll 1 .
\end{equation}
Since satellites move slowly within the radiation belts in comparison to the drift motion of energetic particles, the drift-averaged component associated with a localised injection should therefore be long-lived. Nonetheless, once the satellite displacement along the drift shell satisfies $V_s\tau \gtrsim \sqrt{2}\,\sigma$ relative to the centre of the injection region, the injection signature is no longer sampled and the corresponding information may be regarded as effectively lost. 

\subsubsection{Decorrelation time for $m\neq0$ fluctuations}
We now proceed with the correlation function for the terms $m\neq 0$. In the limit where $T\longrightarrow \infty$, the correlation function of the fluctuating components is given by:
\begin{equation}
\label{Eq:Correlation_mnotzero}
\begin{aligned}
{C}_{m\neq0}(\tau)
&=
\exp\!\left[-\frac{V_s^2\tau^2}{8\sigma^2}\right]\,
\frac{
\displaystyle \sum_{m\neq0} |\psi_m(\kappa)|^2\,e^{-i m A b\,\tau}\,
e^{-2 m^2 A^2\sigma^2\tau^2}\,
\left[\operatorname{erfc}\!\left(
\frac{V_s\tau}{2\sqrt{2}\sigma}
+i\,\sqrt{2}\,mA\sigma\,\tau
\right)\right]
}{
\sum_{m\neq 0} |\psi_m(\kappa)|^2
},
\end{aligned}
\end{equation}
where we simplified the function $\Delta_m(\tau;T)$ in the limit where $T\longrightarrow \infty$ in terms of the complimentary error function $\operatorname{erfc}(x)=1-\operatorname{erf}(x)$: 
\begin{equation}
\Delta_m(\tau;\infty)\simeq
1
-
\operatorname{erf}\!\left(
\frac{V_s\tau}{2\sqrt{2}\sigma}
+i\,\sqrt{2}\,mA\sigma\,\tau
\right) =\operatorname{erfc}\!\left(
\frac{V_s\tau}{2\sqrt{2}\sigma}
+i\,\sqrt{2}\,mA\sigma\,\tau
\right)
\end{equation}

In the following, we seek a closed-form analytical solution to Equation~(\ref{Eq:Correlation_mnotzero}) in the regime where the argument of the exponential prefactor multiplying the sum, namely
\[
x=\frac{V_s\tau}{\sqrt{8}\,\sigma},
\]
is much smaller than unity. In this limit, the complex argument of the error function,
\[
z=x+iy
=
\frac{V_s\tau}{2\sqrt{2}\,\sigma}
+
i\sqrt{2}\,mA\tau,
\]
has a small real part, $x\ll 1$. The second characteristic timescale of the problem is the drift period at the reference drift shell $L=b$,
\[
T_d=\frac{2\pi}{A b}
=
\frac{2\pi}{\langle \dot{\varphi}\rangle_b(b)},
\]
which enters through the imaginary component of $z$, and which we write explicitly as,
\[
y=\sqrt{2}\,\frac{\sigma}{b}\,\frac{\tau}{T_d}.
\]
We are therefore interested in the behaviour of the correlation function on timescales comparable to the drift period, $\tau\gtrsim T_d$, while remaining in the regime
\[
\tau \ll \frac{\sqrt{8}\,\sigma}{V_s},
\]
so that the small $x$ approximation remains valid.

Let's consider the error function $\operatorname{erf}\!(x+iy)$ for the case where
$x\ll 1$ and $y\in\mathbb{R}$ arbitrary. Since the error function is analytic
everywhere, we may perform a Taylor expansion about the point
$z=iy$:
\[
\operatorname{erf}\!(x+iy)
=
\operatorname{erf}\!(iy)
+
x\,\operatorname{erf}'\!(iy)
+
\mathcal{O}(x^{2}),
\qquad x\ll 1.
\]
Using $\operatorname{erf}'\!(z)=\frac{2}{\sqrt{\pi}}e^{-z^{2}}$, we obtain
\[
\operatorname{erf}'\!(iy)=\frac{2}{\sqrt{\pi}}e^{y^{2}}.
\]
Therefore, in the limit $x\ll 1$,
\begin{equation}\label{EQ:First_step_complex_error_function}
\operatorname{erf}\!(x+iy)
=
\operatorname{erf}\!(iy)
+
\frac{2x}{\sqrt{\pi}}e^{y^{2}}
+
\mathcal{O}(x^{2}).
\end{equation}

The first term in Equation~(\ref{EQ:First_step_complex_error_function})
can conveniently be expressed in terms of Dawson's function \cite{abramowitz1968handbook}.
This follows from the identity relating the imaginary error function $\operatorname{erfi} \!(y)$ to the error function with an imaginary argument $\operatorname{erf}\!(iy)$:
\[
\operatorname{erf}\!(iy)= i\,\operatorname{erfi}\!(y),
\]
together with
\begin{equation}\label{EQ:Dawson_function}
\operatorname{erfi}\!(y)
=
\frac{2}{\sqrt{\pi}}\,e^{y^{2}}\,\operatorname{Dawson}\!(y).
\end{equation}
Combining Equations~(\ref{EQ:First_step_complex_error_function}) and
(\ref{EQ:Dawson_function}), we find that
\begin{equation}\label{EQ:Complex_error_function_argument}
1-\operatorname{erf}\!(x+iy)
\simeq
1
-
i\,\frac{2}{\sqrt{\pi}}\,e^{y^{2}}\,\operatorname{Dawson}\!(y)
-
\frac{2x}{\sqrt{\pi}}e^{y^{2}}
+
\mathcal{O}(x^{2}),
\qquad x\ll 1.
\end{equation}
Replacing the error function in Equation~(\ref{Eq:Correlation_mnotzero}) with Equation (\ref{EQ:Complex_error_function_argument}) we find our final expression for the correlation function for all $m\neq 0$ terms: 
\begin{equation}
\label{Eq:Correlation_mnotzero_FINAL_corrected}
C_{m\neq 0}(\tau)
\simeq
\frac{
\displaystyle \sum_{m\neq 0} |\psi_m(\kappa)|^2
\left[
e^{-2m^2A^2\sigma^2\tau^2}\cos(mAb\,\tau)
-\frac{2}{\sqrt{\pi}}\,
\operatorname{Dawson}\!\bigl(\sqrt{2}mA\sigma\tau\bigr)\,
\sin(mAb\,\tau)
\right]
}{
\displaystyle \sum_{m\neq 0} |\psi_m(\kappa)|^2
},
\end{equation}
where we ignored the $\mathcal{O}(x)$ correction which contributes
\[
\Delta C_x(\tau)=-\frac{2x}{\sqrt{\pi}}\,
\frac{\sum_{m\neq 0}|\psi_m|^2 \cos(mAb\tau)}{\sum_{m\neq 0}|\psi_m|^2},
\]
and since the normalised weighted cosine average lies in $[-1,1]$, we have the uniform bound
$|\Delta C_x(\tau)|\le 2x/\sqrt{\pi}$. Hence this term may be neglected for
$x=V_s\tau/(\sqrt{8}\sigma)\ll 1$, and when $x\geq 1$ the exponential decay $\exp(-x^2)$ becomes significant and dampens the correlation. 

The final step in our analysis is to express Equation~(\ref{Eq:Correlation_mnotzero_FINAL_corrected})
in terms of the dimensionless lag $\hat\tau$ normalised by the drift period
$T_d(b)$ at $L=b$:
\begin{equation}
   \hat\tau = \frac{\tau}{T_d(b)}=\frac{Ab}{2\pi}\,\tau.
\end{equation}
For simplicity, we hereafter write $\tau$ for the normalised lag.
The resulting correlation function can therefore be computed from 
\begin{equation}
\boxed{
\label{Eq:Correlation_mnotzero_FINAL_corrected_normalised}
C_{m\neq 0}(\tau)
\simeq
\frac{
\displaystyle \sum_{m\neq 0} |\psi_m(\kappa)|^2
\left[
e^{-8\pi^2 m^2\sigma^2\tau^2/b^2}\cos(2\pi m\tau)
-\frac{2}{\sqrt{\pi}}\,
\operatorname{Dawson}\!\bigl(\sqrt{8}\pi m\sigma\tau/b\bigr)\,
\sin(2\pi m\tau)
\right]
}{
\displaystyle \sum_{m\neq 0} |\psi_m(\kappa)|^2
},}
\end{equation}
Equation~(\ref{Eq:Correlation_mnotzero_FINAL_corrected_normalised}) shows that the correlation time associated with each mode $m\neq 0$, once normalised in terms of the drift period, is a function of the width of the injection $\sigma$, the location of the injection $b$ and the parameter $\kappa$, which quantifies the width in MLT and the amplitudes of the coefficients $\psi_m$.

The first term in the bracket of Equation~(\ref{Eq:Correlation_mnotzero_FINAL_corrected_normalised}) containing a Gaussian oscillates with azimuthal mode number $m$ and decays exponentially once
\begin{equation}
\tau \gtrsim \frac{b}{\sqrt{8}\,\pi\,m\,\sigma}.
\end{equation}
Thus the $m=1$ mode is the last to decay through the Gaussian term, whereas higher-$m$ modes can decay on timescales much shorter than a single drift period, since they decorrelate for
\begin{equation}
\tau \simeq \frac{b}{\sqrt{8}\,\pi\,m\,\sigma}
\approx 0.112\,\frac{b}{m\,\sigma}.
\end{equation}

The second term in the bracket of Equation~(\ref{Eq:Correlation_mnotzero_FINAL_corrected_normalised}) attains its maximum when the argument of the Dawson function satisfies
\begin{equation}
\frac{\sqrt{8}\,\pi\,m\,\sigma}{b}\,\tau \approx 0.924.
\end{equation}
This corresponds to a normalised lag
\begin{equation}
\tau_{\mathrm{peak}}
=
\frac{0.924}{\sqrt{8}\,\pi}\,
\frac{b}{m\,\sigma}
\approx
0.104\,\frac{b}{m\,\sigma}.
\end{equation}
For arguments larger than unity, the Dawson function decays algebraically, with leading-order behaviour proportional to the inverse of its argument (i.e., as $b/(\sqrt{8}\pi m\sigma\tau)$). Consequently, the $m=1$ mode peaks and decays last, whereas higher azimuthal modes reach their maximum earlier and therefore decay sooner. 

\begin{figure*}[t]
    \centering

    \begin{minipage}[t]{0.49\textwidth}
        \centering
        \includegraphics[width=\linewidth]{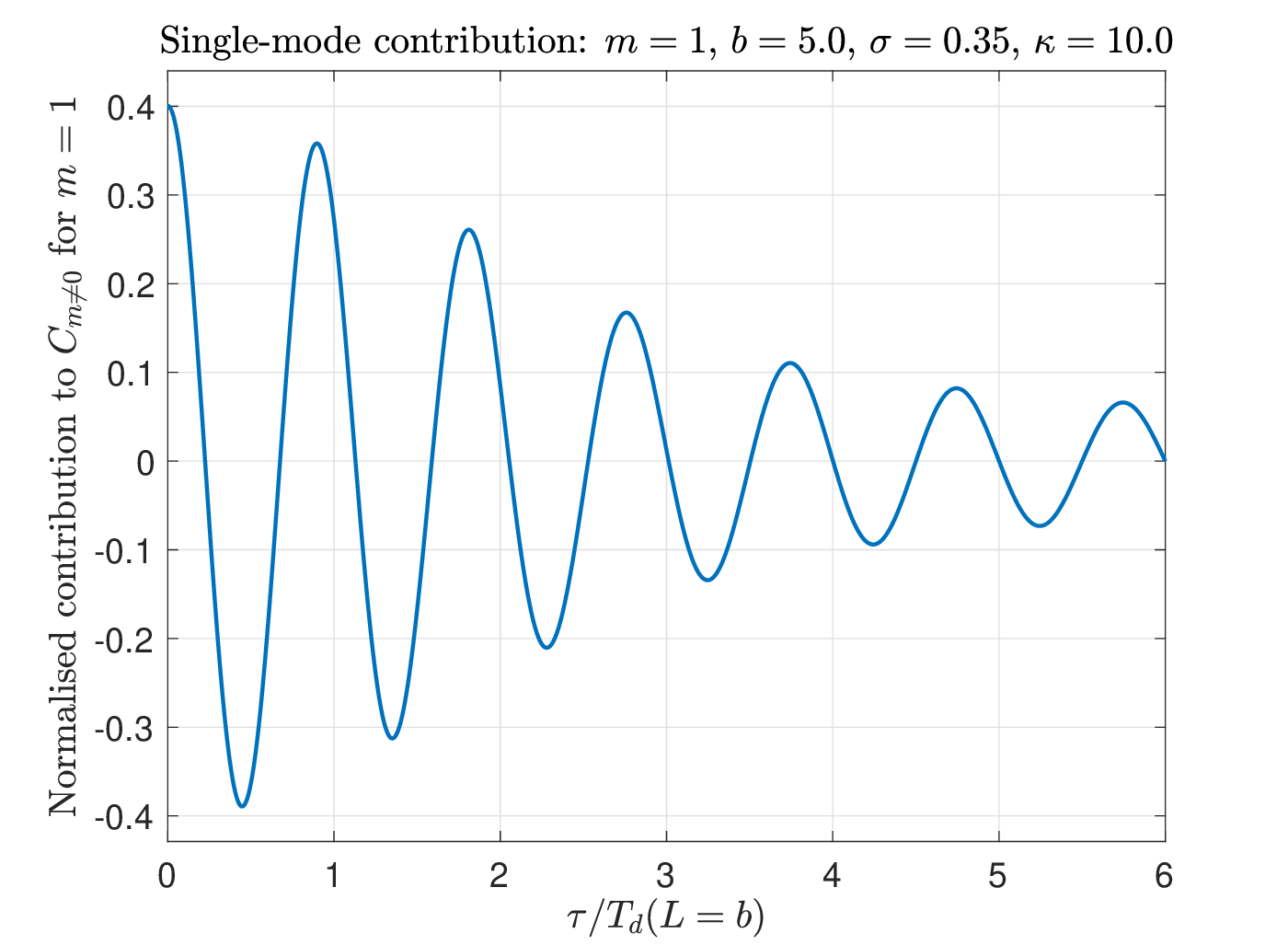}

        (a)
    \end{minipage}
    \hfill
    \begin{minipage}[t]{0.49\textwidth}
        \centering
        \includegraphics[width=\linewidth]{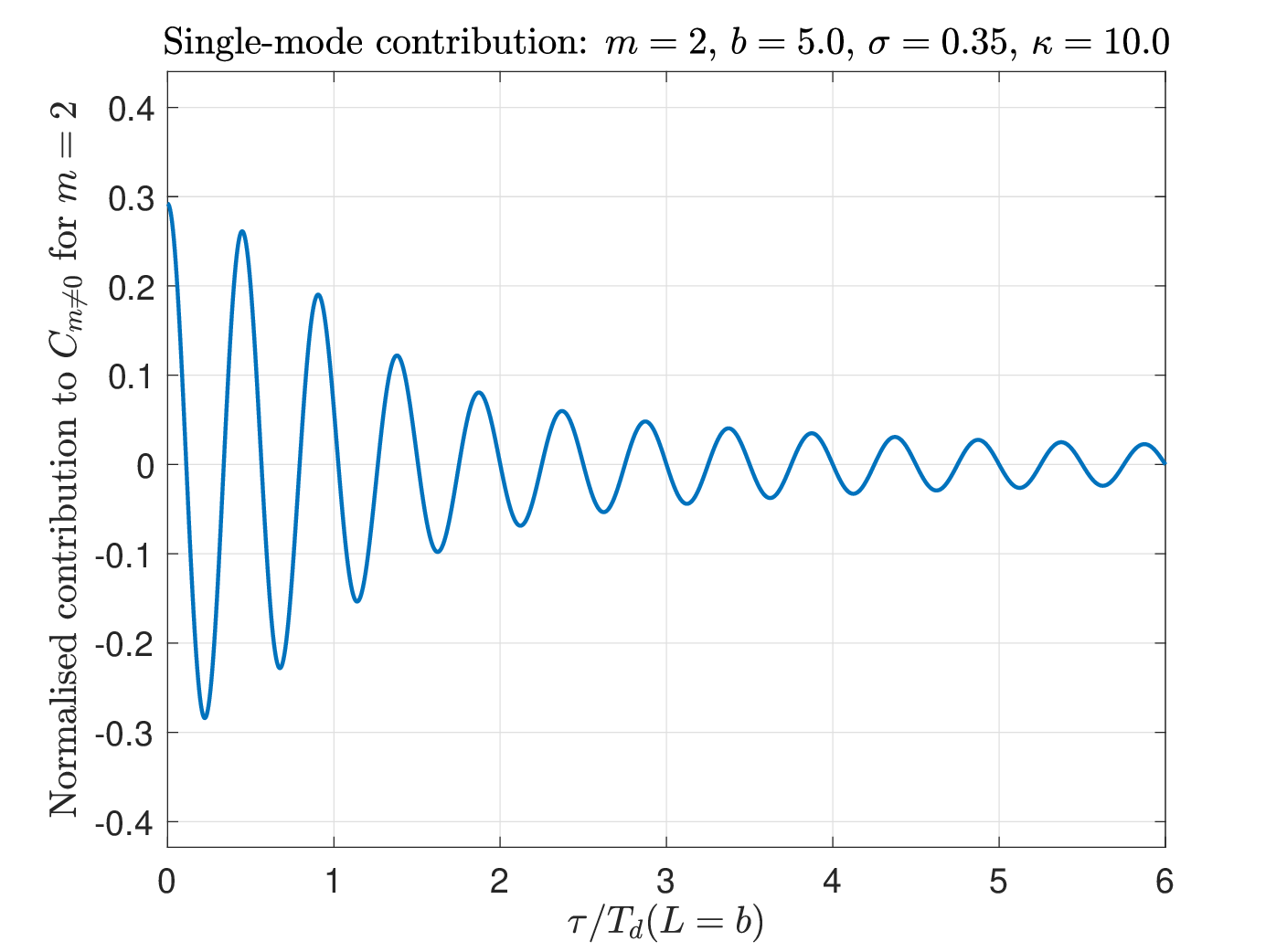}

        (b)
    \end{minipage}

    \vspace{0.5cm}

    \begin{minipage}[t]{0.49\textwidth}
        \centering
        \includegraphics[width=\linewidth]{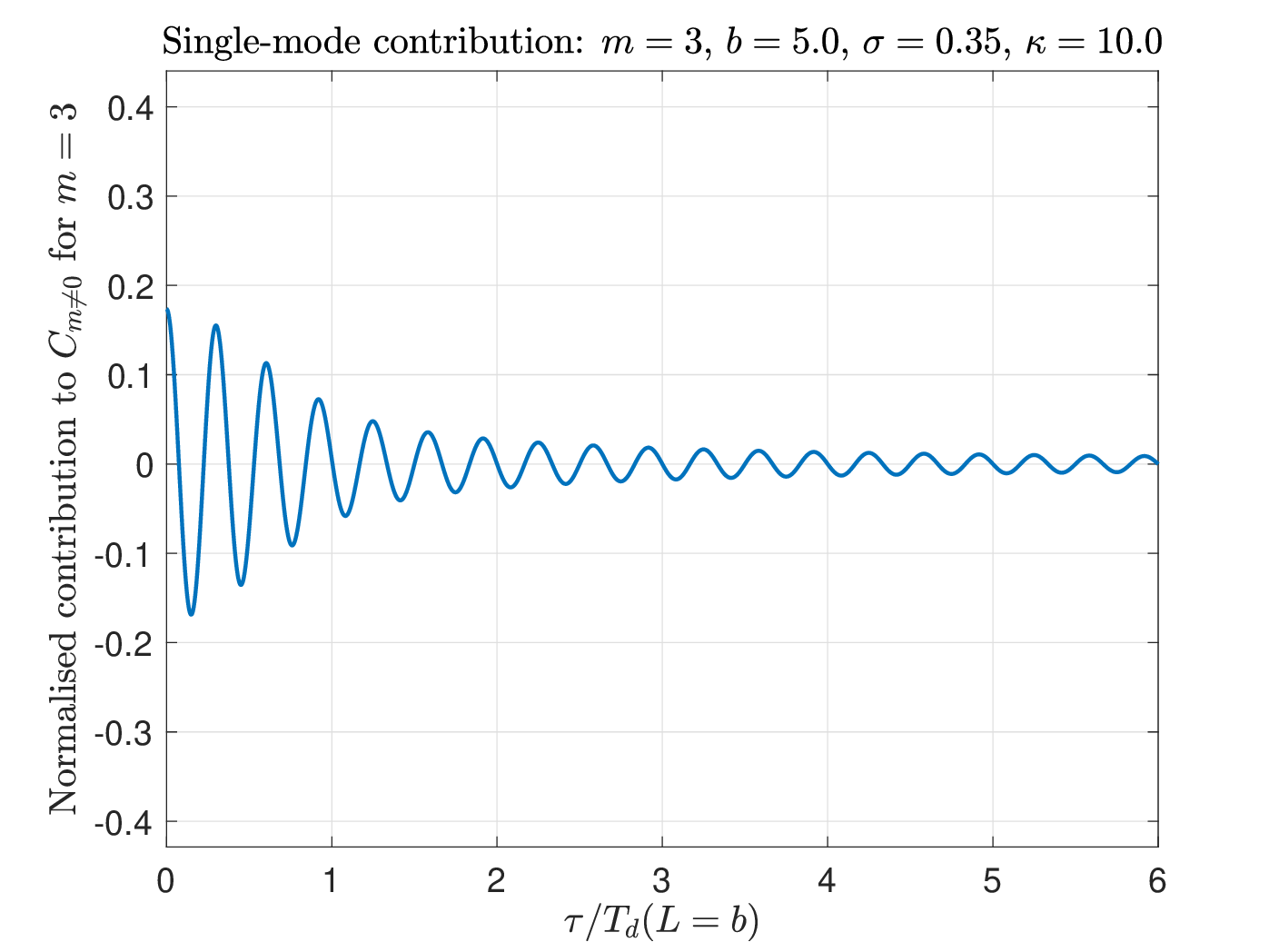}

        (c)
    \end{minipage}
    \hfill
    \begin{minipage}[t]{0.49\textwidth}
        \centering
        \includegraphics[width=\linewidth]{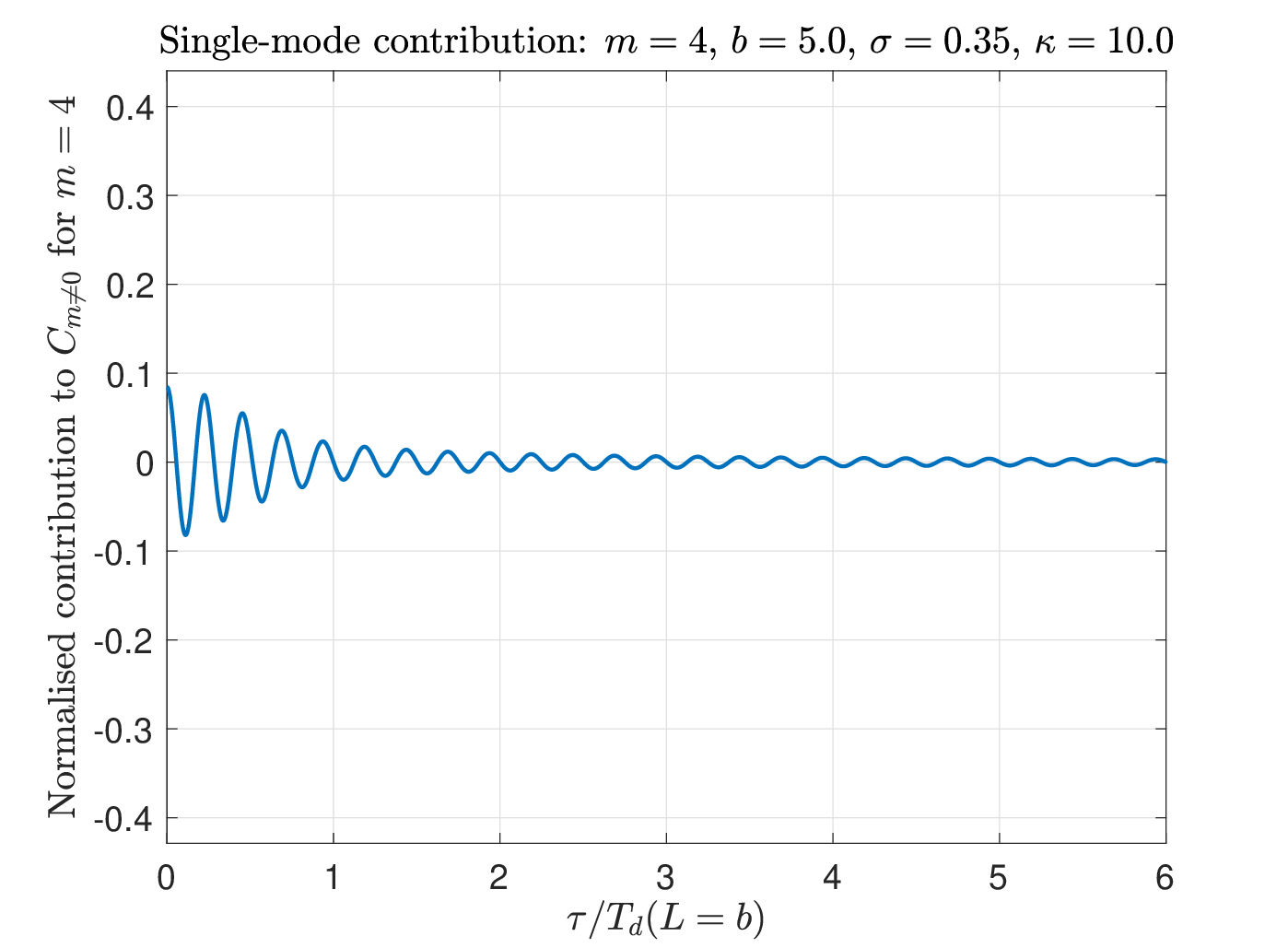}

        (d)
    \end{minipage}

    \caption{Panels (a)--(d) show the normalised contribution to the correlation function given by Eq.~(\ref{Eq:Correlation_mnotzero_FINAL_corrected_normalised}) for the azimuthal modes $m=1$, $m=2$, $m=3$, and $m=4$, respectively. Time is normalised by the drift period $T_d$. The initial injection is centred at $L=5$, has a drift-shell width $\sigma=0.35$, and is localised over approximately $20\%$ of the drift orbit (corresponding to $\kappa=10$ in the von Mises weighting).}
    \label{fig1:four_panel}
\end{figure*}

This behaviour is illustrated in Figure~(\ref{fig1:four_panel}), which shows the normalised contribution to the correlation function for the first four azimuthal modes. The analytical trends discussed above are clearly visible: the $m=1$ mode retains the largest correlation and persists for the longest time, whereas higher-$m$ modes decorrelate increasingly rapidly and contribute only at early lags.

\begin{figure}[t]
    \centering

    \begin{minipage}[t]{0.55\columnwidth}
        \centering
        \includegraphics[width=\linewidth]{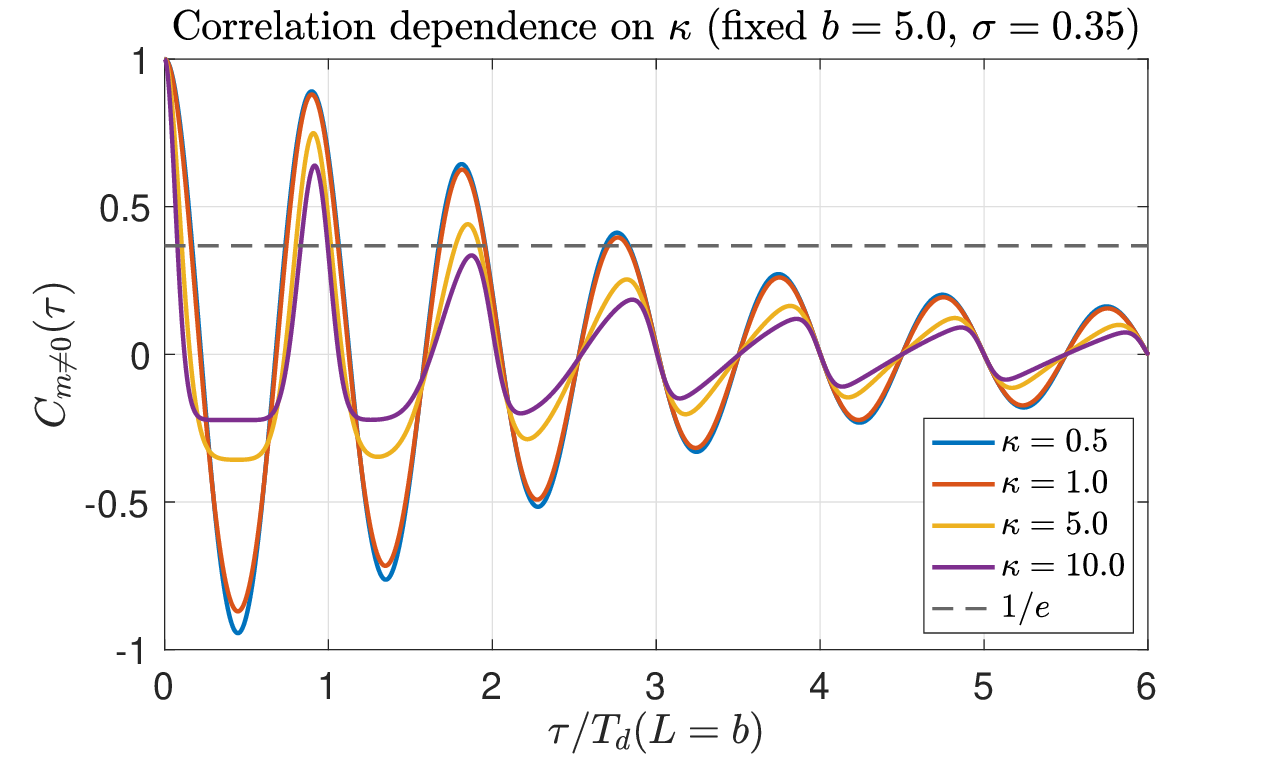}

        {\small (a)}
    \end{minipage}

    \vspace{2mm}

    \begin{minipage}[t]{0.55\columnwidth}
        \centering
        \includegraphics[width=\linewidth]{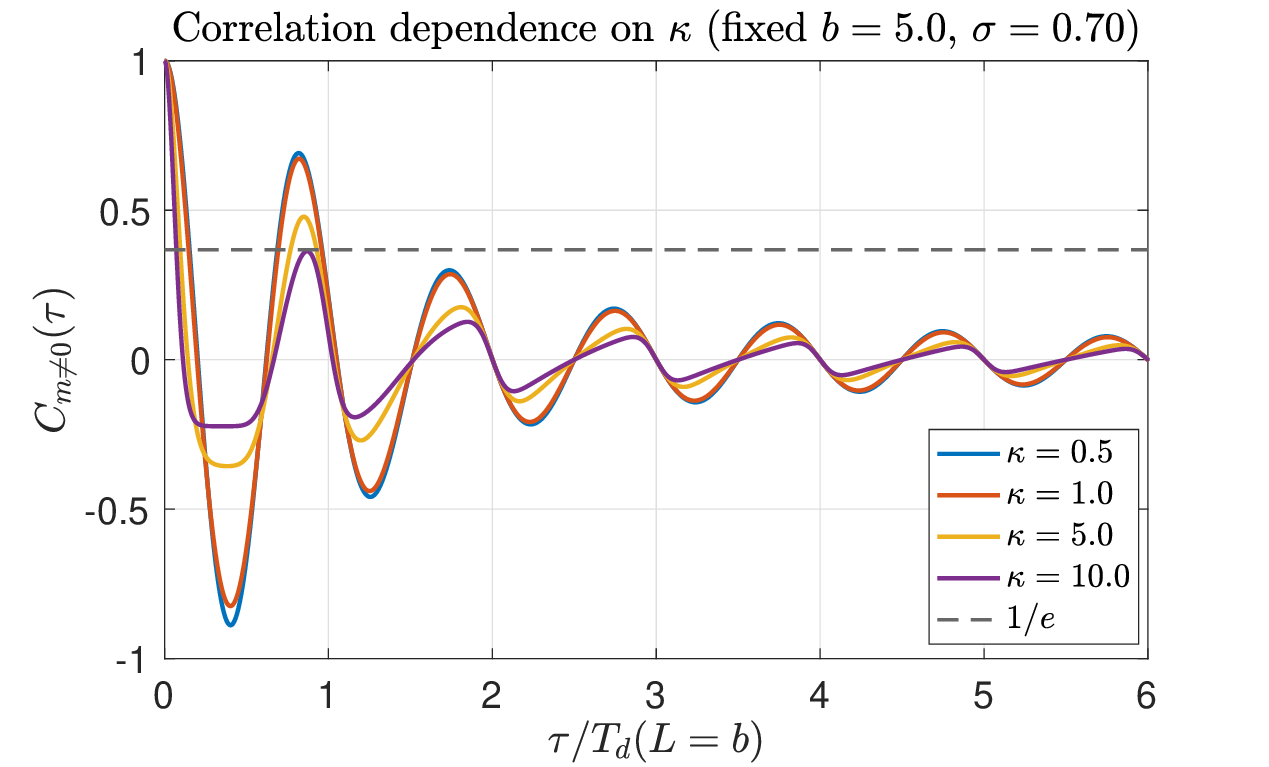}

        {\small (b)}
    \end{minipage}

    \vspace{2mm}

    \begin{minipage}[t]{0.55\columnwidth}
        \centering
        \includegraphics[width=\linewidth]{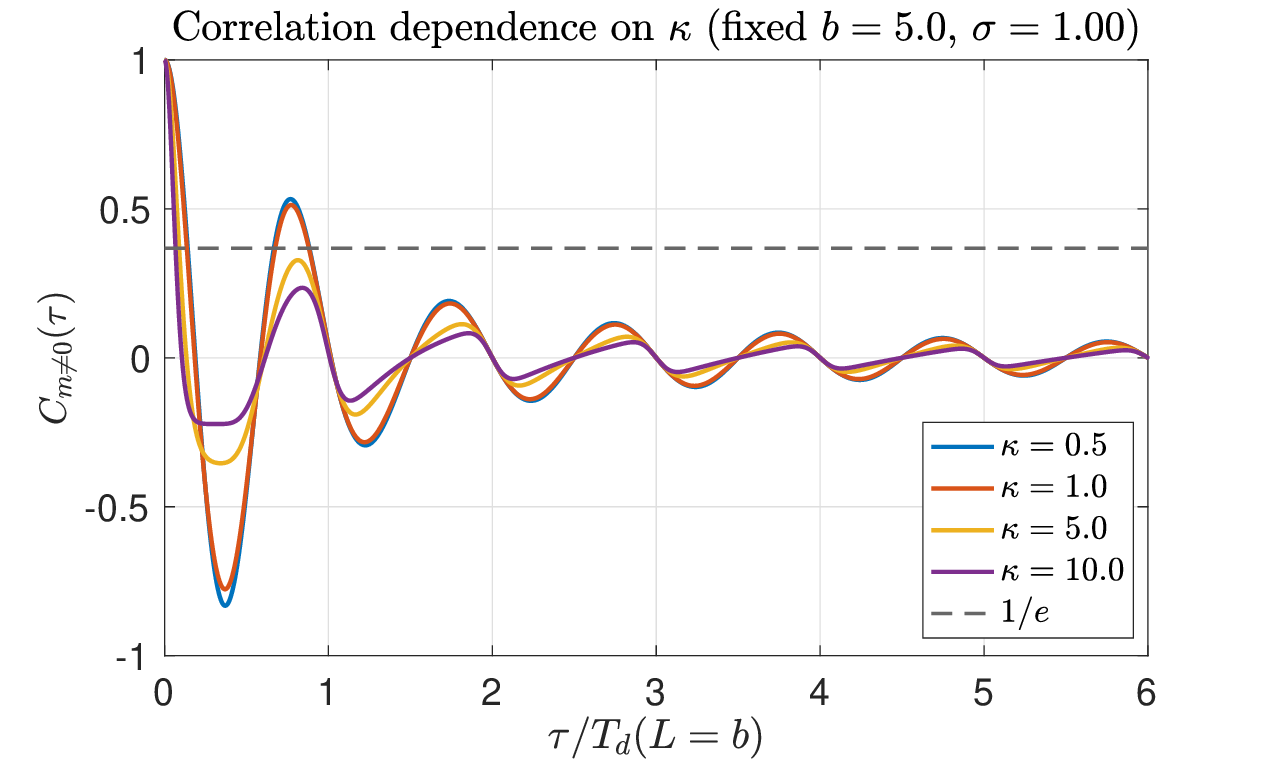}

        {\small (c)}
    \end{minipage}

    \caption{Panels (a)--(c) show the parametric dependence of the total correlation function, given by Eq.~(\ref{Eq:Correlation_mnotzero_FINAL_corrected_normalised}), as a function of time (normalised by the drift period $T_d$). The panels illustrate the dependence on the localisation parameter $\kappa$ and on the ratio $b/\sigma$. In all cases the injected structure is centred at $b=5$. The radial width is $\sigma=0.35$ in panel (a), $\sigma=0.7$ in panel (b), and $\sigma=1$ in panel (c). Within each panel, the curves correspond to $\kappa=[0.5,1,5,10]$. The dashed line indicates the characteristic decorrelation level corresponding to $1/e$. The figure shows that structures that are narrow in drift shells (smaller $\sigma$) but broader in MLT (smaller $\kappa$) exhibit longer decorrelation times. Nevertheless, the decorrelation occurs on timescales of only a few drift periods.}
    \label{fig2:four_panel}
\end{figure}

The correlation is thus composed of two contributions that both decay rapidly for large $m$. Even for $m=1$, however, the characteristic decorrelation time associated with both terms scales as
\begin{equation}
\tau_c \sim 0.11\,\frac{b}{\sigma}.
\end{equation}
For injections occurring between $b=4$ and $b=7$, and for radial widths in the range $\sigma=0.2$--$1$, the corresponding decorrelation time satisfies
\begin{equation}
\tau_c \lesssim \frac{4 T_d}{m}.
\end{equation}
Thus, even the $m=1$ mode is expected to decorrelate after at most approximately 3-4 drift periods. These analytical estimates are confirmed by the numerical evaluation shown in Figure~(\ref{fig2:four_panel}). The numerical solutions follow closely the predicted scaling $\tau_c \propto b/\sigma$ and demonstrate that the decorrelation indeed occurs on timescales of only a few drift periods.

The role of spacecraft sweeping can be further clarified by considering the limiting case where the satellite speed tends to zero, so that the measurements are taken along a single drift shell. In this situation the spacecraft no longer samples neighbouring drift shells with slightly different drift frequencies, and the observational phase--mixing mechanism identified above cannot operate. As shown in Appendix~\ref{app:Vs_zero_limit}, the correlation function in this limit reduces to an expression in which the azimuthal Fourier modes with $m\neq0$ remain purely oscillatory and do not decorrelate. The decorrelation of these modes therefore requires a finite spacecraft sweeping speed, which introduces the differential drift phases necessary to produce phase divergence between successive measurements.

This result has important implications for the interpretation of radiation-belt observations, as discussed in detail in Section \ref{sec4}. In particular, even if injections are initially highly localised in drift phase, the combined effect of drift motion and phase mixing rapidly erases the associated correlations when sampled by an orbiting spacecraft. As a result, structures that may persist in phase space can appear strongly smeared in time-series measurements. Consequently, satellite observations may naturally give the impression of slow, diffusive evolution, even when the underlying dynamics are entirely dominated by collisionless phase mixing associated with the drift of localised injections.

\section{Discussion}\label{sec4}
\subsection{Observational decorrelation of MLT-localised injections}
The results obtained in this work clarify how spatially localised injections appear in measurements from orbiting spacecraft in the radiation belts. Even in the absence of diffusion or dissipation, the combined effects of azimuthal drift and satellite sweeping across neighbouring drift shells produce unavoidable phase mixing. Because particles on adjacent drift shells drift with slightly different angular frequencies, the phase coherence of an initially MLT-localised population is progressively lost. At the same time, a spacecraft samples a continuum of drift shells along its orbit, so that successive measurements correspond to regions characterised by slightly different drift phases. While the difference in drift phases is small, they can diverge substantially due to parabolic time dependence appearing in phase (as shown in Equation (\ref{eq:Phi_m_general})). 

It is useful to emphasise that this effect differs fundamentally from phase mixing considered at fixed spatial locations (such as the one derived in \citet{Lejosne_2024}). In such cases, spatial and temporal dependencies remain separable. By contrast, measurements performed along a moving spacecraft trajectory intrinsically mix spatial and temporal variations, leading to an additional and unavoidable loss of phase coherence that is not captured in static sampling frameworks.


As a result, the temporal signal recorded by a spacecraft rapidly loses the phase information associated with the initial MLT localisation. Our analysis shows that this decorrelation occurs on timescales comparable to only a few drift periods. Consequently, even an injection that is initially narrow in magnetic local time but extended across drift shells will appear, when sampled by dual probes on similar orbits, to become progressively MLT-averaged after only a few drift periods. More generally, observational studies indicate that injections can exhibit a continuum of azimuthal extents, ranging from relatively localised structures to enhancements spanning several hours of magnetic local time \cite{Kavanagh07, Gabrielse19}. In the present framework, broader structures in MLT correspond to smaller values of the concentration parameter $\kappa$ and therefore decorrelate more slowly. Nevertheless, the characteristic observational coherence time generally remains comparable to only a few drift periods. The apparent lifetime of MLT localisation inferred from spacecraft observations therefore reflects an observational coherence time imposed by phase mixing and orbital sampling.

This limitation can also be understood in terms of information sampling. In principle, the drift-phase structure of the distribution function is recoverable from either temporal measurements at fixed radial location, $(t, L=L_0)$, or from instantaneous spatial measurements across drift shells, $(t=t_0, L)$, where spatial and temporal dependencies remain separable. By contrast, a single spacecraft following a trajectory $(t, L(t))$ samples a combination of spatial and temporal variations, effectively mixing the two. The resulting time series therefore does not contain sufficient information to reconstruct the underlying drift-phase structure, even when the dynamics are fully deterministic. This mixing of temporal and spatial fluctuations constitutes a fundamental aspect of the observational phase-mixing process described here.

This rapid decorrelation of spatial fluctuations may also be interpreted as an effective viscosity acting on the observed signal, in the sense that it suppresses fine-scale structure without invoking any dissipative or irreversible process \footnote{Here “effective viscosity” refers to the observational consequence of phase mixing: sampling of spatially localized structures leads to a progressive smoothing of the measured signal, analogous to the action of viscosity on gradients in a fluid. This analogy is purely kinematic and does not imply any underlying dissipation.}. Highly localized phase-space structures can therefore become observationally indistinguishable from azimuthally distributed populations even when their underlying dynamics remain strictly ballistic. The mechanism identified here consequently has important implications for both modelling efforts and the interpretation of spacecraft observations of planetary radiation belts.

\subsection{Model assumptions and limitations}
The model developed in this work addresses a deliberately simple question: what is the fate of a spatially localised injection when observed by an orbiting spacecraft in the absence of wave–particle interactions. In reality, planetary radiation belts are continuously driven by energy and momentum deposition from the solar wind and magnetospheric processes, which generate a broad spectrum of electromagnetic fluctuations. A natural question is therefore whether wave–particle interactions could mitigate the effective viscosity of drift-phase structures identified here and prolong their observational lifetime. In addition, the present formulation assumes that the bounce-averaged azimuthal drift is dominated by gradient and curvature drifts, such that contributions from the convection electric field remain subdominant and do not introduce significant azimuthal asymmetry \footnote{This assumption is well justified for energetic and relativistic electrons in the Earth’s radiation belts, where gradient and curvature drifts dominate over convection-driven $\mathbf{E}\times\mathbf{B}$ motion. Its extension to other regimes, such as the ring current in the Earth’s magnetosphere or energetic particle populations at giant planets, may require the explicit inclusion of convective and corotation electric fields, which introduce azimuthal asymmetries in the drift motion.}.

The answer depends on the particle energy and pitch angle, since particles with different phase-space coordinates interact non-adiabatically with waves of different origin and polarisation. Among these processes, pitch-angle diffusion is particularly relevant because it can operate on timescales comparable to a particle drift period. In this regime, pitch-angle diffusion enters the eigenvalue spectrum of the transport equation and contributes only negative definite terms, thereby accelerating the loss of phase coherence without modifying the fundamental phase-mixing mechanism. In contrast, during strong injection events, wave–particle interactions may proceed on timescales much shorter than the drift period, rapidly restructuring the pitch-angle distribution through nonlinear phase mixing. This is particularly relevant for $\sim$1–100 keV electrons, which have much longer drift periods than energetic and relativistic electrons, while simultaneously experiencing more vigorous pitch-angle scattering by large-amplitude whistler-mode waves (see, e.g., \citet{ArtemyevSSR_Review2025}). In that case, the distribution function may be effectively redistributed prior to significant azimuthal drift, after which the subsequent evolution is again governed by the drift-phase mixing described here, but with modified initial conditions.

To examine this possibility in the regime where the pitch-angle diffusion timescale is comparable to the azimuthal drift period, Appendix~\ref{Appendix_B} quantifies the effect of pitch-angle diffusion on drift-phase structure within a radial transport framework that includes a diffusion term. The analysis shows that pitch-angle diffusion does not sustain or amplify drift-phase structure. Instead, it introduces an additional exponential damping of phase coherence on timescales comparable to the pitch-angle diffusion time \cite{schulz1974, ALBERT_diffusion_solution}. Consequently, wave-driven pitch-angle scattering accelerates the loss of drift-phase coherence rather than compensating for the decorrelation produced by phase mixing \footnote{We might add that eigenfunctions of order $n$, associated with large eigenvalues, decay much faster than the characteristic pitch-angle diffusion time, by a factor proportional to $\pi^2 (1+2n)^2$, and therefore act to rapidly smooth small-scale structures \cite{ALBERT_diffusion_solution}. In particular, they may contribute to efficiently damp high-$m$ modes in the distribution function, which according to the von Mises representations are already smaller by a factor $\psi_m \ll 1$ for large $m$ values. It should thus be stressed that the observational phase-mixing effects described here can appear significantly enhanced in the presence of pitch-angle diffusion.}.

Although regimes in which pitch-angle evolution proceeds on timescales much shorter than the drift period are known to occur, particularly for lower-energy electrons ($<100$ keV) subject to strong whistler-mode activity \cite{ArtemyevSSR_Review2025}, such conditions are not typically realised for the more energetic ($\gtrsim$100 keV) electrons considered here \footnote{A notable exception is provided by nonlinear scattering of relativistic electrons by electromagnetic ion-cyclotron waves \cite{Grach_2022, Mourenas_2024}, which can produce rapid pitch-angle evolution under specific conditions.}. In those cases, very rapid pitch-angle evolution can arise only under exceptional conditions in which the magnetospheric plasma becomes more rarefied than usual, allowing unusually strong wave–particle interactions to develop \cite{Agapitov_2019, Hayley_Allison_2021}. Under such conditions, pitch-angle scattering primarily restructures the distribution before significant azimuthal drift occurs, after which the subsequent evolution is again governed by the drift-phase mixing mechanism described here.

A simplifying assumption of the present analysis also concerns the temporal structure of the source responsible for generating drift-phase structures. The calculations presented here assume an impulsive injection for analytical simplicity. In reality, injections may persist over a substantial fraction of a drift period, exhibit time-dependent amplitudes, or arise from a sequence of successive energization episodes. Because the governing drift-kinetic equation is linear, a finite-duration injection can be represented as a superposition of impulsive injections occurring at different times, each undergoing the same differential drift and observational phase mixing derived in the present work. Temporal variability therefore convolves the observational response with the source history rather than removing the phase-mixing mechanism.

Another simplifying assumption of the present analysis concerns the finite resolution of satellite particle detectors in energy and pitch angle. As discussed by \citet{Lejosne_2024}, the finite width of instrument energy channels implies that particles contributing to a given measurement possess slightly different drift frequencies. This produces an additional phase-mixing process associated with the measurement procedure itself. A similar effect arises from the finite pitch-angle aperture of particle detectors, although the resulting mixing is typically weaker because of the weaker dependence of drift frequency on pitch angle. Instrumental phase mixing therefore acts to reduce the measured amplitude and lifetime of drift-phase structures. Neglecting instrument resolution in the present analysis implies that the decorrelation timescales derived here represent upper bounds on their observational lifetime; in practice, such structures will be observed to decay even more rapidly than predicted \footnote{A natural extension of this work is to follow \citet{Lejosne_2024} and incorporate the effects of finite energy and temporal resolution on the spatially localised injections described by Equation~(\ref{eq:St_full}). This would, for example, enable a quantitative assessment of whether particle injections with a given energy and MLT extent, quantified by the parameter $\kappa$, can be resolved by specific instruments.}.

A further effect omitted from the present analysis is the linear response of trapped particles to ULF fluctuations \cite{Southwood81}. In the non-resonant case, the wave-driven response produces a small-amplitude oscillatory contribution to the distribution function that is effectively a drift echo \cite{osmane2023radial}. In that limit, the phase-mixing mechanism identified here is expected to remain unchanged, since the temporal signal still reflects the superposition of neighbouring drift shells with slightly different drift phases. Equivalently, within the correlation-function framework developed here, this contribution corresponds to the $\kappa=0$ limit of a drift-phase structure generated by the linear response to radial transport.

The situation may differ in the case of drift resonance \cite{Southwood81, Degeling11, Artemyev24}. In this regime, coherent wave–particle interactions can lead to a temporal enhancement of the resonant component of the distribution function. Such amplification may partially compensate the loss of phase coherence caused by differential drift and satellite sweeping, potentially resulting in a longer apparent correlation time than in the purely ballistic case considered here \footnote{It should be stressed that drift resonance also produces phase mixing, but on a timescale set by the inverse bounce frequency within the trapping island, distinct from the drift phase-mixing timescale (see, e.g., \cite{Artemyev24} and references therein). Extending the present framework to drift-resonant interactions would also require accounting for wave properties—such as wave-packet size, amplitude, and polarisation—and thus warrants a dedicated study.}. The decorrelation times obtained in this communication should therefore be interpreted as broadly applicable to freely evolving or non-resonantly forced drift-phase structures, whereas resonantly maintained structures require a separate treatment.

A similar caveat must also be considered for particles interacting nonlinearly with ULF waves \cite{Lili21, Zhou_review_ULF_WPI}. In this regime, nonlinear wave–particle interactions can lead to phase trapping, allowing particles to remain confined within a limited range of magnetic local time as long as the wave potential is sufficiently strong. Under such conditions, the wave may temporarily prevent the dispersion of particles along the drift direction, thereby maintaining drift-phase coherence momentarily, before phase-mixing associated with nonlinear processes or wave decorrelation eventually kick in.

The phase-mixing mechanism described in the present work therefore applies primarily to freely evolving or weakly forced particle populations. In this context, it is worth emphasising that non-diffusive radial transport associated with ULF waves is expected to play a more prominent role in the global evolution of radiation belt electron populations than local transport driven by VLF/ELF waves. This reflects a separation of scales: global changes occur on timescales of a few drift periods, over which azimuthal drift generates increasingly sharp radial gradients in phase-space density that favour radial transport. By contrast, gradients in energy and pitch-angle space typically relax rapidly and are less efficiently maintained, so that local wave-particle interactions driven by very low frequency waves remains effectively diffusive on these timescales. Conversely, the persistence of coherent drift-phase structures beyond the timescales predicted here could provide evidence for nonlinear trapping by ULF waves. In this sense, the theoretical framework developed here may also serve as a diagnostic tool for distinguishing linear from nonlinear ULF wave–particle interactions, and diffusive from non-diffusive radial transport in radiation belt observations.

\subsection{Consequences for radiation belt observations}

\subsubsection{Interpretation of radial diffusion signatures}

Radial diffusion arises from a quasilinear description of the interaction between magnetically trapped particles and electromagnetic fluctuations that violate the third adiabatic invariant \cite{Falthammar65, Parker60, Elkington99}. As in all quasilinear treatments \cite{Diamond10}, it is assumed that the cumulative effect of many waves on the particle population evolves slowly compared with the rapid particle motion, which in this context corresponds to the azimuthal drift of particles around the planet. When the characteristic diffusion timescale approaches the drift period, this fundamental scale separation breaks down. The requirement that diffusion operates on timescales much longer than the drift period underpins both theoretical derivations \cite{Lejosne20, osmane2021radial, osmane2023radial} and numerical implementations of radial diffusion models \cite{Riley92, Sasha_poops_on_radial_diffusion, Li_Mann24}. Without it, the assumptions that allow particle dynamics to be reduced to a Fokker--Planck description in radial coordinate become difficult to justify.

In the present work we have shown that even in the absence of wave–particle interactions, violations of adiabatic invariants, or any diffusive process, satellite measurements would be unable to resolve spatial fluctuations in the distribution function. As phase mixing removes the higher azimuthal modes, leaving only the $m=0$ component, the observed signal becomes indistinguishable from that expected from radial diffusion. This ambiguity is not merely interpretive: diffusion-based models rely on stochastic transport and phase-space mixing assumptions that are not satisfied in the scenario described here. As a consequence, inferring diffusion from such observations may lead to incorrect estimates of acceleration and loss rates associated with radial transport and, ultimately, to biased predictions of long-term radiation-belt flux evolution.

While this evolution is purely a reversible rearrangement of phase space, it nevertheless generates progressively finer filamentary structure in $L$, and therefore increasingly steep radial gradients in phase-space density. Such gradients can render even weak electromagnetic perturbations dynamically important, since the efficiency of wave-driven transport scales with the background radial gradient. In this sense, ballistic phase mixing may naturally precondition the system for subsequent radial transport: the filamentation it produces enhances the effectiveness of linear wave–particle interactions, which can then act to smooth the fine-scale structure and introduce genuine irreversibility. The apparent diffusive signatures discussed above may therefore arise prior to any true, or irreversible, drift shell transport, while simultaneously setting the stage for its eventual onset \footnote{One might argue that the steep radial gradients generated by ballistic motion could enable rapid radial diffusion. However, such gradients also enhance the efficiency of linear wave–particle interactions, whose response scales with the background gradient. As a result, the fine-scale structure produced by ballistic motion is expected to be modified on timescales shorter than those required for a consistent quasilinear (Fokker–Planck) description of radial diffusion to apply.}.

Observational signatures commonly interpreted as evidence for radial diffusion may therefore not uniquely diagnose a genuinely diffusive transport process. Ballistic phase mixing, combined with the unavoidable sampling effects associated with satellite motion, can produce temporal signals that closely resemble those predicted by radial diffusion models. The apparent smoothing of spatial structure in spacecraft measurements may thus arise from the loss of phase coherence across neighbouring drift shells rather than from stochastic transport driven by electromagnetic fluctuations.
Consequently, caution is required when interpreting azimuthally averaged phase-space density profiles or when inferring radial diffusion coefficients directly from spacecraft observations.

Drift echoes and more general drift-phase structures provide another example of the ambiguity inherent in interpreting spacecraft observations in terms of radial diffusion. In some studies, the presence of drift-periodic flux oscillations has been used to estimate radial diffusion coefficients by reproducing the observed oscillations using broadband ULF wave fields and relating the resulting particle displacements to radial transport \cite{https://doi.org/10.1029/2025JA034549}. In others, the relative absence of observable drift-phase structure during storm-time flux enhancements has been interpreted as being entirely consistent with quasilinear radial diffusion driven by broad-band, random-phase ULF fluctuations \cite{Obrien_drift_phase}. In both cases, however, the underlying difficulty is the same: drift-phase structure is a drift-timescale signature of azimuthal phase coherence, whereas radial diffusion is a drift-phase-averaged description of slow stochastic transport operating over many drift periods.

Any electromagnetic perturbation of the magnetically trapped particles, broadband or not, resonant or non-resonant, produces azimuthal phase structure in the distribution function and thus generate drift echoes when sampled by an orbiting spacecraft \cite{osmane2023radial}. Conversely, the results obtained here show that such drift-phase structure is also expected to decorrelate observationally through the combined effects of differential drift and satellite sweeping across neighbouring drift shells, even when the underlying dynamics remain strictly ballistic. The disappearance, or damping of drift echoes and other drift-phase signatures therefore cannot be uniquely interpreted as evidence for radial diffusion. In particular, the lack of strong observable drift-phase structure near \(\sim 1\) MeV reported by \citet{Obrien_drift_phase} and interpreted as evidence of quasi-linear radial diffusion due to broadband ULF wave-particle interactions (see last paragraph of p.15) may admit a simpler explanation: the most energetic particles drift most rapidly and are therefore also the most susceptible to phase mixing, so that their drift-phase structure is lost more rapidly than that of lower-energy seed populations. 

The results obtained here therefore introduce an additional consideration in the interpretation of radiation-belt observations. The damping of drift echoes and related drift-phase signatures in spacecraft data cannot be uniquely attributed to radial diffusion, or to interactions with broadband ULF waves, and may instead arise from this observational phase-mixing process.

\subsubsection{Injection-driven radiation belt enhancements}
Recent studies have documented intermittent, spatially localised injection events that introduce energetic and relativistic particles into the outer radiation belt on timescales of tens of minutes, producing multi-energy flux enhancements that can persist for several hours \cite{Turner17, Reeves_injection}. Building on these observations, it has been proposed that such injections may contribute to the rapid replenishment of the belts \cite{Kim2021_injection, Kim2023_injection}. In this scenario, injections provide a direct mechanism for replenishing the radiation belts on timescales much shorter than those typically associated with diffusive transport.

While the present work does not directly test this injection-driven scenario, our results indicate that even if such injections were responsible for radiation belt enhancements, the resulting observational signatures could nonetheless resemble those expected from radial diffusion. Because phase mixing rapidly erases the higher azimuthal modes of an MLT-localised population, spacecraft measurements would quickly lose sensitivity to the spatial structure of the injection. Within only a few drift periods, the observed distribution would therefore appear azimuthally smoothed, even though the underlying dynamics remain inherently bursty and non-diffusive.

Consequently, injection-driven enhancements may become observationally indistinguishable from slowly evolving diffusive processes when sampled along spacecraft trajectories. The rapid loss of drift-phase coherence demonstrated here therefore complicates attempts to infer the physical origin of radiation belt enhancements from temporal flux measurements alone.

\subsubsection{Phase-space density measurements}
Over the past decades, two broad classes of processes have been identified as key drivers of particle transport and acceleration in Earth’s radiation belts: spatially localised wave–particle interactions and radial diffusion driven by ultra-low-frequency fluctuations \citep{Thorne10}. Although these processes operate across very different scales, they can be understood within a common framework based on adiabatic invariants in nearly periodic Hamiltonian systems \citep{Cary09}.

Within this framework, the central issue is whether particle motion remains adiabatic or whether one or more invariants are violated. Radial diffusion is associated primarily with violations of the third invariant, leading to transport across drift shells and, through the increasing magnetic field strength at lower radial distances, to energisation via the betatron effect. In contrast, wave–particle interactions tend to violate the first and second invariants, producing diffusion in velocity space and more localised changes in the particle distribution.

This distinction underpins the widespread use of phase-space density expressed in terms of adiabatic invariants. Because phase-space density is conserved under adiabatic motion \cite{Kim_Adiabatic_Chan}, it provides a natural coordinate system in which variations due to changing magnetic geometry can be separated from genuinely non-adiabatic processes. As a result, radial diffusion and localised acceleration produce qualitatively different signatures in invariant space, making phase-space density a central diagnostic for interpreting radiation belt observations \citep{Green04, Chen07, Reeves13, Thorne13, Jaynes18, Drozdov_2022, Kalliokoski22b, Olifer24, Hensley_George, Ozeke25}.

The present results introduce an important observational consideration regarding the use of phase-space density in radiation belt studies. Phase-space density is defined locally in phase space and, when evaluated for a fixed set of adiabatic invariants, does not mix information originating from different drift phases and drift shells. However, spacecraft measurements used to construct radial profiles of phase-space density are obtained along trajectories that sweep across a range of magnetic local times and drift shells. When phase-space density is subsequently organised as a function of $L^*$, measurements acquired at different local times are often combined within broad spatial bins and over time intervals that may span several particle drift periods.

Under these conditions, the aggregation of measurements from different drift phases and drift shells can implicitly average spatially localised structures. In light of the phase-mixing mechanism identified here, such averaging can obscure the drift-phase structure of the underlying distribution function. As a result, even when expressed in invariant space, phase-space density profiles may mask spatial localisation and produce signatures that resemble the gradual evolution expected from diffusive transport. This introduces an ambiguity in interpretation: if radial diffusion cannot be reliably identified, then the relative roles of local wave–particle interactions and other acceleration or loss processes cannot be robustly determined. The loss of drift-phase coherence therefore affects not only flux measurements but also the physical interpretation of phase-space density profiles commonly used to diagnose radiation belt dynamics.

\subsection{Broader implications for planetary \& substellar magnetospheres}

\subsubsection{Gas giant radiation belts}
At the giant planets, the application of radial diffusion to radiation belt transport was strongly influenced by observations of absorption signatures produced by moons embedded within the magnetosphere. These signatures appear as persistent flux depletions known as \emph{microsignatures} for energetic electrons and \emph{macrosignatures} for protons and ultrarelativistic electrons above $\sim$10~MeV \cite{Thomsen1977, vanAllen1980Mimas, 1993AdSpR..13j.221S, Roussos07, Andri2011, Kollmann2011jgr, Roussos2018D}. Because these depletions result from irreversible particle losses to moons or rings, their subsequent evolution provides a natural tracer of particle transport in planetary magnetospheres. 

Early analyses, notably those led by Van Allen and collaborators  \cite{Thomsen1977, vanAllen1980Mimas} interpreted the gradual refilling of these absorption features using radial diffusion models, in which quasi-linear theory is used to infer transport rates from the observed flux profiles \cite{1983JGR....88.8947C, 2007JGRA..112.6214R}. This interpretation has played a central role in establishing radial diffusion as the standard framework for describing particle transport in planetary radiation belts. 

However, observations have consistently shown that microsignatures can refill on timescales comparable to, or even shorter than, a single particle drift period \cite{Roussos07, Andri2011, Roussos2018D}. Such rapid evolution is difficult to reconcile with classical quasi-linear radial diffusion, which fundamentally requires many drift periods for particles to lose memory of the electromagnetic fields responsible for radial transport. 
Recent theoretical work by \citet{Osmane2025} has proposed an alternative framework that explicitly accounts for processes operating on drift timescales. In this model, strongly localised loss regions associated with moons can synchronise the azimuthal Fourier modes of the particle distribution function, producing an apparent refilling of depleted regions through phase-space synchronisation rather than diffusive transport. Within this framework, the evolution of microsignatures can possibly emerge from the collective phase dynamics of these Fourier modes, a behaviour mathematically analogous to a generalized Kuramoto model \cite{Osmane2025}.

The phase-mixing mechanism identified in the present work introduces an additional observational complication for the interpretation of these signatures and offers an alternative explanation that does not rely on radial diffusion, complementing the non-diffusive radial transport framework proposed by \citet{Osmane2025}. Absorption features produced by moons are intrinsically localised in drift phase at the moment of their creation. As the depleted particle population drifts azimuthally around the planet, neighbouring drift shells rotate with slightly different angular velocities. When such a structure is sampled along the trajectory of an orbiting spacecraft, successive measurements probe regions characterised by different drift phases rather than a single coherent depletion. The resulting signal therefore undergoes rapid decorrelation through the combined effects of differential drift and spacecraft sweeping across neighbouring drift shells.

Under these conditions, a spatially localised absorption feature may appear to refill gradually in time even in the absence of true diffusive transport. In other words, the observational loss of drift-phase coherence can convert the ballistic evolution of a localised structure into a temporal signature resembling the smooth recovery expected from radial diffusion. If this effect is not explicitly accounted for, transport rates inferred from the apparent refilling of microsignatures may therefore be systematically overestimated. The results presented here thus suggest that some classical interpretations of microsignature evolution, and the radial diffusion coefficients derived from them, may need to be reconsidered, particularly in environments where particle dynamics are strongly structured in magnetic local time.

\subsubsection{Brown dwarfs magnetospheres}
Recent observations have revealed the existence of radiation belts around ultracool brown dwarfs, extending the phenomenon of magnetically trapped energetic particles far beyond the solar system \cite{Kao23_BD, Climent_BD}. In these systems, the particle populations are inferred indirectly from synchrotron emission rather than from \textit{in situ} measurements. Remarkably, recent work has shown that radiation belts spanning planets and brown dwarfs follow a common universal power law relating the spatial distribution of energetic electrons to synchrotron losses and radial transport \cite{Turner_BD_2026}. In this framework, the inward transport of particles toward regions of stronger magnetic field leads to non-adiabatic energisation while synchrotron radiation provides the dominant loss mechanism. Importantly, the transport required to produce this universal scaling does not need to be diffusive; any process capable of moving particles radially while approximately preserving the first adiabatic invariant can reproduce the observed behaviour.

In this context, the results presented here highlight the need for caution when interpreting radial transport in such poorly constrained environments. Since the electromagnetic fluctuations responsible for particle acceleration and transport in brown dwarf magnetospheres remain essentially unknown, the assumption that radial transport operates in a quasi-linear diffusive regime is difficult to justify \textit{a priori}. More generally, our results demonstrate that observational signatures resembling diffusive evolution can arise from phase mixing of spatially localised structures even in the absence of true diffusion. For unresolved systems such as brown dwarf radiation belts, where particle dynamics can only be inferred indirectly from global emission, distinguishing between diffusive and non-diffusive transport mechanisms may therefore prove particularly challenging.

\section{Conclusion}\label{Conclusion}

For more than six decades, the dynamics of energetic particles in planetary radiation belts have largely been interpreted within the framework of radial diffusion. In this work we have identified a previously unrecognised source of epistemic uncertainty in the interpretation of spacecraft observations of these systems, that is, an uncertainty arising from incomplete knowledge of the system rather than from intrinsic stochastic variability \cite{DerKiureghian2009}. Even when the underlying particle dynamics are purely ballistic, the combination of azimuthal drift and satellite sweeping across neighbouring drift shells produces an unavoidable observational phase-mixing process. Because particles on adjacent drift shells drift with slightly different angular frequencies, and because spacecraft sample a continuum of drift shells along their trajectories, temporal measurements progressively lose the phase coherence associated with the original spatial localisation of the particle population.

The principal consequence is that information about the spatial and temporal structure of the distribution function is rapidly erased in the observational signal on timescales of a few drift periods. Structures that are initially localised in magnetic local time evolve into temporally decorrelated signatures when sampled by orbiting spacecraft. From an observational point of view, the radiation belts therefore appear as a nearly viscous medium of energetic particles, even though the underlying dynamics may remain collisionless and non-diffusive, with the apparent transport arising from phase mixing rather than from a stochastic diffusion process.

This result has several implications for the interpretation and modelling of radiation belt dynamics. First, it suggests that the empirical success of radial diffusion models may in part reflect the observational phase-mixing identified here. Because spacecraft measurements naturally lose information about the drift-phase structure of particle populations, temporal signals may mimic the signatures expected from diffusive transport even when no physical diffusion is present. This ambiguity is not merely interpretive: diffusion-based models rely on stochastic transport and phase-space mixing assumptions that are not satisfied in the scenario described here. As a consequence, inferring diffusion from such observations may lead to incorrect estimates of acceleration and loss rates associated with radial transport and, ultimately, to biased predictions of long-term radiation-belt flux evolution. Second, it highlights a limitation of commonly used diagnostics based on phase-space density. Although phase-space density is typically evaluated locally on individual drift shells and within restricted magnetic local time sectors, the aggregation of measurements over timescales comparable to or exceeding particle drift periods can obscure spatially localised enhancements such as the injection events reported by \citet{Kim2021_injection, Kim2023_injection}.

The mechanism described here also has implications for the interpretation of radiation belts throughout the solar system and beyond. Radial diffusion has long served as the standard framework for describing energetic particle transport in planetary magnetospheres because of its conceptual simplicity and its success in reproducing many observational features \cite{vanAllen1980Mimas, Roussos07, Jaynes18, Lejosne20}. However, several environments already exhibit behaviour that cannot be fully captured by purely diffusive models. In the radiation belts of Jupiter and Saturn, for example, spatially localised injections and absorption by moons introduce strongly non-axisymmetric structures that require a more detailed dynamical description. At the other extreme, recently discovered radiation belts around ultra-cool brown dwarfs appear to operate in regimes where intense synchrotron losses and extremely rapid rotation can modify particle trapping on timescales comparable to, or shorter than, a single drift period \cite{Turner_BD_2026}.

Taken together, these considerations suggest that the interpretation of radiation belt observations must account not only for the underlying particle dynamics but also for the observational filtering imposed by spacecraft trajectories. Observational phase mixing provides a natural mechanism through which deterministic drift-phase structures may become observationally indistinguishable from stochastic transport when sampled by sparsely distributed spacecraft. In addition to reinterpreting existing measurements, the framework developed here also provides a means of assessing which spacecraft configurations and orbital separations are capable of resolving spatially localised particle populations before observational phase mixing obscures them, offering a quantitative tool for assessing whether existing or future spacecraft configurations possess sufficient spatio-temporal coverage to resolve evolving drift-phase structures before observational phase mixing obscures them. In this context, combining measurements from multi-spacecraft constellations and equatorial missions on elliptical orbits, which sample a wide range of $L$-shells over several hours \cite{Mauk2013, Miyoshi2018, Morley2016, Olifer24, Hensley_George}, with low-altitude satellites that traverse the same region on timescales of minutes and provide near-instantaneous snapshots \cite{ELFIN_Angelopoulos, CIRBE, Hayley18, Sun24}, may offer a particularly powerful approach. Such complementary observations could enable a direct comparison between instantaneous spatial structure and temporally evolving signals, thereby providing a means to disentangle genuine temporal evolution from apparent decorrelation produced by spacecraft sampling and differential drift. Future efforts may also explore whether data-assimilation techniques, Kalman filtering \cite{Castillo_Shprits}, or physics-informed machine-learning approaches \cite{Camporeale2022} are capable of recovering fine-scale drift-phase structure from sparse observations, or whether observational phase mixing fundamentally limits such reconstructions. In parallel, future modelling efforts that explicitly incorporate spacecraft sampling and phase-mixing effects will enable a quantitative separation of intrinsic particle dynamics from observational artefacts, allowing diffusive and non-diffusive behaviour to be unambiguously distinguished. Ultimately, these results highlight how measurement geometry can fundamentally shape the physical interpretation of magnetised dilute plasma environments, from the Earth’s radiation belts to planetary magnetospheres and the recently discovered belts of ultra-cool brown dwarfs.

\begin{acknowledgments}
This research was supported by the International Space Science Institute (ISSI) in Bern, through the ISSI International Team project Beyond Diffusion: Advancing Earth’s Radiation Belt Models with Nonlinear Dynamics (ISSI Team project \# 25-640). XA and AVA were supported by NASA grant \# 80NSSC22K1634 and and XA was additionally supported by NSF grant \# 2108582. OA would like to acknowledge the United Kingdom Research and Innovation (UKRI) Natural Environment Research Council (NERC) Independent Research Fellowship NE/V013963/1 and NE/V013963/2. JA was supported by AFOSR grant
25RVCOR003 and the Air Force Research Laboratory. MH acknowledges funding from the GACR Grant 25-18095X and from the Johannes Amos Comenius Programme (P JAC), project \# CZ$.02.01.01/00/22\_008/0004605$, Natural and anthropogenic georisks. We thank Greg Cunningham, Alexander Drozdov, Milla Kalliokoski, Melodie Kao, Peter Kollmann, Leon Olifer, Daniel Ratliff, Elias Roussos, and Lucile Turc for helpful and insightful discussions. Open access for this publication was generously supported by Helsinki University Library.\\
\end{acknowledgments}

\appendix

\section{Derivation of the bounced-averaged drift kinetic equation}\label{_Appendix_A}
\subsection{Dipole magnetic field model}\label{secA1}
We consider a centred magnetic dipole with magnetic moment $M=B_ER_E^3$. In spherical coordinates
$(r,\theta,\varphi)$, with colatitude $\theta$ and azimuthal angle $\varphi$, the field components are given by:
\begin{equation}
B_r=\frac{2M\cos\theta}{r^3},\qquad
B_\theta=\frac{M\sin\theta}{r^3},\qquad
B_\varphi=0,
\end{equation}
and the magnetic field magnitude by
\begin{equation}
B(r,\theta)=\frac{M}{r^3}\sqrt{1+3\cos^2\theta}.
\end{equation}
From here, we switch to the magnetic latitude $\lambda=\frac{\pi}{2}-\theta$, which is more commonly used in radiation belts' studies, and thus $\sin(\theta)=\cos(\theta)$ and $\cos(\theta)=\sin(\lambda)$. 

\noindent The magnetic field lines can be traced by solving the field-line equation (see, e.g. \citet{longcope2005topological})
\begin{equation}
\frac{dr}{B_r}=\frac{rd\theta}{B_\theta},
\end{equation}
which, upon integration, yields
\begin{equation}
r,\sin^2\theta=\text{constant}.
\end{equation}
Thus, when particles move along a field line between coordinates $(r_1, \theta_1)$ and  $(r_2, \theta_2)$, the quantity $r \sin^2(\theta)$ remains constant.  It is conventional to write the constant in terms of McIlwain's drift-shell parameter $L$, which corresponds to the equatorial crossing distance normalised with the planetary radius. Thus the field line equation is given by: 
\begin{equation}
r(\theta)=L R_E \sin^2(\theta),
\end{equation}
or when written in terms of magnetic latitude, as: 
\begin{equation}
r(\lambda)=L R_E \cos^2(\lambda),
\end{equation}
The magnetic field components, written in terms of the drift shell and the magnetic latitude are given by: 
\begin{equation}
r=\frac{2B_E}{L^3} \frac{\sin(\lambda)}{\cos^6(\lambda)}\qquad B_\lambda=-\frac{B_E}{L^3} \frac{1}{\cos^5(\lambda)} \qquad B_\varphi=0 \ 
\label{eq:dipole_B_L_lambda}
\end{equation}
and the magnetic magnitude by
\begin{equation}
B(L,\lambda)=\frac{B_E}{L^3}
\frac{\sqrt{1+3\sin^2\lambda}}{\cos^6\lambda}.
\label{eq:dipole_magnitude_B_L_lambda}
\end{equation}
Since, we use the orthogonal basis $\{\boldsymbol{\hat{e}}_r, \boldsymbol{\hat{e}}_\lambda, \boldsymbol{\hat{e}}_\varphi\}$, instead of $\{\boldsymbol{\hat{e}}_r, \boldsymbol{\hat{e}}_\theta, \boldsymbol{\hat{e}}_\varphi\}$, the component $B_\lambda=\boldsymbol{B}\cdot\boldsymbol{\hat{e}}_\lambda=\boldsymbol{B}\cdot(-\boldsymbol{\hat{e}}_\theta)=-B_\theta$. In order to compute the guiding centres, we need an expression for the local magnetic field direction, defined as the unit vector $\boldsymbol{b}=\boldsymbol{B}/B$. Using the dipole field components expressed in terms of the magnetic latitude $\lambda$, the field-line direction depends only on $\lambda$ and can be written as
\begin{equation}
    \boldsymbol{b}(\lambda)=\frac{2\sin(\lambda) \boldsymbol{\hat{e}}_r-\cos(\lambda)\boldsymbol{\hat{e}_\lambda}}{\sqrt{1+3\sin^2(\lambda)}}
\end{equation}

\noindent Using the Lamé coefficients $(h_r, h_\lambda, h_{\varphi})=(1,\ r, \ r\cos(\lambda))$, one can verify, as a useful sanity check, that the magnetic field defined above is divergence-free for $r\neq 0$: 
\begin{eqnarray}
    \nabla\cdot\boldsymbol{B}&=&\frac{1}{h_rh_\lambda h_\varphi}\bigg{[}\frac{\partial}{\partial r}(h_\lambda h_\varphi B_r)+\frac{\partial}{\partial \lambda}(h_r h_\varphi B_\lambda)+\frac{\partial}{\partial \varphi}(h_r h_\lambda B_\varphi) \bigg{]} \nonumber\\ 
    &=&\frac{1}{r^2\cos(\lambda)}\left[\frac{\partial}{\partial r}(r^2\cos(\lambda) B_r)+\frac{\partial}{\partial \lambda}(r\cos(\lambda) B_\lambda)+\frac{\partial}{\partial \varphi}(r B_\varphi)\right] \nonumber \\ 
    &=&\frac{1}{r^2}\frac{\partial}{\partial r}(r^2 B_r)+\frac{1}{r^2\cos(\lambda)}\frac{\partial}{\partial \lambda}(r\cos(\lambda) B_\lambda)  \\
    &=&\frac{1}{r^2}\frac{\partial }{\partial r}\left(\frac{2B_ER_E^3\sin(\lambda)}{r}\right)-\frac{1}{r^2\cos(\lambda)}\frac{\partial}{\partial \lambda}\left(\frac{B_ER_E^3\cos^2(\lambda)}{r^2}\right) \nonumber \\
    &=&0, \ (\text{for} \ r\neq 0).\nonumber
\end{eqnarray}

\noindent With the inhomogeneous magnetic field expressed in terms of Equations (\ref{eq:dipole_B_L_lambda}) and the basis $\{\boldsymbol{\hat{e}}_r, \boldsymbol{\hat{e}}_\lambda, \boldsymbol{\hat{e}}_\varphi\}$, one can compute the guiding centre orbits.


\subsection{Guiding centre trajectories}\label{secB1}
In this work, we use the relativistic guiding-centre equations, first derived by \citet{Vandervoort} (see also \citet{northrop1963adiabatic}, \citet{Hazeltine18}, \citet{Cary09} and \citet{Son_Relativistic_GC}, especially the later for a short but elegant Hamiltonian derivation), to compute particle trajectories:
\begin{eqnarray} 
\label{relativistic_guiding_center}
{{\langle\boldsymbol{{\dot{r}}}\rangle}_\phi}&\simeq& v_\parallel\mathbf{b}+\frac{\boldsymbol{E}\times \mathbf{b}}{B}-\frac{v_\parallel^2}{\Omega_s}\mathbf{(b\cdot\nabla) b}\times\mathbf{b}{-\frac{\mu}{\gamma q_s B}{\nabla B}\times \mathbf{b}}\nonumber \\ 
&=& v_\parallel\mathbf{b} + \boldsymbol{v_E}+\boldsymbol{v}_{\boldsymbol{\kappa}}+\boldsymbol{v}_{\nabla B}
\end{eqnarray}
\noindent In Equation~(\ref{relativistic_guiding_center}), the guiding-centre trajectory is determined by the electric and magnetic fields, \(\boldsymbol E\) and \(\boldsymbol B\), and by the first adiabatic invariant \(\mu_s\) given by
\begin{equation}
\mu_s=\frac{p_\perp^2}{2m_s B},
\end{equation}
where \(p_\perp=m_s\gamma v_\perp\) is the relativistic perpendicular momentum. The Larmor frequency for species \(s\) is
\begin{equation}
\Omega_s=\frac{q_s B}{\gamma m_s},
\end{equation}
which includes the relativistic Lorentz factor 
\[
\gamma=\sqrt{1+\frac{p_\perp^2}{m_s^2c^2}+\frac{p_\parallel^2}{m_s^2c^2}}.
\]
Equation ~(\ref{relativistic_guiding_center}) indicates that, in addition to the parallel motion along the local magnetic field, the particle experiences the $E$ cross $B$ drift $\boldsymbol{v_E}$, the curvature drift $\boldsymbol{v_\kappa}$ and the magnetic grad-B drift $\boldsymbol{v}_{\nabla B}$. Equation ~(\ref{relativistic_guiding_center}) must be supplemented with an evolution equation for the parallel velocity $v_\parallel$ or equivalently, the relativistic energy $ \mathcal{E}$:
\begin{equation}
    \mathcal{E}=m c^2 \gamma=m c^2 \sqrt{1+\frac{p_\parallel^2}{m^2 c^2}+\frac{2\mu B}{m c^2}}. 
\end{equation}
Following the notation of \citet{Hazeltine18}, the evolution equation for $\mathcal{E}=\gamma m c^2 $ can be written as: 
\begin{equation}\label{relativistic_energy}
    \frac{d \mathcal{E}}{dt} =q \boldsymbol{E}\cdot {{\langle\boldsymbol{{\dot{r}}}\rangle}_\phi}+\mu \mathbf{b}\cdot \nabla \times \boldsymbol{E}.
\end{equation}
For the problem considered here, the electric field is assumed to be negligible, and therefore the relativistic energy is also conserved. 
Then accounting for $\mu$ and $\mathcal{E}$ conservation, the evolution equation for the relativistic parallel momentum is given by: 
\begin{eqnarray}
    \label{EQ:parallel_guiding_centre_equation}
    \frac{dp_\parallel}{dt}&=&-\frac{\mu}{\gamma v_\parallel}{{\langle\boldsymbol{{\dot{r}}}\rangle}_\phi}\cdot\nabla B \nonumber \\
    &=&-\frac{\mu}{\gamma}\nabla B \cdot \left(\boldsymbol{b}-\frac{v_\parallel}{\Omega_s}(\boldsymbol{b}\cdot\nabla\boldsymbol{b})\times\boldsymbol{b}\right),
\end{eqnarray}
where we made use of the fact that the Lagrangian derivative of the magnetic field magnitude is entirely given by the convective term, i.e., $\frac{dB}{dt}={{\langle\boldsymbol{{\dot{r}}}\rangle}_\phi}\cdot\nabla B$, since $\partial B/\partial t=0$.

\noindent The guiding-centre orbits given by Equations (\ref{relativistic_guiding_center}) and (\ref{EQ:parallel_guiding_centre_equation}) can therefore be simplified to: 
\begin{equation}
\boxed{
\begin{aligned}
\big\langle \dot{\boldsymbol r} \big\rangle_\phi
&\simeq
v_\parallel \boldsymbol b
-\frac{v_\parallel^2}{\Omega_s}\,(\boldsymbol b\cdot\nabla)\boldsymbol b\times\boldsymbol b
-\frac{\mu}{\gamma q_s B}\,\nabla B\times \boldsymbol b,
\\[4pt]
\big\langle \dot v_\parallel \big\rangle_\phi
&\simeq
-\frac{\mu}{m\gamma^2}\,
\nabla B\cdot
\left(
\boldsymbol b
-\frac{v_\parallel}{\Omega_s}(\boldsymbol b\cdot\nabla\boldsymbol b)\times\boldsymbol b
\right).
\end{aligned}
}
\label{EQ:Final_relativistic_guiding_centres}
\end{equation}


\noindent Equations (\ref{EQ:Final_relativistic_guiding_centres}) are averaged over the gyrophase $\phi$ and are valid in the limit of long wavelengths relative to the Larmor radius $\rho=v_\perp/\Omega_s$ and low-frequency fluctuations relative to the Larmor frequency $\Omega_s$, conditions under which the first adiabatic invariant $\mu$ is conserved \footnote{We reiterate that in the present model the electric field $\boldsymbol E$ and non-adiabatic processes are neglected. Moreover, for the axisymmetric dipole geometry considered here one has $\boldsymbol b\cdot(\nabla\times\boldsymbol b)=0$, so the Ba\~nos correction for guiding-centres \cite{banos_1967}, which is often ignored, but can be significant in the presence of parallel currents, here vanishes and the first adiabatic invariant is exactly conserved for our magnetic field model, $d\mu/dt=0$, to guiding-centre order \cite{Hazeltine73}.}.

\paragraph{Grad-$B$ drift $\boldsymbol{v}_{\nabla B}$.}
Let's compute the guiding centre orbits as per Equations (\ref{EQ:Final_relativistic_guiding_centres}) and given for the time-independent inhomogeneous magnetic field given by Equation (\ref{eq:dipole_B_L_lambda}). Let's start with computing the grad-B drift:
\begin{eqnarray}
\nabla B
&=&
\hat{\boldsymbol e}_r\,\frac{\partial B}{\partial r}
+
\hat{\boldsymbol e}_\lambda\,\frac{1}{r}\frac{\partial B}{\partial \lambda}
+
\hat{\boldsymbol e}_\varphi\,\frac{1}{r\cos\lambda}\frac{\partial B}{\partial\varphi} \nonumber \\
&=&-\frac{3B_E R_E^3}{r^4} \sqrt{1+3\sin^2(\lambda)}\hat{\boldsymbol e}_r+\frac{3B_E R_E^3}{r^4} \frac{\sin(\lambda)\cos(\lambda)}{\sqrt{1+3\sin^2(\lambda)}}\hat{\boldsymbol e}_\lambda\, 
\end{eqnarray}
For convenience, we define
\begin{equation}
D(\lambda)= \sqrt{1+3\sin^2\lambda},
\label{EQ_Famous_D_lambda}
\end{equation}
so that the gradient $\nabla B$ can be written as: 
\begin{equation}
\nabla B=-\frac{3B_E R_E^3}{r^4} D(\lambda)\boldsymbol {\hat{e}}_r+\frac{3B_E R_E^3}{r^4} \frac{\sin(\lambda)\cos(\lambda)}{D(\lambda)}\boldsymbol{\hat{e}}_\lambda\
\end{equation}
and with it, the magnetic drift $\boldsymbol{v}_{\nabla B}$ can be computed
\begin{eqnarray}
\label{EQ:Final_gradB_drift}
\boldsymbol{v}_{\nabla B}
&=&-\frac{\mu_s}{\gamma q_s B}\nabla B \times \boldsymbol{b} \nonumber \\
&=&
-\frac{\mu_s}{\gamma q_s B}\left[\left(-\frac{3B_E R_E^3}{r^4}D\right)\left(-\frac{\cos\lambda}{D}\right)
-\left(\frac{3B_E R_E^3}{r^4}\frac{\sin\lambda\cos\lambda}{D}\right)\left(\frac{2\sin\lambda}{D}\right)\right]\boldsymbol{\hat{e}}_\varphi
\nonumber\\
&=&-\frac{\mu_s}{\gamma q_s B} \frac{3B_E R_E^3}{r^4}\,
\frac{\cos\lambda\left(1+\sin^2\lambda\right)}{1+3\sin^2\lambda}\;\boldsymbol{\hat{e}}_\varphi \nonumber \\
&=&-\frac{3\mu_s}{\gamma q_s L R_E} \frac{1+\sin^2(\lambda)}{\cos(\lambda)} \frac{1}{D^2(\lambda)}\boldsymbol{\hat{e}}_\varphi
\end{eqnarray}
As a result, particles confined to the equatorial plane ($\lambda=0$) experience a grad-$B$ drift given by
\begin{eqnarray}
\boldsymbol{v}_{\nabla B}
&=&
-\frac{3\mu_s}{\gamma q_s L R_E}\,\boldsymbol{\hat e}_\varphi .
\end{eqnarray}
At all longitudes, negative charges ($q_s<0$) drift eastward ($+\boldsymbol{\hat e}_\varphi$), whereas positive charges ($q_s>0$) drift westward ($-\boldsymbol{\hat e}_\varphi$).
\paragraph{Curvature drift $\boldsymbol{v_\kappa}$}
We start with the computation of the curvature vector $\boldsymbol\kappa=(\boldsymbol b\cdot\nabla)\boldsymbol b$. Since the local unit vector $\boldsymbol b$ has no $\varphi$-component and is axisymmetric, the dot product of $\boldsymbol{b}$ and the gradient operator is given by: 
\begin{equation}
\boldsymbol b\cdot\nabla
=
\frac{2\sin(\lambda)}{D}\,\frac{\partial}{\partial r}
-\frac{\cos(\lambda)}{r D}\,\frac{\partial}{\partial \lambda}.
\end{equation}
Moreover, since the unit vectors $\boldsymbol{\hat{e}}_r, \boldsymbol{ \hat{e}}_\lambda$ do not depend on $r$, $\frac{\partial \boldsymbol b}{\partial r}=\boldsymbol{0}$, and the curvature vector can be simplified to: 
\begin{equation}
(\boldsymbol b\cdot\nabla)\boldsymbol b
=
-\frac{\cos(\lambda)}{r D}\,\frac{\partial\boldsymbol b}{\partial\lambda}.
\end{equation}
Differentiating and retaining the $\lambda$-dependence of the basis vectors, we find:
\begin{eqnarray}
\frac{\partial\boldsymbol b}{\partial\lambda}
&=&
\frac{3\left(1+\sin^2\lambda\right)}{\left(1+3\sin^2\lambda\right)^{3/2}}
\left(\cos\lambda\,\hat{\boldsymbol e}_r+2\sin\lambda\,\hat{\boldsymbol e}_\lambda\right).
\end{eqnarray}
Therefore, the curvature vector is given by: 
\begin{eqnarray}
\boldsymbol\kappa
=(\boldsymbol b\cdot\nabla)\boldsymbol b
&=&
-\frac{3\cos\lambda\left(1+\sin^2\lambda\right)}
{r\left(1+3\sin^2\lambda\right)^2}
\left(\cos\lambda\,\boldsymbol{ \hat{e}}_r+2\sin\lambda\,\boldsymbol{\hat{e}}_\lambda\right).
\end{eqnarray}
and the cross product of $\boldsymbol{\kappa}$ with $\boldsymbol{b}$ is given by: 
\begin{eqnarray}
\boldsymbol\kappa\times\boldsymbol b
&=&
\frac{3\cos\lambda\left(1+\sin^2\lambda\right)}
{r\left(1+3\sin^2\lambda\right)^{3/2}}\,
\boldsymbol{\hat{e}}_\varphi,
\end{eqnarray}
and the curvature drift is therefore given by: 
\begin{eqnarray}
\label{EQ:Final_v_curvature}
    \boldsymbol{v_\kappa}&=&-\frac{v_\parallel^2}{\Omega_s}\boldsymbol\kappa\times\boldsymbol b \nonumber \\
    &=&-\frac{v_\parallel^2}{\Omega_s}\frac{3\cos\lambda\left(1+\sin^2\lambda\right)}
{r\left(1+3\sin^2\lambda\right)^{3/2}}\,
\boldsymbol{\hat{e}}_\varphi \nonumber \\ 
&=&-\frac{m_s\gamma v_\parallel^2 r^2}{q_s B_E R_E^3}\frac{3\cos\lambda\left(1+\sin^2\lambda\right)}
{\left(1+3\sin^2\lambda\right)^{2}}\,
\boldsymbol{\hat{e}}_\varphi \nonumber \\
&=&-3\frac{m_s\gamma v_\parallel^2 L^2}{q_s B_E R_E}\frac{\cos^5\lambda\left(1+\sin^2\lambda\right)}
{\left(1+3\sin^2\lambda\right)^{2}}\,
\boldsymbol{\hat{e}}_\varphi
\end{eqnarray}
Similarly to the grad-$B$ drift, at all longitudes negative charges ($q_s<0$) drift eastward ($+\boldsymbol{\hat e}_\varphi$), whereas positive charges ($q_s>0$) drift westward ($-\boldsymbol{\hat e}_\varphi$) due to the curvature drift. However, for equatorially trapped particles, the curvature drift vanishes, since $v_\parallel=0$. 

\paragraph{Parallel velocity $v_\parallel$}
Since the electric field is set to $\boldsymbol E=\boldsymbol 0$, the Lorentz factor $\gamma$ (equivalently the kinetic energy $E_k=m_sc^2(\gamma-1)$) and the first adiabatic invariant $\mu$ are conserved. The parallel momentum magnitude can therefore be obtained exactly from
\begin{equation}
\label{EQ:Final_v_parallel}
|p_\parallel(\lambda)|
=
\sqrt{m_s^2c^2(\gamma^2-1)-2m_s\mu\,B(L,\lambda)},
\qquad
|v_\parallel(\lambda)|
=
\frac{|p_\parallel(\lambda)|}{\gamma m_s},
\end{equation}
with
\begin{equation}
B(L,\lambda)=\frac{B_E}{L^3}\frac{D(\lambda)}{\cos^6\lambda}.
\end{equation}
To track the direction of motion between mirror points, we write
\begin{equation}
p_\parallel(\lambda,t)=\sigma(t)\,|p_\parallel(\lambda)|,
\qquad
v_\parallel(\lambda,t)=\sigma(t)\,|v_\parallel(\lambda)|,
\end{equation}
where $\sigma=\pm 1$ is fixed by the initial condition and reverses sign at each bounce (i.e. at $\lambda=\lambda_m$ where $p_\parallel=0$).

\paragraph{Parallel acceleration $\langle \dot{v}_\parallel\rangle_\phi$}
For the sake of completeness, we compute the parallel acceleration given in (\ref{EQ:Final_relativistic_guiding_centres}), even though the parallel velocity is known exactly. Since the vector $\boldsymbol{\kappa}\times\boldsymbol{b}$ is perpendicular to the gradient of the magnetic field magnitude, the parallel acceleration is entirely dictated by the mirror force, i.e., 

\begin{equation}
    \big\langle m\gamma \dot{v}_\parallel \big\rangle_\phi= \big\langle\dot{p}_\parallel \big\rangle_\phi
\simeq
-\frac{\mu}{\gamma}\,
\boldsymbol b\cdot\nabla B
\end{equation}

We first proceed by computing the dot product in parallel force equation:
\begin{eqnarray}
\boldsymbol b\cdot\nabla B
&=&
\left(\frac{2\sin\lambda}{D}\hat{\boldsymbol e}_r-\frac{\cos\lambda}{D}\hat{\boldsymbol e}_\lambda\right)\cdot
\left(
-\frac{3B_E R_E^3}{r^4}D\,\hat{\boldsymbol e}_r
+\frac{3B_E R_E^3}{r^4}\frac{\sin\lambda\cos\lambda}{D}\,\hat{\boldsymbol e}_\lambda
\right)
\nonumber\\[4pt]
&=&
\frac{2\sin\lambda}{D}\left(-\frac{3B_E R_E^3}{r^4}D\right)
\left(\hat{\boldsymbol e}_r\cdot\hat{\boldsymbol e}_r\right)
-\frac{\cos\lambda}{D}\left(\frac{3B_E R_E^3}{r^4}\frac{\sin\lambda\cos\lambda}{D}\right)
\left(\hat{\boldsymbol e}_\lambda\cdot\hat{\boldsymbol e}_\lambda\right)
\nonumber\\[4pt]
&=&
-\frac{6B_E R_E^3}{r^4}\sin\lambda
-\frac{3B_E R_E^3}{r^4}\frac{\sin\lambda\cos^2\lambda}{D^2}.
\end{eqnarray}
Factoring out common terms and using $D^2(\lambda)=1+3\sin^2\lambda$ we obtain the equivalent compact form
\begin{eqnarray}
\boldsymbol b\cdot\nabla B
&=&
-\frac{3B_E R_E^3}{r^4}\,\sin\lambda
\left(
2+\frac{\cos^2\lambda}{1+3\sin^2\lambda}
\right)
\nonumber\\[4pt]
&=&
-\frac{3B_E R_E^3}{r^4}\,
\sin\lambda\,
\frac{3+5\sin^2\lambda}{1+3\sin^2\lambda} \nonumber \\
&=&
-\frac{3B_E}{R_EL^4}\,
\frac{\sin\lambda}{\cos^8(\lambda)}\,
\frac{3+5\sin^2\lambda}{1+3\sin^2\lambda}
\label{Eq:parallel_gradient_B}
\end{eqnarray}
Using Equation (\ref{Eq:parallel_gradient_B}), we write the parallel force as: 
\begin{equation}
\label{EQ:Final_p_parallel}
    \big\langle\dot{p}_\parallel \big\rangle_\phi=\frac{\mu}{\gamma}\frac{3B_E}{R_EL^4}\,
\frac{\sin\lambda}{\cos^8(\lambda)}\,
\frac{3+5\sin^2\lambda}{1+3\sin^2\lambda}
\end{equation}
It should be mentioned here that the sign of the parallel force is consistent with our convention that
$p_\parallel>0$ corresponds to motion along $\boldsymbol b$.
With the present definition of $\boldsymbol b$, one has
$\boldsymbol b=-\hat{\boldsymbol e}_\lambda$ at the equator, so that
$p_\parallel>0$ describes southward motion (toward decreasing magnetic latitude).
For $\lambda>0$ (northern hemisphere), the magnetic field strength increases
away from the equator, and one finds $\boldsymbol b\cdot\nabla B<0$.
Hence $\dot p_\parallel=-(\mu/\gamma)\,\boldsymbol b\cdot\nabla B>0$,
meaning that a particle moving southward is accelerated toward the equator,
whereas a particle moving northward is decelerated until it mirrors. Finally, since the parallel force vanishes at the magnetic equator ($\lambda=0$), particles that are initially confined to the equatorial plane remain equatorially trapped.

\subsection{Drift-kinetic equation}\label{secC2}
Equations (\ref{EQ:Final_gradB_drift}, \ref{EQ:Final_v_curvature}, 
\ref{EQ:Final_v_parallel}, \ref{EQ:Final_p_parallel}), together with the magnetic-field magnitude given by Equation (\ref{eq:dipole_magnitude_B_L_lambda}), fully determine the kinetic evolution of a collection of guiding centres in a dipolar magnetic field. 
In the absence of external sources and electric fields, these relations provide a closed drift-kinetic description of the system \cite{Hazeltine73, Hazeltine18, osmane2023radial}. Thus, the drift-kinetic equation for the distribution function of guiding centres is written, in conservative form, as: 
\begin{equation}
    \frac{\partial (Bf)}{\partial t}+\boldsymbol{\nabla}\cdot(B\langle \boldsymbol{\dot{r}}\rangle_\phi f) + \frac{\partial }{\partial v_\parallel}(B \langle\dot{v}_\parallel\rangle_\phi f)=0. 
\end{equation}
Since Liouville's theorem, for the conservation of phase-space density, states that: 
\begin{equation}
\label{EQ:Liouville}
      \frac{\partial (B)}{\partial t}+\boldsymbol{\nabla}\cdot(B\langle \boldsymbol{\dot{r}}\rangle_\phi ) + \frac{\partial }{\partial v_\parallel}(B \langle\dot{v}_\parallel\rangle_\phi )=0, 
\end{equation}
the drift-kinetic equation can be reduce to the following evolution equation for the gyrophase-averaged distribution function $f=\langle f\rangle_\phi$:
\begin{equation}
\boxed{
    \frac{\partial f}{\partial t}+\langle \boldsymbol{\dot{r}}\rangle_\phi\boldsymbol{\cdot\nabla}f +\langle\dot{v}_\parallel\rangle_\phi \frac{\partial f}{\partial v_\parallel}=0,} 
\end{equation}
where the guiding-centre velocity is given by
\begin{eqnarray}
    \langle \boldsymbol{\dot{r}}\rangle_\phi&=&v_\parallel \boldsymbol{b}+v_\varphi\,\hat{\boldsymbol e}_\varphi \nonumber \\
&=&
v_\parallel \boldsymbol{b}+\left(v_{\kappa,\varphi}+v_{\nabla B,\varphi}\right)\hat{\boldsymbol e}_\varphi, 
\end{eqnarray}
with the curvature and grad-$B$ contributions combining to produce an azimuthal drift whose latitude dependence is given by
\begin{equation}
\label{EQ:Final_azimuthal_drfit_}
v_\varphi(\lambda)
=
-\frac{3}{q_s}\left[
\frac{m_s\gamma v_\parallel^2 L^2}{B_E R_E}\,
\frac{\cos^5\lambda\left(1+\sin^2\lambda\right)}{D^4(\lambda)}
+
\frac{\mu_s}{\gamma L R_E}\,
\frac{1+\sin^2\lambda}{\cos\lambda\,D^2(\lambda)}
\right].
\end{equation}
The guiding-centre velocity must then be complemented with the parallel acceleration along the magnetic field line and given by Equation (\ref{EQ:Final_p_parallel}). 

Liouville's theorem can be shown to be satisfied for the magnetic field model and guiding centres given by Equations (\ref{eq:dipole_magnitude_B_L_lambda}, \ref{EQ:Final_gradB_drift}, \ref{EQ:Final_v_curvature}, 
\ref{EQ:Final_v_parallel}, \ref{EQ:Final_p_parallel}). While conservation of phase-space density may appear trivial, it is rarely verified, and its violation  has significant consequences for modelling.  The divergence of the magnetic field, $\boldsymbol{\nabla\cdot B}$, enters explicitly through the term $\boldsymbol{\nabla\cdot}(B v_\parallel \boldsymbol{b}) = v_\parallel \boldsymbol{\nabla\cdot B}$.
Moreover, the projection of Faraday's law along the local magnetic field arises in the presence of time-dependent electromagnetic fields as
$\partial B/\partial t + \boldsymbol{\nabla\cdot}(\boldsymbol{E}\times\boldsymbol{b})=\boldsymbol{b \cdot}(\partial \boldsymbol{B}/\partial t + \boldsymbol{\nabla}\times\boldsymbol{E})$, in Equation (\ref{EQ:Liouville}).
Therefore, any violation of Maxwell's equations, in computational or analytical models, directly compromises phase-space conservation.

\subsection{Scale separation of bounce and drift motion}

Our objective is to quantify the evolution of drift-timescale structures in the distribution function. While we retain particles undergoing bounce motion away from the equator, we exploit the widely disparate timescales of the bounce and drift motion to reduce the dimensionality of the drift-kinetic equation. 

The drift-kinetic equation may be written as
\begin{equation}
\label{EQ:Drift_Kinetic_BEfore_average}
\frac{\partial f}{\partial t}
+ v_\parallel\,\boldsymbol b\cdot\nabla f
+ v_\varphi\,\hat{\boldsymbol e}_\varphi\cdot\nabla f
+ \langle \dot v_\parallel\rangle_\phi \frac{\partial f}{\partial v_\parallel}
=0,
\end{equation}
where the perpendicular drift is purely azimuthal. Since
\begin{equation}
\hat{\boldsymbol e}_\varphi\cdot\nabla
=
\frac{1}{r(\lambda)\cos\lambda}\frac{\partial}{\partial\varphi}
=
\frac{1}{L R_E\cos^3\lambda}\frac{\partial}{\partial\varphi},
\end{equation}
the drift term becomes
\begin{equation}
v_\varphi\,\hat{\boldsymbol e}_\varphi\cdot\nabla f
=
\dot\varphi(\lambda)\,\frac{\partial f}{\partial\varphi},
\qquad
\dot\varphi(\lambda)= \frac{v_\varphi(\lambda)}{r(\lambda)\cos\lambda}.
\end{equation}

Along a field line, we introduce the arc length $s$, related to the parallel velocity by
$ds = v_\parallel dt$, so that
\begin{equation}
\boldsymbol b\cdot\nabla = \frac{\partial}{\partial s}.
\end{equation}
The parallel streaming term then reads
\begin{equation}
v_\parallel\,\boldsymbol b\cdot\nabla f
=
v_\parallel\,\frac{\partial f}{\partial s}.
\end{equation}
If $f$ varies over a characteristic parallel scale $\ell_\parallel$ (of order the distance between mirror points for trapped particles outside the loss cone), then
\begin{equation}
v_\parallel\,\boldsymbol b\cdot\nabla f
\sim
\frac{v_\parallel}{\ell_\parallel}\,f
=
\omega_b\,f,
\qquad
\omega_b\sim\frac{v_\parallel}{\ell_\parallel},
\end{equation}
where $\omega_b$ is the characteristic bounce frequency.

Using the closed-form expression for $v_\varphi(\lambda)$, the local azimuthal drift frequency becomes
\begin{equation}
\dot\varphi(\lambda)
=
-\frac{3}{q_s L R_E^2}\left[
\frac{m_s\gamma v_\parallel^2 L^2}{B_E}\,
\frac{\cos^2\lambda\left(1+\sin^2\lambda\right)}{D^4(\lambda)}
+
\frac{\mu}{\gamma}\,
\frac{1+\sin^2\lambda}{\cos^4\lambda\,D^2(\lambda)}
\right].
\end{equation}
For order-unity azimuthal structure ($\partial_\varphi f\sim f$), the drift term scales as
\begin{equation}
v_\varphi\,\hat{\boldsymbol e}_\varphi\cdot\nabla f
=
\dot\varphi(\lambda)\,\frac{\partial f}{\partial\varphi}
\sim
\dot\varphi(\lambda)\,f
=
\omega_d(\lambda)\,f,
\qquad
\omega_d(\lambda)= |\dot\varphi(\lambda)|.
\end{equation}

The separation between bounce and drift is therefore quantified by
\begin{equation}
\frac{|v_\parallel\,\boldsymbol b\cdot\nabla f|}
{|v_\varphi\,\hat{\boldsymbol e}_\varphi\cdot\nabla f|}
\sim
\frac{\omega_b}{\omega_d(\lambda)}
\sim
\frac{v_\parallel/\ell_\parallel}{|\dot\varphi(\lambda)|}.
\end{equation}
Since $\ell_\parallel\sim \mathcal{O}(L R_E)$, one has
$\omega_b\sim v_\parallel/(L R_E)$,
whereas
$\omega_d\sim v_\varphi/(L R_E)$.
Guiding-centre ordering, i.e., Larmor radius $\rho$ being much smaller than the characteristic length scale of the magnetic field variation $l_\parallel^{-1}\simeq|\nabla B|/B$ \cite{northrop1963adiabatic}, implies
$v_\parallel\gg v_\varphi$, and therefore
\begin{equation}
v_\parallel\,\boldsymbol b\cdot\nabla f
\gg
v_\varphi\,\hat{\boldsymbol e}_\varphi\cdot\nabla f,
\end{equation}
i.e. bounce motion advects the distribution function on a much shorter timescale than azimuthal drift.

\vspace{0.3cm}

We now consider the convective term $\dot v_\parallel \,\partial f/\partial v_\parallel$.
In the absence of electric fields, the parallel dynamics is entirely governed by the mirror force,
\begin{equation}
\dot v_\parallel
=
-\frac{\mu}{\gamma^2 m_s}\,\boldsymbol b\cdot\nabla B.
\end{equation}
Since $B$ varies over the same characteristic length $\ell_\parallel$,
\begin{equation}
|\boldsymbol b\cdot\nabla B|
\sim
\frac{B}{\ell_\parallel},
\end{equation}
and using $\mu=p_\perp^2/(2m_s B)$ gives
\begin{equation}
|\dot v_\parallel|
\sim
\frac{v_\perp^2}{2\,\ell_\parallel}.
\end{equation}
For trapped particles away from the loss cone \footnote{Particles near the loss cone can irreversibly be lost due to wave-particle interactions, and as explained in more detail below, the bounce-averaging procedure would prove inadequate for equatorial pitch-angles near the loss cone if the distribution function is altered within one bounce period.},
$v_\perp\sim v_\parallel\sim v$, so that
\begin{equation}
|\dot v_\parallel|
\sim
\frac{v^2}{\ell_\parallel}.
\end{equation}
Since $\tau_b\sim \ell_\parallel/v$, it follows that
\begin{equation}
|\dot v_\parallel|
\sim
\frac{v}{\tau_b},
\end{equation}
showing that the mirror force modifies $v_\parallel$ on the bounce timescale.
Consequently,
\begin{equation}
\dot v_\parallel\,\frac{\partial f}{\partial v_\parallel}
\sim
\frac{v}{\tau_b}\,\frac{f}{v}
=
\frac{f}{\tau_b},
\end{equation}
so that the parallel-acceleration term is likewise a bounce-timescale operator.
\begin{equation}
v_\parallel\,\boldsymbol b\cdot\nabla f
\simeq 
\dot{v}_\parallel\frac{\partial f}{\partial v_\parallel},
\end{equation}
\vspace{0.3cm}

The above ordering motivates a multiple-timescale expansion of the form
\begin{equation}
f = g_0 + \varepsilon g_1 + \cdots,
\qquad
\varepsilon = \frac{\omega_d}{\omega_b} \ll 1,
\end{equation}
where $g_0$ evolves on the fast bounce timescale and satisfies
\begin{equation}
\frac{\partial g_0}{\partial t}
+ v_\parallel\,\boldsymbol b\cdot\nabla g_0
+ \langle \dot v_\parallel\rangle_\phi
\frac{\partial g_0}{\partial v_\parallel}
=0,
\end{equation}
while $g_1$ describes the slow drift evolution,
\begin{equation}
\frac{\partial g_1}{\partial t}
+ v_\varphi\,\hat{\boldsymbol e}_\varphi\cdot\nabla g_1
=0.
\end{equation}
This separation justifies removing the fast bounce oscillation through a bounce average, yielding an effective transport equation for the slowly evolving bounce-averaged distribution.

\subsection{Bounce-averaged drift-kinetic equation}

We now present the procedure to remove the fast bounce motion between mirror points of any quantity $Q$, as
\begin{equation}
\langle Q\rangle_b = \frac{1}{\tau_b}\int_0^{\tau_b} Q(t)\,dt,
\end{equation}
where $\tau_b$ is the bounce period. Along a dipole field line one may equivalently write the bounce average as an integral along the latitude $\lambda$ as: 
\begin{equation}
\langle Q\rangle_b
=
\frac{4}{\tau_b}\int_0^{\lambda_m}
Q(\lambda)\,
\frac{1}{|v_\parallel(\lambda)|}\,
\frac{ds}{d\lambda}\,d\lambda,
\qquad
\frac{ds}{d\lambda}=L R_E\cos\lambda\,D(\lambda),
\label{EQ:Def_bounceaveraged_1}
\end{equation}
with $\lambda_m$ the mirror latitude and $D(\lambda)=\sqrt{1+3\sin^2\lambda}$.
The bounce period then follows as
\begin{equation}
\label{EQ:Final_bounce_period}
\tau_b
=
4\int_0^{\lambda_m}
\frac{1}{|v_\parallel(\lambda)|}\,
\frac{ds}{d\lambda}\,d\lambda
=
4L R_E\int_0^{\lambda_m}
\frac{\cos\lambda\,D(\lambda)}{|v_\parallel(\lambda)|}\,d\lambda.
\end{equation}
Starting from the drift-kinetic equation (\ref{EQ:Drift_Kinetic_BEfore_average}), we define the bounce-averaged distribution function
\begin{equation}
F(\varphi, L, \alpha_{eq, }t) = \langle f\rangle_b,
\label{EQ:Definition_bounce_averaged_distribution}
\end{equation}
and assume a clear separation of timescales, such that the bounce motion is much faster than the azimuthal drift and any secular evolution of $F$.
Under this ordering, the parallel streaming and parallel force terms average out,
\begin{equation}
\left\langle v_\parallel\,\boldsymbol b\cdot\nabla f\right\rangle_b \simeq 0, \qquad \left\langle \dot{v}_\parallel\ \frac{\partial f}{\partial v_\parallel}\right\rangle_b \simeq 0
\end{equation}
and the bounce-averaged drift-kinetic equation reduces to
\begin{equation}
\label{EQ:Final_Bounce_averaged_Drift_Kinetic_}
\frac{\partial F}{\partial t}
+
\left\langle
\frac{v_\varphi(\lambda)}{r(\lambda)\cos\lambda}
\right\rangle_b
\frac{\partial F}{\partial \varphi}
=0,
\end{equation}
where the coefficient multiplying $\partial/\partial\varphi$ is the bounce-averaged azimuthal drift frequency
\begin{equation}
\langle \dot{\varphi}\rangle_b
=
\left\langle
\frac{v_\varphi(\lambda)}{r(\lambda)\cos\lambda}
\right\rangle_b.
\end{equation}
Using $r(\lambda)=L R_E\cos^2\lambda$ and $ds/d\lambda=L R_E\cos\lambda\,D(\lambda)$, this may be written explicitly as
\begin{equation}
\begin{aligned}
\label{Eq:Final_bounce_averaged_drift}
\langle \dot{\varphi}\rangle_b
&=
\frac{4}{\tau_b}\int_0^{\lambda_m}
\frac{v_\varphi(\lambda)}{r(\lambda)\cos\lambda}\,
\frac{1}{|v_\parallel(\lambda)|}\,
\frac{ds}{d\lambda}\,d\lambda  \\
&=
\frac{4}{\tau_b}\int_0^{\lambda_m}
v_\varphi(\lambda)\,
\frac{D(\lambda)}{\cos^2\lambda}\,
\frac{1}{|v_\parallel(\lambda)|}\,d\lambda.
\end{aligned}
\end{equation}


In the following, it will prove judicious to write the drift frequency in terms of the equatorial pitch-angle $\alpha_{\rm eq}$ and the kinetic energy $E_k=m_sc^2(\gamma-1)$. 

We start from the local (non--bounce-averaged) azimuthal drift speed (\ref{EQ:Final_azimuthal_drfit_}) and note that
conservation of the first adiabatic invariant $\mu_s$  implies the pitch-angle mapping
\begin{equation}
\sin^2\alpha(\lambda)
=\sin^2\alpha_{\rm eq}\,\frac{B(L,\lambda)}{B_{\rm eq}(L)}
=\sin^2\alpha_{\rm eq}\,\frac{D(\lambda)}{\cos^6\lambda}.
\label{eq:alpha_map}
\end{equation}
Hence, for fixed values of the couple $(\gamma,\alpha_{\rm eq})$,
\begin{equation}
v_\parallel^2(\lambda)
=v^2\cos^2\alpha(\lambda)
=v^2\left[1-\sin^2\alpha_{\rm eq}\,\frac{D(\lambda)}{\cos^6\lambda}\right].
\label{eq:vpar_alphaeq}
\end{equation}
Equivalently, one may express the same replacement in terms of $\mu_s$ and $B(L,\lambda)$:
\begin{equation}
v_\perp^2(\lambda)=\frac{2\mu_s\,B(L,\lambda)}{m_s\gamma^2},
\qquad
v_\parallel^2(\lambda)=v^2-\frac{2\mu_s\,B(L,\lambda)}{m_s\gamma^2},
\label{eq:vpar_mu}
\end{equation}
which is identical to \eqref{eq:vpar_alphaeq} once $\mu_s$ is evaluated at the equator. At the equator $v_{\perp,{\rm eq}}^2=v^2\sin^2\alpha_{\rm eq}$ and $B=B_{\rm eq}(L)$, so
\begin{equation}
\mu_{\rm eq}
=\frac{m_s\gamma^2 v^2\sin^2\alpha_{\rm eq}}{2B_{\rm eq}(L)}
=\frac{m_s\gamma^2 v^2\sin^2\alpha_{\rm eq}}{2}\,\frac{L^3}{B_E}.
\label{eq:mu_eq}
\end{equation}
Therefore,
\begin{equation}
\frac{\mu_{\rm eq}}{\gamma L R_E}
=\frac{m_s\gamma v^2\sin^2\alpha_{\rm eq}}{2}\,\frac{L^2}{B_E R_E}.
\label{eq:mu_term_scaling}
\end{equation}

We now insert \eqref{eq:vpar_alphaeq} and \eqref{eq:mu_term_scaling} into $v_\varphi$ in Equation (\ref{EQ:Final_azimuthal_drfit_})
\begin{equation}
v_\varphi(L,\lambda)
=
-\frac{3 m_s\gamma v^2}{q_s}\,
\frac{L^2}{B_E R_E}\;
\mathcal{V}(\lambda;\alpha_{\rm eq}),
\label{eq:vphi_factored}
\end{equation}
where the dimensionless kernel is
\begin{align}
\mathcal{V}(\lambda;\alpha_{\rm eq})
&=
\left[1-S\,\frac{D(\lambda)}{\cos^6\lambda}\right]
\frac{\cos^5\lambda\left(1+\sin^2\lambda\right)}{D^4(\lambda)}
+\frac{S}{2}\,
\frac{1+\sin^2\lambda}{\cos\lambda\,D^2(\lambda)}.
\label{eq:V_unsimplified}
\end{align}
A compact algebraic form is obtained by expanding and collecting the function $S_{\rm eq}= \sin^2\alpha_{\rm eq}$ gives:
\begin{equation}
\mathcal{V}(\lambda;\alpha_{\rm eq})
=
\left(1+\sin^2\lambda\right)\left[
\frac{\cos^5\lambda}{D^4(\lambda)}
+\frac{S_{\rm eq}}{2}\,
\frac{D(\lambda)-2}{\cos\lambda\,D^3(\lambda)}
\right].
\label{eq:V_compact}
\end{equation}
Using $v_\varphi=r\cos\lambda\,\dot\varphi$ with $r(L,\lambda)=L R_E\cos^2\lambda$, the drift frequency can be written as: 
\begin{equation}
\dot\varphi(L,\lambda)
=\frac{v_\varphi}{L R_E\cos^3\lambda}
=
-\frac{3 m_s\gamma v^2}{q_s}\,
\frac{L}{B_E R_E^2}\;
\frac{\mathcal{V}(\lambda;\alpha_{\rm eq})}{\cos^3\lambda}.
\label{eq:phidot_local}
\end{equation}
And the bounce period can be written as:
\begin{equation}
\tau_b
=
4\int_{0}^{\lambda_m}\frac{ds}{|v_\parallel(\lambda)|}
=
4\int_{0}^{\lambda_m}
\frac{L R_E \cos\lambda\,D(\lambda)}
{v\,\sqrt{1-S_{\rm eq}\,D(\lambda)/\cos^6\lambda}}\,d\lambda.
\label{eq:taub}
\end{equation}
Thus, the bounce-averaged drift frequency is
\begin{equation}
\langle \dot\varphi\rangle_b
=
\frac{4}{\tau_b}\int_0^{\lambda_m}\dot\varphi(L,\lambda)\,\frac{ds}{|v_\parallel(\lambda)|}.
\label{eq:phidot_b_def}
\end{equation}
Substituting \eqref{eq:phidot_local} and \eqref{eq:vpar_alphaeq} gives
\begin{equation}\label{eq:phidot_b_factored}
\begin{aligned}
\langle \dot\varphi\rangle_b
&=
-\frac{3 m_s\gamma v}{q_s}\,
\frac{L}{B_E R_E^2}\;
\frac{
\displaystyle
\int_{0}^{\lambda_m}
\frac{D(\lambda)}{\cos^2\lambda}\;
\frac{\mathcal{V}(\lambda;\alpha_{\rm eq})}
{\sqrt{1-S_{\rm eq}\,D(\lambda)/\cos^6\lambda}}\,d\lambda
}{
\displaystyle
\int_{0}^{\lambda_m}
\cos\lambda\,D(\lambda)\;
\frac{1}{\sqrt{1-S_{\rm eq}\,D(\lambda)/\cos^6\lambda}}\,d\lambda
}
.
\end{aligned}
\end{equation}

All $L$-dependence has been extracted in the prefactor proportional to $L/(B_E R_E^2)$; the remaining ratio of integrals
depends only on the equatorial pitch-angle $\alpha_{\rm eq}$ through $S=\sin^2\alpha_{\rm eq}$. We finally express the bounce drift frequency $\langle \dot\varphi\rangle_b$ in a more compact form as: 
\begin{equation}\label{eq:phidot_b_factored}
\boxed{
\begin{aligned}
\langle \dot\varphi\rangle_b
&=
-\frac{3 m_s c^2 L}{q_sB_E R_E^2}\,
\ \frac{\frac{E_k}{m_sc^2}\left(\frac{E_k}{m_s c^2}+2\right)}{\frac{E_k}{m_sc^2}+1} \ \mathcal{M}(\alpha_{\rm eq}) 
.
\end{aligned}
}
\end{equation}
where the function $\mathcal{M}$ is given by:
\begin{equation}
 \mathcal{M}(\alpha_{\rm eq})=   \frac{
\displaystyle
\int_{0}^{\lambda_m}
\frac{D(\lambda)}{\cos^2\lambda}\;
\frac{\mathcal{V}(\lambda;\alpha_{\rm eq})}
{\sqrt{1-S_{\rm eq}\,D(\lambda)/\cos^6\lambda}}\,d\lambda
}{
\displaystyle
\int_{0}^{\lambda_m}
\cos\lambda\,D(\lambda)\;
\frac{1}{\sqrt{1-S_{\rm eq}\,D(\lambda)/\cos^6\lambda}}\,d\lambda
}.
\end{equation}
Equation (\ref{eq:phidot_b_factored}) is the bounce drift frequency with an explicit dependence on the equatorial pitch-angle and the kinetic energy $E_k=m_sc^2(\gamma-1)$ of the particle.

The last ingredient required to solve Equation~(\ref{eq:phidot_b_factored}) is the determination of the mirror latitude $\lambda_m$, obtained by imposing $v_\parallel(\lambda_m)=0$, i.e.
\begin{equation}
1-S_{\rm eq}\,\frac{D(\lambda_m)}{\cos^6\lambda_m}=0
\quad\Longleftrightarrow\quad
S_{\rm eq}=\frac{\cos^6\lambda_m}{D(\lambda_m)}.
\label{eq:mirror_condition}
\end{equation}
Introducing the variable $x=\cos^2\lambda_m$, so that $\sin^2\lambda_m=1-x$ and $D(\lambda_m)=\sqrt{4-3x}$, Equation~(\ref{eq:mirror_condition}) yields
\begin{equation}
x^6+3S_{\rm eq}^2 x-4S_{\rm eq}^2=0.
\label{eq:mirror_polynomial}
\end{equation}
Equation~(\ref{eq:mirror_polynomial}) is a sixth-degree polynomial in $x=\cos^2\lambda_m$ for a fixed equatorial pitch-angle $\alpha_{\rm eq}$. For arbitrary $\alpha_{\rm eq}$, the admissible root satisfying $0<x\le1$ can be obtained numerically in a straightforward manner.

The set of Equations (\ref{EQ:Final_Bounce_averaged_Drift_Kinetic_}), (\ref{eq:V_compact}),  (\ref{eq:phidot_b_factored}) and (\ref{eq:mirror_condition})  are therefore the set of equations we use to characterise the radial transport ballistic motion in the Earth's radiation belts. It should however be emphasised that the above results rely on several assumptions that may be readily violated under even moderately disturbed geomagnetic conditions. In particular, electromagnetic fluctuations capable of breaking the second adiabatic invariant can modify the distribution function on timescales comparable to the bounce motion \cite{Kamaletdinov25, Chaston25}. Similarly, particles near the loss cone, which may scatter into the upper atmosphere, do not satisfy the assumptions underlying the bounce-averaged formulation \cite{BLUM202049}. A discussion of these limitations and their impact on the results are discussed in Section \ref{sec4}

\section{Pitch-angle diffusion in the bounce-averaged drift-kinetic equation and properties of the associated eigenvalue problem}\label{Appendix_B} 
In this appendix we first detail the incorporation of pitch-angle diffusion into the bounce-averaged drift-kinetic equation and outline the limits of validity of this approach within a radial transport framework. The drift-kinetic formulation adopted throughout this work relies on conservation of the first adiabatic invariant, $\mu = \frac{p_\perp^2}{2m_s B}$, which allows the reduction of the full kinetic description to guiding-centre phase-space coordinates and slowly evolving invariants. Under this ordering the fast Larmor motion is eliminated and the dynamics are organised according to a hierarchy of timescales.

Pitch-angle diffusion, however, arises from cyclotron-resonant interactions that locally violate conservation of the first adiabatic invariant \cite{Kennel1966velocity, Cunningham23}. If such scattering becomes sufficiently strong, the basic assumptions underlying the drift-kinetic reduction cease to be valid. In particular, if the characteristic pitch-angle diffusion timescale
\[
\tau_{\alpha} \sim \frac{1}{D_{\alpha\alpha}},
\]
where $D_{\alpha\alpha}$ denotes the pitch-angle diffusion coefficient, becomes comparable to or shorter than the bounce period $\tau_b$, i.e.
\begin{equation}
\tau_{\alpha} \lesssim \tau_b,
\end{equation}
particles no longer execute well-defined bounce orbits between successive scattering events. In this regime the bounce averaging procedure becomes ill-posed and a full kinetic treatment retaining the fast Larmor motion is undeniably required.

In contrast, the present framework, detailed in Appendix~\ref{Appendix_bounce_averaged_diffusion}, assumes that pitch-angle diffusion operates on timescales long compared with the bounce motion and sufficiently slowly that the second and third adiabatic invariants remain approximately conserved over a drift period. Formally, we require the ordering
\begin{equation}
\tau_b \ll \tau_{\alpha},
\end{equation}
together with the assumption that radial transport acts on timescales not shorter than the diffusion time associated with pitch-angle scattering. This ordering is well justified for energetic and relativistic electrons interacting with chorus, hiss, and electromagnetic ion-cyclotron waves according to quasilinear theory, for which the characteristic diffusion timescales remain long compared with the bounce period \cite{Summers07,Mourenas12}. We note, however, that during strong injection events, lower-energy ($\sim$1–100 keV) electrons may undergo nonlinear interactions with large-amplitude whistler-mode waves \cite{Albert_2002, Bortnik_2008, Tao_2012,osmane2016connection, Zhang_2020}, leading to pitch-angle evolution on timescales much shorter than the drift period; in such cases, the present ordering does not apply, and the distribution is rapidly restructured prior to the drift-phase mixing considered here \cite{ArtemyevSSR_Review2025}.

Under these conditions, corresponding to the regime in which pitch-angle diffusion operates on timescales comparable to or longer than the azimuthal drift period for energetic and relativistic electrons, the bounce motion remains well defined and the distribution function may be expressed in terms of bounce-averaged phase-space variables. Pitch-angle diffusion and loss-cone precipitation can then be incorporated as a quasilinear collisional operator acting on the bounce-averaged distribution function. The resulting evolution equation no longer satisfies Liouville's theorem, since the diffusion operator irreversibly mixes phase space and allows particle loss, but the reduced drift-kinetic description remains self-consistent within the stated ordering.

Once pitch-angle diffusion is included, the radial transport equations can be decomposed into an eigenvalue--eigenvector problem. In Appendix~\ref{appendix:eigenvalue_proof} we demonstrate that, irrespective of the parametric dependence of the pitch-angle diffusion coefficient, the associated eigenvalues are strictly non-positive. Treatments of the pitch-angle diffusion equation in terms of eigenvalues and eigenvectors are not new (see, e.g., Section~V.2 of \citet{schulz1974}, as well as \citet{ALBERT_diffusion_solution} and \citet{OBRIEN20081738}). However, the result presented here establishes that the eigenvalues are necessarily non-positive provided the pitch-angle diffusion coefficient remains positive definite, and therefore does not rely on any specific parametric form such as those adopted in earlier studies.

\subsection{Bounce-averaged pitch-angle diffusion operator}
\label{Appendix_bounce_averaged_diffusion}
We seek the bounce average of the standard pitch-angle diffusion operator \cite{Lyons} written in the \emph{local} pitch angle $\alpha(\lambda)$,
\begin{equation}
\mathcal{L}_\alpha[f]
\;=\;
\frac{1}{\sin\alpha}\frac{\partial}{\partial \alpha}
\left(
D_{\alpha\alpha}\,\sin\alpha\,\frac{\partial f}{\partial \alpha}
\right),
\label{eq:local_pitch_angle_diff_operator}
\end{equation}
and we wish to express the resulting term as an operator acting on the {bounce-averaged} distribution function written in terms of the {equatorial} pitch angle $\alpha_{\rm eq}$. The first step is therefore to relate $\alpha(\lambda)$ to $\alpha_{\rm eq}$ along a bounce orbit, using Equation (\ref{eq:alpha_map}). Assuming conservation of the first adiabatic invariant $\mu$ between scattering events, and fixed speed $v$ (or fixed kinetic energy), one has
\begin{equation}
\sin\alpha(\lambda)=\sqrt{g(\lambda)}\,\sin\alpha_{\rm eq}.
\label{eq:alpha_lambda_vs_alpha_eq}
\end{equation}
where we defined $g(\lambda)= \frac{D(\lambda)}{\cos^6\lambda}$ in terms of (\ref{EQ_Famous_D_lambda}). In the next step we compute the Jacobian between $\alpha$ and $\alpha_{\rm eq}$ by differentiating \eqref{eq:alpha_lambda_vs_alpha_eq} at fixed $\lambda$:
\begin{equation}
2\sin\alpha\,\cos\alpha\,\frac{\partial \alpha}{\partial \alpha_{\rm eq}}
=
g(\lambda)\,2\sin\alpha_{\rm eq}\,\cos\alpha_{\rm eq}.
\end{equation}
Using $\sin\alpha=\sqrt{g}\,\sin\alpha_{\rm eq}$ gives
\begin{equation}
\frac{\partial \alpha}{\partial \alpha_{\rm eq}}
=
\sqrt{g(\lambda)}\,
\frac{\cos\alpha_{\rm eq}}{\cos\alpha(\lambda)}
\label{eq:jacobian_alpha_alphaeq}
\end{equation}
We denote the Jacobian of the transformation as 
\begin{equation}
J(\lambda;\alpha_{\rm eq})
=
\frac{\partial \alpha}{\partial \alpha_{\rm eq}}
\end{equation}

The crucial step in the derivation comes right here. The bounce-averaged drift--kinetic closure requires a clear separation between the fast bounce motion and the slow pitch-angle scattering, i.e.\ that the characteristic pitch-angle diffusion timescale $\tau_{\alpha}$ is long compared with the bounce period $\tau_b$, i.e., 
$\tau_b \ll \tau_{\alpha}$.
Under this ordering, particles complete many bounce oscillations between successive scattering events, so that to leading order in $\tau_b/\tau_{\alpha}$ the distribution function may be represented by its bounce-averaged component
$F(\varphi,L,\alpha_{\rm eq},t)$ defined in Equation~(\ref{EQ:Definition_bounce_averaged_distribution}),
with the dependence on latitude entering only through the kinematic mapping between the local pitch angle $\alpha(\lambda)$ and the orbit label $\alpha_{\rm eq}$.

Consequently, when rewriting the local operator $\partial/\partial \alpha$ in terms of the equatorial variable $\alpha_{\rm eq}$, the derivative acts (to leading order) on $F$. In particular, at fixed $\lambda$ the chain rule gives
\begin{equation}
\left.\frac{\partial f}{\partial \alpha}\right|_{\lambda}
=
\frac{\partial F}{\partial \alpha_{\rm eq}}
\left.\frac{\partial \alpha_{\rm eq}}{\partial \alpha}\right|_{\lambda}
=
J^{-1}(\lambda;\alpha_{\rm eq})\,\frac{\partial F}{\partial \alpha_{\rm eq}},
\label{eq:df_dalpha_chainrule}.
\end{equation}
The operator  $\mathcal{L}_\alpha[f]$ in \eqref{eq:local_pitch_angle_diff_operator} can therefore be rewritten as:
\begin{equation}
\mathcal{L}_\alpha[f]
=
\frac{1}{\sin\alpha}\,J^{-1}\,
\frac{\partial}{\partial \alpha_{\rm eq}}
\left(
D_{\alpha\alpha}\,\sin\alpha\,J^{-1}\,
\frac{\partial F}{\partial \alpha_{\rm eq}}
\right).
\label{eq:operator_transformed_intermediate}
\end{equation}
Using Equation (\ref{eq:alpha_lambda_vs_alpha_eq}) and $J^{-1}=\cos\alpha/(\sqrt{g}\cos\alpha_{\rm eq})$, we obtain the compact form
\begin{equation}
\mathcal{L}_\alpha[f]
=
\frac{\cos\alpha(\lambda)}{g(\lambda)\,\sin\alpha_{\rm eq}\,\cos\alpha_{\rm eq}}
\,
\frac{\partial}{\partial \alpha_{\rm eq}}
\left[
D_{\alpha\alpha}(\lambda,\alpha)\,
\sin\alpha_{\rm eq}\,
\frac{\cos\alpha(\lambda)}{\cos\alpha_{\rm eq}}\,
\frac{\partial F}{\partial \alpha_{\rm eq}}
\right].
\label{eq:operator_transformed_alphaeq_local}
\end{equation}
Using the bounce average definition (\ref{EQ:Def_bounceaveraged_1}) and the bounce period expression (\ref{EQ:Final_bounce_period}),
we obtain
\begin{equation}
\big{\langle} \mathcal{L}_\alpha[f]\big{\rangle}_b
=
\frac{4}{\tau_b}\int_0^{\lambda_m}
\mathcal{L}_\alpha[f](\lambda)\,
\frac{1}{|v_\parallel(\lambda)|}\,
\frac{ds}{d\lambda}\,d\lambda.
\label{eq:bounce_avg_operator_def}
\end{equation}
At this point we {define} an effective equatorial diffusion coefficient by collecting the factors that multiply $\partial F/\partial \alpha_{\rm eq}$ and write the bounce-averaged operator in the same functional form as \eqref{eq:local_pitch_angle_diff_operator} but in $\alpha_{\rm eq}$:
\begin{equation}
\left\langle \mathcal{L}_\alpha[f]\right\rangle_b
\;\equiv\;
\frac{1}{\sin\alpha_{\rm eq}}
\frac{\partial}{\partial \alpha_{\rm eq}}
\left(
\langle D_{\alpha\alpha}\rangle_b\,
\sin\alpha_{\rm eq}\,
\frac{\partial F}{\partial \alpha_{\rm eq}}
\right),
\label{eq:desired_bounce_avg_form}
\end{equation}
with
\begin{equation}
\boxed{
\langle D_{\alpha\alpha}\rangle_b
\;=\;
\frac{4}{\tau_b}\int_0^{\lambda_m}
D_{\alpha\alpha}(\lambda,\alpha)\,
\frac{\cos^2\alpha(\lambda)}{g(\lambda)\,\cos^2\alpha_{\rm eq}}\,
\frac{1}{|v_\parallel(\lambda)|}\,
\frac{ds}{d\lambda}\,d\lambda
}.
\label{eq:Db_alphaeq_alphaeq}
\end{equation}
Here, $\alpha$ is a function of $(\lambda,\alpha_{\rm eq})$ and $g(\lambda)= \frac{D(\lambda)}{\cos^6\lambda}=\frac{\sqrt{1+3\sin^2(\lambda)}}{\cos^6\lambda}$, and the integral can be solved for any arbitrary pitch-angle diffusion, derived within a quasi-linear framework \cite{Kennel1966velocity, Allanson22, Cunningham23} or one that incorporates nonlinear wave-particle interactions \cite{Artemyev2018}.

We note that the factor $\cos^2\alpha/(g\,\cos^2\alpha_{\rm eq})$ in \eqref{eq:Db_alphaeq_alphaeq} is precisely the square of the variable-change Jacobian in \eqref{eq:jacobian_alpha_alphaeq}:
\begin{equation}
\left(\frac{\partial \alpha_{\rm eq}}{\partial \alpha}\right)^2
=
\frac{\cos^2\alpha(\lambda)}{g(\lambda)\,\cos^2\alpha_{\rm eq}}.
\end{equation}
Thus, Equation \eqref{eq:Db_alphaeq_alphaeq} is identical to the expression derived in \citet{Lyons} (p. 3461) and may be read as a bounce average of the locally transformed diffusion coefficient,
\begin{equation}
\langle D_{\alpha\alpha}\rangle_b
=
\left\langle
D_{\alpha\alpha}\left(\frac{\partial \alpha_{\rm eq}}{\partial \alpha}\right)^2
\right\rangle_b.
\end{equation}

In order to account for the pitch-angle diffusion taking place on timescales comparable to the drift period, we combine Equations (\ref{EQ:Final_Bounce_averaged_Drift_Kinetic_}) with Equation (\ref{eq:bounce_avg_operator_def}) 
\begin{equation}
\label{EQ:bounced_average_full_1}
\boxed{
\frac{\partial F}{\partial t}
+
\langle \dot{\varphi}\rangle_b
\frac{\partial F}{\partial \varphi}
=\frac{1}{\sin\alpha_{\rm eq}}
\frac{\partial}{\partial \alpha_{\rm eq}}
\left(
\langle D_{\alpha\alpha}\rangle_b\,
\sin\alpha_{\rm eq}\,
\frac{\partial F}{\partial \alpha_{\rm eq}}
\right),
}
\end{equation}
where the bounced-average drift frequency is given by Equation (\ref{eq:phidot_b_factored}) and the bounced-averaged pitch-angle diffusion given by Equation (\ref{eq:Db_alphaeq_alphaeq}). Equation~(\ref{EQ:bounced_average_full_1}) retains both azimuthal drift motion and pitch-angle diffusion on equal footing, thereby allowing their associated timescales to be comparable within a unified framework. Since we assume that the bounce motion constitutes the fastest dynamical timescale in the system, a reduced description obtained by averaging over the bounce phase and expressing the distribution in terms of the equatorial pitch angle $\alpha_{\rm eq}$ is appropriate and well justified for energetic and relativistic electrons interacting with a wide range of electromagnetic fluctuations according to quasilinear theory \cite{Summers07,Mourenas12}.

\subsection{Proof that pitch-angle diffusion cannot increase the effective lifetime of structures in drift phase}
\label{appendix:eigenvalue_proof}
In the computation of the correlation function in Section~\ref{sec3}, we found that drift-phase structures localized in MLT or in drift shells, or in both, exhibit an effective lifetime of only a few drift periods. Consequently, on such timescales pitch-angle diffusion may also act to modify the bounce-averaged distribution function, and must therefore be expected to influence the associated decorrelation time. In this appendix, we demonstrate that the pitch-angle diffusion can only result in additional apparent decay of the correlation function, and thus, that the computation and estimates for the decorrelation time in Section~\ref{sec3} can be taken as maximum values for the effective lifetime in the absence of significant pitch-angle diffusion. 

We use Equation~(\ref{EQ:bounced_average_full_1}) to compute solutions for 
$F(L,\varphi,\alpha_{\rm eq},t)$ by formulating the problem in terms of eigenvalues and eigenvectors. The equatorial pitch angle is defined over the interval between the loss-cone angle $\alpha_{LC}$ and $\pi/2$, i.e.\ $\alpha_{\rm eq}\in[\alpha_{LC},\pi/2]$. For completeness, we also allow for an optional sink term proportional to $-\nu(\alpha_{\rm eq})$ to represent precipitation losses. As will be shown below, however, the presence or absence of this sink term (for $\nu>0$) does not alter our conclusions. 

Equation~(\ref{EQ:bounced_average_full_1}), complemented with a sink term proportional to $-\nu F$, and after Fourier decomposition along the azimuthal phase, i.e., $F=\sum_m F_m e^{im\varphi}$, results in the following equation: 
\begin{equation}
\label{EQ:bounced_average_full_2}
\frac{\partial F_m}{\partial t}
=
\left(im\langle \dot{\varphi}\rangle_b
 -\nu \right)F_m
+\frac{1}{\sin\alpha_{\rm eq}}
\frac{\partial}{\partial \alpha_{\rm eq}}
\left(
\langle D_{\alpha\alpha}\rangle_b\,
\sin\alpha_{\rm eq}\,
\frac{\partial F_m}{\partial \alpha_{\rm eq}}
\right).
\end{equation}

We seek separated solutions for each azimuthal Fourier coefficient $F_m$ of the form
\begin{equation}
F_m(L,\alpha_{\rm eq},t)
=
\sum_{n} a_{n}(L)\,\Phi_{n}(\alpha_{\rm eq})\,
\exp\!\left(\Lambda_{n}\,t\right),
\label{eq:Fm_eigen_expansion}
\end{equation}
where $\Phi_{n}$ and $\Lambda_{n}$ are, respectively, the eigenfunctions and eigenvalues associated with the pitch-angle operator (including the optional sink term), and the $a_n$ coefficients encode initial structure along $L$ and are determined by the projection of the initial Fourier coefficient $F_m(t=0, L)$ onto the pitch-angle eigenfunctions. 

For a fixed $m$ value (suppressed in the notation), the pairs $(\Phi_n,\Lambda_n)$ satisfy
\begin{equation}
\Lambda_{n}\,\Phi_{n}
=
\left(i m\langle \dot{\varphi}\rangle_b-\nu\right)\Phi_{n}
+\frac{1}{\sin\alpha_{\rm eq}}\frac{d}{d\alpha_{\rm eq}}
\left(
\langle D_{\alpha\alpha}\rangle_b\,\sin\alpha_{\rm eq}\,
\frac{d\Phi_{n}}{d\alpha_{\rm eq}}
\right),
\label{eq:eigenproblem_alpha}
\end{equation}
with $\alpha_{\rm eq}\in[\alpha_{LC},\pi/2]$. In order to solve the eigenvalue problem \eqref{eq:eigenproblem_alpha} we require two boundary conditions for $\Phi_n$. The first boundary condition is for the absorbing loss cone,
\begin{equation}
\Phi_n(\alpha_{LC})=0,
\label{eq:bc_absorbing_lc}
\end{equation}
which enforces complete removal of particles upon entering the loss cone. And the second one enforces equatorial symmetry around the magnetic equator:
\begin{equation}
\left.\frac{d\Phi_n}{d\alpha_{\rm eq}}\right|_{\alpha_{\rm eq}=\pi/2}=0.
\label{eq:bc_equatorial_symmetry}
\end{equation}
We note that the real part of the operator in Equation~(\ref{eq:eigenproblem_alpha}) constitutes a canonical Sturm--Liouville operator with weight function $\sin(\alpha_{\rm eq})$ \cite{boyce2017elementary}. It follows that the real component of the operator is self-adjoint in the corresponding weighted Hilbert space, a property that may be exploited to advantage \cite{Masujima}. 

A common approach for this class of problems is to prescribe a specific functional form for the bounce-averaged pitch-angle diffusion coefficient and to determine the associated eigenfunctions and eigenvalues that satisfy the imposed boundary conditions  \cite{schulz1974, OBRIEN20081738, ALBERT_diffusion_solution}. In the present context, however, our aim is not to obtain a detailed spectral solution for a particular diffusion model. Rather, we seek to address a simpler and more general question: whether the inclusion of pitch-angle diffusion can increase the effective lifetime of a spatially localised injection or drift echo. The only assumption we require is that the diffusion coefficient be positive definite, namely $\langle D_{\alpha\alpha}\rangle_b>0$, a condition that must hold irrespective of the specific nature of the underlying wave--particle interactions.

Before proceeding with the proof, we introduce a weighted inner product over the interval 
$[\alpha_{LC},\pi/2]$,
\begin{equation}
\langle F,G\rangle
=
\int_{\alpha_{LC}}^{\pi/2} F(\alpha_{\rm eq})\,G^*(\alpha_{\rm eq})\,\sin\alpha_{\rm eq}\,d\alpha_{\rm eq},
\qquad
\|F\|^2 = \langle F,F\rangle.
\label{eq:weighted_inner_product}
\end{equation}
This plays a role analogous to the inner product used in quantum mechanics to define the norm of a wavefunction. The weight $\sin\alpha_{\rm eq}$ arises naturally from the pitch-angle coordinate and is also associated with the Sturm–Liouville structure of the diffusive operator. 

Consider an arbitrary eigenmode of the form 
$F(\alpha_{\rm eq},t)=\Phi(\alpha_{\rm eq})\,e^{\Lambda t}$ 
satisfying Equation~(\ref{eq:eigenproblem_alpha}) together with the boundary conditions 
(\ref{eq:bc_absorbing_lc})--(\ref{eq:bc_equatorial_symmetry}). 
Writing the eigenvalue in terms of a real and imaginary part, i.e., $\Lambda=\Lambda_R+i\Lambda_I$, the weighted norm evolves as
\begin{align}
\|F\|^2
&=
\int_{\alpha_{LC}}^{\pi/2}
|F(\alpha_{\rm eq},t)|^2\,
\sin\alpha_{\rm eq}\,d\alpha_{\rm eq} \nonumber \\
&=
\int_{\alpha_{LC}}^{\pi/2}
|\Phi(\alpha_{\rm eq})|^2
\,e^{(\Lambda+\Lambda^*)t}\,
\sin\alpha_{\rm eq}\,d\alpha_{\rm eq} \nonumber \\
&=
e^{2\Lambda_R t}
\int_{\alpha_{LC}}^{\pi/2}
\Phi(\alpha_{\rm eq})\,\Phi^*(\alpha_{\rm eq})\,
\sin\alpha_{\rm eq}\,d\alpha_{\rm eq}
=
e^{2\Lambda_R t}\,\|\Phi\|^2 .
\end{align}
It follows that
\begin{equation}
\frac{d}{dt}\|F\|^2
=
2\Lambda_R\,e^{2\Lambda_R t}\,\|\Phi\|^2 .
\label{eq:eigen_norm_growth}
\end{equation}
Equation~(\ref{eq:eigen_norm_growth}) shows that the sign of $\Lambda_R$ is determined by the temporal behaviour of the weighted norm: if $\frac{d}{dt}\|F\|^2\le 0$ for all $t$, then necessarily $\Lambda_R\le 0$. We now extract $\frac{d}{dt}\|F\|^2$ directly from the evolution equation, without assuming any particular form for $\langle D_{\alpha\alpha}\rangle_b$ beyond positivity. Using the inner product \eqref{eq:weighted_inner_product},
\begin{equation}
\frac{d}{dt}\|F\|^2
=
\frac{d}{dt}\langle F,F\rangle
=
\langle \partial_t F,F\rangle+\langle F,\partial_t F\rangle
=
2\,\operatorname{Re}\,\langle F,\partial_t F\rangle.
\label{eq:norm_identity}
\end{equation}
Substituting \eqref{EQ:bounced_average_full_2} into \eqref{eq:norm_identity} yields
\begin{align}
\frac{d}{dt}\|F\|^2
&=
2\,\operatorname{Re}
\int_{\alpha_{LC}}^{\pi/2}
F^*(\alpha_{\rm eq},t)
\Bigg[
\left(im\langle\dot{\varphi}\rangle_b-\nu\right)F
\nonumber\\
&\qquad\qquad
+\frac{1}{\sin\alpha_{\rm eq}}
\frac{\partial}{\partial\alpha_{\rm eq}}
\left(
\langle D_{\alpha\alpha}\rangle_b
\,\sin\alpha_{\rm eq}\,
\frac{\partial F}{\partial\alpha_{\rm eq}}
\right)
\Bigg]
\sin\alpha_{\rm eq}\,d\alpha_{\rm eq}
\nonumber\\
&=
2\,\operatorname{Re}\Bigg\{
im\int_{\alpha_{LC}}^{\pi/2}
\langle\dot{\varphi}\rangle_b\,|F|^2
\,\sin\alpha_{\rm eq}\,d\alpha_{\rm eq}
\nonumber\\
&\qquad
-\int_{\alpha_{LC}}^{\pi/2}
\nu\,|F|^2
\,\sin\alpha_{\rm eq}\,d\alpha_{\rm eq}
\nonumber\\
&\qquad
+\int_{\alpha_{LC}}^{\pi/2}
F^*\,
\frac{\partial}{\partial\alpha_{\rm eq}}
\left(
\langle D_{\alpha\alpha}\rangle_b
\,\sin\alpha_{\rm eq}\,
\frac{\partial F}{\partial\alpha_{\rm eq}}
\right)
d\alpha_{\rm eq}
\Bigg\}.
\label{eq:energy_step1}
\end{align}

The first term in \eqref{eq:energy_step1} is purely imaginary and therefore drops out upon taking the real part. Integrating the last term by parts gives
\begin{eqnarray}
\int_{\alpha_{LC}}^{\pi/2}
F^*\,
\frac{\partial}{\partial\alpha_{\rm eq}}
\left(
\langle D_{\alpha\alpha}\rangle_b\,\sin\alpha_{\rm eq}\,
\frac{\partial F}{\partial\alpha_{\rm eq}}
\right)
d\alpha_{\rm eq}
&=&
\left[
F^*\,
\langle D_{\alpha\alpha}\rangle_b\,\sin\alpha_{\rm eq}\,
\frac{\partial F}{\partial\alpha_{\rm eq}}
\right]_{\alpha_{LC}}^{\pi/2} \nonumber \\
&-&
\int_{\alpha_{LC}}^{\pi/2}
\langle D_{\alpha\alpha}\rangle_b\,\sin\alpha_{\rm eq}\,
\left|\frac{\partial F}{\partial\alpha_{\rm eq}}\right|^2
d\alpha_{\rm eq} \nonumber.
\label{eq:energy_step2}
\end{eqnarray}
The boundary term in the above equation vanishes under \eqref{eq:bc_absorbing_lc} and \eqref{eq:bc_equatorial_symmetry}. Hence,
\begin{equation}
\frac{d}{dt}\|F\|^2
=
-2\int_{\alpha_{LC}}^{\pi/2}
\nu\,|F|^2\,\sin\alpha_{\rm eq}\,d\alpha_{\rm eq}
-
2\int_{\alpha_{LC}}^{\pi/2}
\langle D_{\alpha\alpha}\rangle_b\,\sin\alpha_{\rm eq}\,
\left|\frac{\partial F}{\partial\alpha_{\rm eq}}\right|^2
d\alpha_{\rm eq}
\;\le\;0.
\label{eq:energy_identity_final}
\end{equation}

Finally, for the eigenmode $F=\Phi e^{\Lambda t}$, combining \eqref{eq:energy_identity_final} with \eqref{eq:eigen_norm_growth} yields
\begin{equation}
\Lambda_R\,\|\Phi\|^2
=
-\int_{\alpha_{LC}}^{\pi/2}
\nu\,|\Phi|^2\,\sin\alpha_{\rm eq}\,d\alpha_{\rm eq}
-
\int_{\alpha_{LC}}^{\pi/2}
\langle D_{\alpha\alpha}\rangle_b\,\sin\alpha_{\rm eq}\,
\left|\frac{d\Phi}{d\alpha_{\rm eq}}\right|^2
d\alpha_{\rm eq},
\label{eq:proof_lambda_negative}
\end{equation}
and since $|\Phi|^2>0$ for any non-trivial eigenfunction, it follows that $\Lambda_R\le 0$, with strict inequality whenever either $\nu>0$ or $d\Phi/d\alpha_{\rm eq}\not\equiv 0$. Therefore pitch-angle diffusion cannot introduce exponentially growing modes and cannot enhance the effective lifetime of drift-phase structures, nor compensate for the decorrelation produced by phase mixing we determined in Section \ref{sec3}. To the contrary, the presence of pitch-angle diffusion with or without absorbing boundary would introduce an additional decorrelation factor proportional to $\exp(-|\Lambda_R| \tau)$ that would become significant on timescales comparable to the pitch-angle diffusion characteristic timescale \cite{schulz1974, ALBERT_diffusion_solution}.

\section{Zero-sweeping limit of the autocorrelation function}
\label{app:Vs_zero_limit}

We now consider the limit $V_s\longrightarrow 0$, corresponding to a spacecraft fixed on a single drift shell. In this limit, Equation ~\eqref{eq:Cbar_closed_L0b_simpler} appears to take the indeterminate form \(0/0\), since both the numerator through \(\Delta_m(\tau;T)\) and the denominator through the error function vanish linearly with \(V_s\). This can be evaluated either by L’Hôpital’s rule or, equivalently, by expanding the numerator and denominator to leading order in the limit $V_s\longrightarrow 0$. We adopt the latter approach,
Starting from Equation~\eqref{eq:Cbar_closed_L0b_simpler} 
\begin{equation}
{C}(\tau;T)
=
\exp\!\left[-\frac{V_s^2\tau^2}{8\sigma^2}\right]\,
\frac{T}{T-\tau}\,
\frac{
\displaystyle \sum_m |\psi_m(\kappa)|^2\,e^{-i m A b\,\tau}\,
e^{-2 m^2 A^2\sigma^2\tau^2}\,
\Delta_m(\tau;T)
}{
\displaystyle \operatorname{erf}\!\left(\frac{V_s T}{\sqrt{2}\sigma}\right)
\sum_m |\psi_m(\kappa)|^2
},
\label{eq:C_app_start}
\end{equation}
with
\begin{equation}
\Delta_m(\tau;T)=
\operatorname{erf}\!\left(
\frac{V_s}{\sqrt{2}\sigma}\left(T-\frac{\tau}{2}\right)
+i\,\sqrt{2}\,mA\sigma\,\tau
\right)
-
\operatorname{erf}\!\left(
\frac{V_s\tau}{2\sqrt{2}\sigma}
+i\,\sqrt{2}\,mA\sigma\,\tau
\right),
\label{eq:Delta_app_start}
\end{equation}
we introduce the shorthand $ z_m = i\,\sqrt{2}\,mA\sigma\,\tau$.
Then Eq.~\eqref{eq:Delta_app_start} becomes
\[
\Delta_m(\tau;T)
=
\operatorname{erf}(z_m+\varepsilon_1)
-
\operatorname{erf}(z_m+\varepsilon_2),
\]
where
\[
\varepsilon_1=
\frac{V_s}{\sqrt{2}\sigma}\left(T-\frac{\tau}{2}\right),
\qquad
\varepsilon_2=
\frac{V_s\tau}{2\sqrt{2}\sigma}.
\]
For \(V_s\longrightarrow 0\), both \(\varepsilon_1\) and \(\varepsilon_2\) are small, so a first-order Taylor expansion of the error function yields
\[
\operatorname{erf}(z+\varepsilon)
=
\operatorname{erf}(z)
+
\frac{2}{\sqrt{\pi}}e^{-z^2}\varepsilon
+
\mathcal{O}(\varepsilon^2).
\]
Applying this to \(\Delta_m\) gives
\[
\Delta_m(\tau;T)
=
\frac{2}{\sqrt{\pi}}e^{-z_m^2}(\varepsilon_1-\varepsilon_2)
+\mathcal{O}(V_s^2).
\]
Since
\[
\varepsilon_1-\varepsilon_2
=
\frac{V_s}{\sqrt{2}\sigma}(T-\tau),
\]
we obtain
\begin{equation}
\Delta_m(\tau;T)
=
\frac{2}{\sqrt{\pi}}
e^{-z_m^2}
\frac{V_s}{\sqrt{2}\sigma}(T-\tau)
+\mathcal{O}(V_s^2).
\label{eq:Delta_smallVs}
\end{equation}
Using \(z_m=i\sqrt{2}\,mA\sigma\,\tau\), one has
\[
-z_m^2 = 2m^2A^2\sigma^2\tau^2,
\]
so Eq.~\eqref{eq:Delta_smallVs} may be written as
\begin{equation}
\Delta_m(\tau;T)
=
\frac{2}{\sqrt{\pi}}
e^{2m^2A^2\sigma^2\tau^2}
\frac{V_s}{\sqrt{2}\sigma}(T-\tau)
+\mathcal{O}(V_s^2).
\label{eq:Delta_smallVs_explicit}
\end{equation}

The denominator of Eq.~\eqref{eq:C_app_start} is treated similarly. For small argument,
\[
\operatorname{erf}(x)
=
\frac{2}{\sqrt{\pi}}x+\mathcal{O}(x^3),
\]
hence
\begin{equation}
\operatorname{erf}\!\left(\frac{V_s T}{\sqrt{2}\sigma}\right)
=
\frac{2}{\sqrt{\pi}}
\frac{V_sT}{\sqrt{2}\sigma}
+\mathcal{O}(V_s^3).
\label{eq:erf_smallVs}
\end{equation}

Substituting Eqs.~\eqref{eq:Delta_smallVs_explicit} and \eqref{eq:erf_smallVs} into Eq.~\eqref{eq:C_app_start}, the factors proportional to \(V_s\) cancel, and one finds
\begin{equation}
  \lim_{V_s\longrightarrow0}  \frac{T}{T-\tau}
\frac{\Delta_m(\tau;T)}
{\operatorname{erf}\!\left(\frac{V_s T}{\sqrt{2}\sigma}\right)}
\longrightarrow
e^{2m^2A^2\sigma^2\tau^2}.
\end{equation}
This exactly cancels the factor \(e^{-2m^2A^2\sigma^2\tau^2}\) already present in the numerator of Eq.~\eqref{eq:C_app_start}. Moreover, $\exp\!\left[-\frac{V_s^2\tau^2}{8\sigma^2}\right]\longrightarrow 1$ as
$V_s\longrightarrow 0.$ Therefore the normalised autocorrelation has the finite limit
\begin{equation}
\lim_{V_s\longrightarrow 0} C(\tau;T)
=
\frac{
\displaystyle \sum_m |\psi_m(\kappa)|^2\,e^{-imAb\tau}
}{
\displaystyle \sum_m |\psi_m(\kappa)|^2
}.
\label{eq:C_Vs0_final}
\end{equation}

Equation~\eqref{eq:C_Vs0_final} is independent of the observation window \(T\). Physically, when the spacecraft no longer sweeps across neighbouring drift shells, the additional decorrelation induced by shell-to-shell phase differences disappears entirely, and the autocorrelation reduces to the weighted phase average associated with a single drift shell.

\nocite{Hamlin1961,Grach_2022,Mourenas_2024, banos_1967}
\bibliography{apssamp}

\end{document}